\documentclass[12pt,letterpaper,usenames,dvipsnames]{article}
\usepackage{jheppub}
\usepackage{enumitem}

\allowdisplaybreaks[2]  %allow page breaks in multi-line eqns, [1]=as little as poss; ... [4]=most permissive
\usepackage{pifont}% http://ctan.org/pkg/pifont
\usepackage{bm} %bold math=
\usepackage{mathrsfs}   % fancy script letters = \mathscr
\usepackage{dsfont}  %double struck math = \mathds
\usepackage{ytableau}   % Young diagrams
       
\usepackage{float}
\usepackage{hhline}
\usepackage{mathtools}
\usepackage{multirow} % multiple row elements in a table

\graphicspath{{figures/}}	 % set directory for figures
\usepackage{tikz}
\usetikzlibrary{arrows,calc,snakes,shapes,shapes.arrows,
decorations.markings,decorations.pathmorphing}
\newcounter{sarrow}

\tikzset{>=triangle 45}
\tikzstyle{gr}=[draw,circle,green!50!black,fill=green!50!black,scale=.6]
\usepackage{tikz-3dplot}

\usepackage{array} % more control over arrays and tables
\usepackage{multirow} % multiple row elements in a table
\usepackage{colortbl} % colored rows and columns in tables
\usepackage{arydshln} % dashed lines in arrays and tables
\usepackage{stfloats}  % ??
\usepackage{tabstackengine}% controls width of matrix columns
    \setstacktabbedgap{1ex} % sets column widths to be the same
\usepackage{outlines} % easy nested lists
\usepackage{amsthm} % to have unnumbered theorem environments.
\newtheorem{definition}{Definition}
\newtheorem{conjecture}{Conjecture}
\newtheorem{result}{Result}

\usepackage{xcolor}   % \color{...}, colored text
\def\ccg{\cellcolor{green!07}}

\def\ccy{\cellcolor{yellow!20}}
\def\ccr{\cellcolor{red!07}}

\def\rcg{\rowcolor{green!07}}

\def\bar{\overline}
\def\til{\widetilde}
\def\hat{\widehat}
\def\del{{\partial}}
\def\delb{{\bar\del}}
 
\def\vev#1{{\langle{#1}\rangle}} 
\newcommand{\beq}{\begin{equation}}
\newcommand{\eeq}{\end{equation}}
\newcommand{\bpm}{\begin{pmatrix}}
\newcommand{\epm}{\end{pmatrix}}
\newcommand{\bsm}{\begin{smallmatrix}}
\newcommand{\esm}{\end{smallmatrix}}
\newcommand{\bspm}{\left(\begin{smallmatrix}}
\newcommand{\espm}{\end{smallmatrix}\right)}
\def\nn{\nonumber}

\def\^{\wedge}
\def\I{\mathds{1}}
\def\Tr{{\rm\, Tr}}
\def\Im{{\rm\, Im}}

\def\U{{\rm U}}
\def\Uot{{\til{\U(1)}}}
\def\SU{{\rm SU}}
\def\O{{\rm O}}
\def\SO{{\rm SO}}
\def\SL{{\rm SL}}
\def\GL{{\rm GL}} 
\def\Sp{{\rm Sp}}

\def\spdnz{{\Sp_\DD(2n,\Z)}}
\def\suf{\mathfrak{su}}

\def\uf{\mathfrak{u}}

\def\C{\mathbb{C}} 

\def\DD{\mathbb{D}} 
\def\H{\mathbb{H}}
\def\J{\mathbb{J}}
 
\def\P{\mathbb{P}}

\def\R{\mathbb{R}} 

\def\Z{\mathbb{Z}}

\def\ba{{\bf a}}
\def\ab{{\bar a}}
\def\bA{{\bf A}}

\def\Bb{{\bar B}}

\def\Qb{{\bar Q}}

\def\bR{{\bf R}}

\def\bZ{{\bf Z}}

\def\cC{{\mathcal C}}

\def\cF{{\mathcal F}}
\def\cH{{\mathcal H}}
\def\cI{{\mathcal I}}

\def\cL{{\mathcal L}}
\def\cM{{\mathcal M}}
\def\cN{{\mathcal N}}

\def\cS{{\mathcal S}}
\def\cT{{\mathcal T}}
\def\cV{{\mathcal V}}

\def\a{{\alpha}}

\def\bal{{\boldsymbol\a}}

\def\b{{\beta}}

\def\g{{\gamma}}
\def\G{{\Gamma}}
\def\d{{\delta}}
\def\D{{\Delta}}
\def\e{{\epsilon}}

\def\z{{\zeta}}

\def\l{{\lambda}}
\def\bl{{\boldsymbol\l}}
\def\L{{\Lambda}}
\def\m{{\mu}}

\def\x{{\xi}}
\def\r{{\rho}}

\def\S{{\Sigma}}
\def\t{{\tau}}

\def\vf{{\varphi}}

\def\w{{\omega}}

\title{On the moduli spaces of 4d $\cN=3$ SCFTs I: triple special K\"ahler structure}
\author[1]{Philip C. Argyres,}
\author[2]{Antoine Bourget,}
\author[3]{Mario Martone}
\affiliation[1]{University of Cincinnati,
Physics Department, Cincinnati OH 45221}
\affiliation[2]{Theoretical Physics Group, The Blackett Laboratory, Imperial College London, Prince Consort
Road London, SW7 2AZ, UK}
\affiliation[3]{University of Texas, Austin, Physics Department, Austin TX 78712}
\emailAdd{philip.argyres@gmail.com}
\emailAdd{a.bourget@imperial.ac.uk}
\emailAdd{mariomartone@utexas.edu}

\abstract{We initiate a systematic analysis of moduli spaces of vacua of four dimensional $\cN=3$ SCFTs. Our analysis is based on the one hand on the properties of $\cN=3$ chiral rings --- which we review in detail and contrast with chiral rings of theories with less supersymmetry --- and on the other hand on constraints coming from low-energy supersymmetry. This leads us to introduce a new type of geometric structure, which characterizes $\cN=3$ SCFT moduli spaces, and that we call \emph{triple special K\"ahler} (TSK). A rank-$n$ TSK moduli space has complex dimension $3n$, and is singular at complex co-dimension 3 subspaces where charged states become massless. The structure of singularities defines a stratification of the TSK space in terms of lower-dimensional TSK manifolds.}

\preprint{Imperial/TP/2019/AB/03}
  
\begin{document}
\maketitle 

\section{Motivation, summary, and open questions}\label{sec1}

Supersymmetric theories have been extensively studied in the past four decades. This has partly been because their enhanced symmetries allow exact computations which provide useful windows into strong coupling physics.  Conformal invariance further enhances the ability to perform exact calculations.  This has resulted in great recent progress in understanding the properties of, in particular, four dimensional superconformal field theories (SCFTs) with various amount of supersymmetry. 

Supersymmetry and conformal invariance also equip the moduli space of vacua of these theories with rigid geometric structures \cite{Seiberg:1997ax, Cecotti:2015wqa, Argyres:2018zay, Caorsi:2018zsq}. The geometric structure becomes increasingly constraining as the number of supersymmetry charges increases.  Depending on the number of supercharges, this suggests the possibility of a systematic ``bottom-up'' classification of 4d SCFTs by constructing all their possible moduli space geometries.

In this and a follow up \cite{PAMN3II} paper, we will discuss this program in the case of $\cN=3$ SCFTs.  Our aim is to assemble various results on $\cN=3$ SCFTs and their moduli spaces from the literature, precisely formulate the mathematical properties of the moduli space of vacua of $\cN=3$ SCFTs, and critically assess the prospects for carrying out a classification of all such possible moduli space geometries. In this paper we will analyze the relations between these moduli spaces and properties of operators at the conformal vacuum, and introduce the notion of TSK manifolds and its generalization to singular TSK spaces. Before summarizing these results, we pause to motivate the study of $\cN=3$ SCFTs.

For $\cN=1$ supersymmetry (four supercharges), such an approach is not particularly fruitful, basically because the moduli spaces are constrained only to be (singular) K\"ahler spaces, which is too large a class of geometries to get a useful handle on.  Furthermore, upon deformation by relevant $\cN=1$-preserving operators, the moduli spaces of these SCFTs are generically lifted, leaving only discrete vacua (where often the supersymmetry is spontaneously broken). 
Nevertheless matching of the moduli spaces of different $\cN=1$ theories has been a very valuable tool to argue for non trivial $\cN=1$ dualities \cite{Seiberg:1994bz, Seiberg:1997ax}. 

By contrast, the moduli spaces of $\cN=2$ SCFTs are much more tightly constrained.  All known examples have continuous moduli spaces of vacua which are combinations of singular hyperk\"ahler (Higgs branch, HB for short) or special K\"ahler (Coulomb branch, CB for short) geometries, and upon deformation the CB is not lifted, but instead is deformed.  Also HB and CB geometries are much more rigid than $\cN=1$ K\"ahler geometries as they are closely related to certain algebraic structures. Indeed, since the groundbreaking work of Seiberg and Witten \cite{Seiberg:1994rs, Seiberg:1994aj}, large portions of the landscape of moduli spaces of $\cN=2$ SCFTs have been illuminated by algebraic techniques.  Citing only a selection of the most recent efforts:  systematic studies of three dimensional rational Gorenstein graded isolated singularities which give rise, via geometric engineering, to consistent $\cN=2$ SCFT Coulomb branches \cite{Xie:2015rpa,Chen:2016bzh,Wang:2016yha,Chen:2017wkw}, a beautiful algebraic characterization of the geometry of the Higgs branches of $\cN=2$ theories \cite{Beem:2017ooy}, a successful program of systematic classification of rank-1 $\cN=2$ SCFTs through the study of allowed Coulomb branches and their deformations \cite{Argyres:2015ffa, Argyres:2015gha, Argyres:2016xua, Argyres:2016xmc, Caorsi:2018ahl}.  These efforts have brought remarkable progress in our understanding of these theories.  For instance, it is now known that the operators parametrizing Coulomb branches which have no complex singularities can only have rational scaling dimensions, and for a given rank only a finite set of possibilities are allowed \cite{Argyres:2018zay, Caorsi:2018zsq, Argyres:2018urp}.  It has also been established, contrary to an earlier conjecture, that the Coulomb branch of $\cN=2$ theories can have complex structure singularities and not just metric ones \cite{Bourget:2018ond, Argyres:2018wxu}.  

Despite this remarkable progress, the space of $\cN=2$ SCFTs is very large \cite{Chacaltana:2015bna, Chacaltana:2017boe, Chacaltana:2018zag} and it is clear that it is still largely unexplored.  A systematic classification of all possible $\cN=2$ CB geometries requires new insights possibly tying together CB and HB constraints.   In fact, even at rank 2, it is not clear that the rules for what kinds of metric singularities on the CB are physically allowed are currently known.

Before discussing the $\cN=3$ case, let us briefly mention a few words about the moduli space of $\cN=4$ SCFTs. This case is extremely constrained. In fact it is believed that the space of allowed $\cN=4$ theories is filled by the Lagrangian theories \cite{Seiberg:1997ax} and their discrete gaugings \cite{Bourget:2018ond, Argyres:2018wxu} along with a choice of line operators \cite{Aharony:2013hda}. It is worth mentioning that the list of $\cN=2$ rank-1 Coulomb branch geometries contains two entries corresponding to $\cN=4$ theories \cite{Argyres:2015gha, Argyres:2016yzz, Caorsi:2019vex}. While this matter has not been settled completely yet, there is increasing evidence that one of the two geometries corresponds to the ``standard'' $\SU(2)$ $\cN=4$ theory while the other to an $\cN=4$ relative theory with a non-maximal spectrum of line operators \cite{Caorsi:2019vex}. 

We thus arrive at the moduli spaces of $\cN=3$ SCFTs \cite{Ferrara:1998zt, Aharony:2015oyb, Garcia-Etxebarria:2015wns, Aharony:2016kai, Nishinaka:2016hbw, Garcia-Etxebarria:2016erx, Lemos:2016xke, Bourton:2018jwb, Bonetti:2018fqz}.  As we will review below, a rank-$n$ theory has a $3n$-complex-dimensional moduli space which is a Coulomb branch. It has a flat generalization of a special K\"ahler structure which we call a ``triple special K\"ahler'' (TSK) structure. It contains a $\C\P^2$ of inequivalent complex structures compatible with the metric are induced by the $\cN=3$ supersymmetry.  Surprisingly, it also has an additional isolated `special'' complex structure. The global geometry of this moduli space is tightly constrained by the $\cN=3$ superconformal algebra, mainly through its symmetries which are spontaneously broken on the moduli space.  These imply that $\cM$ is a non-compact, metrically complete collection of cones with a common tip, and each cone is TSK with non-analyticities in complex co-dimension 3. We will provide a detailed description of this structure below. Additional restrictions on $\cN=3$ moduli spaces could follow from associativity (crossing symmetry) of the $\cN=3$ superconformal local operator algebra.  Preliminary studies of these constraints have been carried out in \cite{Nishinaka:2016hbw, Lemos:2016xke, Cornagliotto:2017dup, Bonetti:2018fqz}.  But these studies have not identified any additional restriction on $\cN=3$ moduli space geometries beyond those discussed below.

It is natural to guess that TSK spaces have an orbifold structure $\cM\cong\C^{3r}/\G$ for some discrete group $\G$ acting linearly on $\C^{3r}$. Somewhat surprisingly, in \cite{PAMN3II} we will show that TSK structure alone is not enough to enforce this conclusion and in fact we will discuss some counter-examples. Furthermore we will discuss that even if one assumes the orbifold structure then other assumptions need to be made to fully characterize the type of finite group $\G$ acting on the affine space. Under these assumptions $\G$ is constrained to be a crystallographic complex reflection group (CCRG) admitting a principal polarization \cite{Caorsi:2018zsq, Argyres:2019ngz, PAMN3II} and a complete classification is possible. The work in \cite{PAMN3II} will also address how to lift the extra assumption and attempt a systematic analysis of the orbifold geometries.

The paper is organized as follows. In the next two sections we systematically analyze the connection between the geometry of the moduli space of vacua and the algebra of CFT operators which can have a non-vanishing vacuum expectation value. The latter are constrained by the (unitary, positive energy) representation theory of the $\cN=3$ superconformal algebra. Section \ref{sec2.3} sets some groundwork by describing the situation for $\cN\leq2$ while section \ref{sec2.3.3} discusses in detail the $\cN=3$ case as well as the extra structures that appear in the $\cN=4$ case. Section \ref{sec2.1} starts the systematic analysis of the conditions on the moduli space geometry of an $\cN=3$ SCFT implied by the assumption of unbroken $\cN=3$ supersymmetry. This sets the stage for section \ref{sec2-TSK} where the definition of a \emph{TSK manifold} is provided and the properties that follow are studied in detail. We conclude in section \ref{TSKspace} by discussing the singularity structure of the moduli space of $\cN=3$ SCFTs which will lead to a definition of a \emph{TSK space}. We conjecture that it applies to any moduli space of vacua with unbroken $\cN=3$ supersymmetry.

\section{Chiral rings and moduli spaces of vacua}\label{sec2.3}

We start by analyzing the structure of the $\cN=3$ moduli space of vacua from the perspective of the SCFT operator algebra. This analysis does not capture all of the metric structure of the moduli space and can therefore be considered to be coarser than the one which will lead to the definition of the \emph{triple special K\"ahler} (TSK) structure below. But the ring structure that certain operators (conjecturally) need to satisfy to acquire a vacuum expectation value captures information on the singularities of the moduli space as a complex variety which are difficult to access from the TSK perspective. Thus both are useful for characterizing the moduli space geometry.

The relation between the structure of the operator algebra and the moduli space of vacua of four dimensional SCFTs seems to be special. In particular, as we will review shortly, the problem of when a given operator can acquire a vev can be conjecturally formulated in terms of a set of precise conditions on the operator algebra in the four dimensional case. These heavily rely on the complex structure that these moduli spaces inherit by virtue of supersymmetry, and, relatedly, on the shortening conditions satisfied by the BPS operators whose vevs parametrize the space. In five ($\cN=1$) and six ($\cN=(1,0)$) dimensions there are well-known examples of moduli space of vacua parameterized by real coordinates which carry no action of the R-symmetry group \cite{Seiberg:1996bd, Seiberg:1996qx, Intriligator:1997pq, Hanany:1997gh}. These are clearly counter-examples to the set of conjectural conditions that we summarize below. More recently, it was pointed out that in three dimensions with $\cN=1$ supersymmetry, time reversal symmetry can ensure the existence of real moduli space of vacua whose corresponding operators neither belong to short multiplets nor satisfy chiral ring relations \cite{Gaiotto:2018yjh}. 

Another feature that singles out four dimensions from the rest is the fact that the gauge coupling is dimensionless. This in turns allows us to study a large class of SCFTs reliably at arbitrarily weak coupling and so to make precise statements relating the SCFT operator algebra to the coordinate ring of the moduli space.

In the following we will first review some of the issues involved in making a connection between the local operator content of four-dimensional CFTs and their moduli space geometries. The goal is both to introduce the central notion of chiral rings as a, conjecturally, necessary and sufficient condition for the existence of a moduli space of vacua in four dimensions as well as to describe the structures on chiral rings of $\cN<3$ theories. The $\cN=3$ supersymmetric case will be analyzed in the next section. There, elaborating on the material presented in this section, we will highlight the new structures which arise with $\cN=3$ supersymmetry.

\subsection{Chiral ring generalities}\label{sec2.3.1}

We start with the empirical observation that all known examples of moduli spaces in 4d CFTs are associated with the existence of supersymmetry and of chiral rings. With this in mind, in this section we make an attempt at singling out a minimal set of assumptions for the existence of such moduli spaces. Note that supersymmetry is not part of the assumptions; see section \ref{sectionN=0}. This leads to a collection of results and conjectures, which are mostly known to experts but do not necessarily appear gathered in a single place in the literature.  Concerning notations, we indicate the (super)translationally invariant vacua with the generic letter $u$ and the set of all allowed such values, that is the moduli space of vacua, with $\cM$. When we refer to the value of the vev of a specific operator $\vf_a$ at $u$ we will use the shorthand notation 
\begin{equation}
\label{notationua}
    u_a :=\langle \vf_a \rangle_u \,. 
\end{equation}

We now state our main assumptions. Throughout this paper, we consider CFTs in four dimensions which satisfy two conditions: 
\begin{enumerate}
    \item[\textbf{A.} ] The theory has a $\U(1)$ conserved charge $R$ such that the scaling dimension $\D$ of every local operator in the CFT satisfies
\begin{equation}\label{alg0.2}
\D\ge |R| \, . 
\end{equation}
We denote generically by $\varphi_a$ the operators that saturate the bound with $\D = R$ and call them \emph{chiral}, and $\bar{\varphi}_a$ those which saturate the bound with $\D = -R$, and call them \emph{anti-chiral}. 
\item[\textbf{B.} ] In any (super)translationally invariant vacuum $u$, the vevs of products of chiral operators which can acquire a vev are independent of their spatial separation
\begin{equation}\label{alg0.1}
\del_x \vev{\vf_a (x) \vf_b(0)}_u = 0 \, , 
\end{equation}
and similarly for anti-chiral operators. 
\end{enumerate}

A crucial consequence of these assumptions is that the operators which saturate the bound \eqref{alg0.2} $\varphi_a$ satisfy an operator product expansion (OPE) of the form
\begin{equation}\label{alg0}
\vf_a (x) \vf_b(0) \sim \sum\limits_c C^c_{ab}\ \vf_c(0) + \text{(regular terms vanishing as $x\to0$)} \, . 
\end{equation} 
Indeed $\U(1)$ charge conservation implies their OPEs have no singular terms as their space-time separation vanishes, $x\to0$, and that the leading terms (independent of $x$) are other such primaries. The regular terms in \eqref{alg0} can either be descendants or conformal primaries which do not saturate \eqref{alg0.2}. In superconformal theories, the $\vf_a$ are usually components of superconformal primaries of short multiplets with positive $\U(1)$ charge. The shortening condition is a consequence of the fact that they saturate the bound \eqref{alg0.2} (more below). This is the reason for calling them the \emph{chiral} operators. Similarly, the $\bar{\varphi}_a$ operators satisfy an analogous OPE and set of shortening conditions, thus the name \emph{anti-chiral}.

The limit of the algebra \eqref{alg0} as $x\to0$ gives
\begin{equation}\label{ring0}
\vf_a \vf_b = \sum\limits_c C^c_{ab}\ \vf_c.
\end{equation}
This defines a ring structure, called the \emph{chiral ring}, and this leads to the following
\begin{conjecture}
The set of chiral operators $\vf_a$ which can acquire a vev, form an infinite basis of a ring thought of as a vector space over $\C$. \eqref{ring0} are an infinite set of relations defining the ring product.  The $C^c_{ab}$ are constrained by graded commutativity and associativity of the ring product.
\end{conjecture}

\begin{conjecture}
The set of all operators which can acquire a vev are products of chiral and anti-chiral operators, and they satisfy no further relation in addition to the aforementioned chiral ring relations. 
\end{conjecture}

The OPE \eqref{alg0} is valid in general only in the conformal vacuum. Applying \eqref{alg0} inside an expectation value in a putative vacuum $0 \neq u \in \cM$ which breaks conformal invariance spontaneously, is a priori valid only if $|x|$ is much less than the smallest length scale associated to the vevs $\vev{\vf_a}_u := u_a$ at the $u \in\cM$ vacuum. This is where the condition \textbf{B}, see \eqref{alg0.1}, comes in: it implies that the vev of the left side of \eqref{alg0} at $u\in\cM$ is independent of $x$, and so cluster decomposition in the $|x|\to\infty$ limit implies $\vev{\vf_a(x)\vf_b(0)}_u = \vev{\vf_a}_u \vev{\vf_b}_u = u_a u_b$, using the notation (\ref{notationua}). Assigning non-zero complex vacuum expectation values $u_a \neq 0$ is consistent with the OPE algebra \eqref{alg0} as long as
\begin{conjecture}
The vevs $u_a\in\cM$ obey the holomorphic ring relations
\begin{align}\label{ring1}
u_a u_b = \sum\limits_c C^c_{ab} u_c ,
\end{align}
and any other ``regular'' superconformal primaries appearing on the right side of \eqref{alg0} are assigned zero vev.
\end{conjecture}  

Hermitian conjugation gives the anti-chiral ring relations
\begin{align}\label{ring2}
\bar\vf^a \bar\vf^b &= \sum\limits_c \bar C_c^{ab}\  \bar\vf^c, 
\end{align}
made up of operators satisfying $\D = -R$, and assigning consistent complex vevs gives the anti-holomorphic coordinate ring of $\cM$. The chiral and anti-chiral primary operators satisfy a non-trivial operator product algebra,
\begin{align}\label{alg1}
\vf_a (x) \bar\vf^b(0) &\sim \frac1{|x|^{2\D_b}}
\Bigl[ \textstyle{\sum_c} D_a^{bc}\ \vf_c(0)
 + \text{reg.} \Bigr] & &(\D_a \ge \D_b).
\end{align}
$\U(1)$ charge conservation implies that the only chiral primaries which can appear on the right side of \eqref{alg1} are those with dimension $\D_c = \D_a - \D_b$. 
It is known how to compute a subset of the $D^{bc}_a$ in $\cN=2$ gauge SCFTs, see for instance \cite{Baggio:2015vxa, Gerchkovitz:2016gxx}. But the $D_a^{bc}$ do not enter directly into the moduli space coordinate ring. This is because the $|x|\to0$ and $|x|\to\infty$ limits of vacuum expectation values of the left side of \eqref{alg1} --- at points on $\cM$ away from the superconformal vacuum --- are not related in any simple way; cf.\  \cite{Karananas:2017zrg}.

Another conjecture is that conditions \textbf{A} and \textbf{B} are also sufficient for flat directions corresponding to $\vev{\vf_a}$ to exist. That is, there exists a vacuum with spontaneously broken conformal invariance corresponding to every assignment $\vev{\vf_a}=u_a\in\C$ which satisfies the ring relations \eqref{ring1}.  Since the only consistent assignment of vev to a nilpotent element is zero, this sufficiency conjecture implies 
\begin{conjecture}
The reduced ring corresponding to \eqref{ring0} --- i.e., the quotient of the chiral ring by its nilradical --- is the holomorphic coordinate ring of $\cM$ in a particular complex structure.\footnote{In the presence of extended supersymmetry, different choices of complex structures are identified by different choices of the supercharge which annihilates the $\varphi_a$'s. Below we will explain this point in detail.} Explicitly
\beq\label{holoring}
\C\{\cM\}:=\C[u_1,...,u_n]/\cI(\cM)
\eeq
where $\cI(\cM):=\langle u_a u_b -\sum_c C^c_{ab}u_c\rangle$, is the ideal generated by the holomorphic ring relations \eqref{ring1}.
\end{conjecture}
\noindent It is important to the note that not all holomorphic coordinate rings of $\cM$ arise this way. In fact, we will find examples of holomorphic coordinate rings with respect to the ``special'' complex structure of a TSK space for which there does not correspond a chiral ring in the SCFT. This point will be discussed in detail below in section \ref{sec5.2.5-kahler}. 

The $\U(1)$ charges available for playing the role of $R$ in the $\D\ge |R|$ constraint in CFTs are $\U(1)$ factors in a Cartan subalgebra (maximal torus) of the compact bosonic symmetries, which are Lorentz rotations and R-symmetries.\footnote{Since flavor symmetries are not part of the superconformal symmetry, unitarity conditions do not bound dimensions in terms of flavor charges.  It is conceivable that constraints from crossing symmetry in specific theories could relate dimensions to flavor symmetry representations enforcing a relation like \eqref{alg0.2}.}  These $\U(1)$ charges are thus the weights of the Lorentz and R-symmetry irreducible representations (irreps) of the fields; we will present a systematic review below.  

Finally, the chiral operators $\vf_a$ appearing in \eqref{alg0} need not in principle be Lorentz scalars, and, indeed, in 4-dimensional SCFTs chiral operators with $[j_L>0,j_R=0]$ do occur. If such Lorentz non-singlets contribute to the reduced chiral ring, they can consistently be assigned non-zero constant vevs, thus spontaneously breaking Lorentz invariance without breaking supertranslation invariance. It is not known whether such moduli spaces can occur, though it is generally assumed that they do not. Lorentz invariance is automatically unbroken if Lorentz non-singlet chiral operators in the spectrum of local operators are in the nilradical of the chiral ring.  Otherwise, given our current knowledge, it would be necessary to impose ``by hand'' that this class of operators do not get vevs, and the sufficiency conjecture mentioned above would be violated.

\subsection{Chiral rings for $\cN<3$ CFTs}\label{sec2.3.2}

It is helpful to see this general discussion in action in the context of 4 dimensional CFTs with various amounts of supersymmetry. To this end let us  first review the constraints which come from superconformal representation theory \cite{Cordova:2016emh}.

The superconformal algebra in 4 dimensions with $\cN$ supersymmetries is $\suf(2,2|\cN)$ for $\cN\leq 3$ and $\mathfrak{psu}(2,2|4)$ for $\cN=4$. We focus here on the case $\cN\leq 3$. The bosonic subalgebra is 
\begin{equation}
    \suf(2,2|\cN) \supset \mathfrak{so}(4,2) \oplus \suf(\cN)_R \oplus \uf(1)_r .
\end{equation}
Here $\mathfrak{so}(4,2)$ is the conformal algebra, with maximal compact subalgebra $\suf(2)_{\mathrm{Left}} \oplus \suf(2)_{\mathrm{Right}} \oplus \mathfrak{so}(2)_{\Delta}$. A representation is specified by its Lorentz spins\footnote{We follow the conventions of \cite{Cordova:2016emh} and normalize the Lorentz spins to be integers, so that they are the Dynkin labels of the $\suf (2)_{\mathrm{Left}} \oplus \suf (2)_{\mathrm{Right}}$ irreps.} $[j_L,j_R]$, its conformal dimension $\D$, its $\suf(\cN)_R$ highest weight $\bR=(R_1 , \cdots , R_{\cN-1})$ and its $\uf(1)_r$ charge $r$, and, following \cite{Cordova:2016emh}, we denote it by 
\begin{equation}\label{eqA2}
    [j_L,j_R]_{\Delta}^{(\mathbf{R} , r)} \,. 
\end{equation}
Conjugation acts as 
\begin{equation}
   \left( [j_L,j_R]_{\Delta}^{(\mathbf{R} , r)} \right)^{\dagger} = [j_R,j_L]_{\Delta}^{(\overline{\mathbf{R}} , -r)} \,. 
\end{equation}
The Poincar\'e supercharges transform in two irreducible representations, 
\begin{equation}
\label{supercharges}
    Q \in [1,0]_{1/2}^{(\mathbf{fund}, -1)} \qquad  \Qb \in [0,1]_{1/2}^{(\overline{\mathbf{fund}}, +1)} \, ,  
\end{equation}
and similarly for the $S$-supercharges, 
\begin{equation}
\label{superchargesS}
    S \in [1,0]_{-1/2}^{(\overline{\mathbf{fund}}, +1)} \qquad  \bar{S} \in [0,1]_{-1/2}^{(\mathbf{fund}, -1)} \, .   
\end{equation}

For unitary SCFTs in four dimensions, operators $\vf_a$ saturate a $\D=R$ bound only if they are annihilated by at least one set of anti-chiral --- i.e., Lorentz labels $[j_L,j_R]=[0,1]$ --- supercharges, $\bar Q^{(I)}$. This implies that the superconformal multiplets to which the $\vf_a$ belong are special in that they satisfy shortening conditions. We will call these multiplets \emph{chiral multiplets} and, as already mentioned above, the $\vf_a$ \emph{chiral fields}. Furthermore we will call the set of $\vf_a$'s saturating the $\D=R$ bound and all annihilated by the same $\bar Q^{(I)}$ a set of \emph{co-chiral fields}. In a SCFT with extended supersymmetry the set of co-chiral fields depends on the choice of the supercharge $\bar Q^{(I)}$, $I=1,...,\cN$ defining the chiral ring.  The assumption that the vacua $u_a\in\cM$ do not spontaneously break the supertranslation symmetry implies that the vacua are annihilated by the supercharges.  This, together with the supertranslation algebra $\{Q^{(I)},\bar Q^{(I)}\} = P$, then implies \eqref{alg0.1} as a supersymmetric Ward identity and thus:
\begin{result}\label{Chiral}
Chiral rings appear naturally in 4d SCFTs and the operators $\vf_a$ satisfying \eqref{ring0} belong to a set of co-chiral fields. That is, they are the components of chiral (short) multiplets which are all annihilated by a chosen supercharge.
\end{result}

Supertranslation symmetry also implies that the regular terms on the right side of \eqref{alg0} are annihilated by the $\bar Q^{(I)}$, and since they are not chiral primaries, they must be $\bar Q^{(I)}$-exact, i.e., superconformal descendants of chiral primaries. By the assumed supertranslation invariance it follows that
\begin{result}
In any consistent vacua $u\in\cM$ superconformal descendants automatically have zero vev.
\end{result} 

It is worth clarifying the differences between the different choices of the supercharge defining the chiral ring. First notice that operator products of chiral multiplet primaries close on other chiral multiplets, but may have singular terms in their OPE, because, while the $\U(1)_r$ charges of multiplets are additive, the Dynkin labels (highest weights) of $\SU(\cN)_R$ irreps are not. This is another way to see what was stated in Result \ref{Chiral}, which is that only special primary components of chiral multiplets enjoy operator product algebras of the form \eqref{alg0} and thus form a chiral ring. These components are precisely those annihilated by the chosen supercharge defining the set of co-chiral fields.

\begin{table}[]
     \centering
     \begin{tabular}{|c||c|c|c|}
     \hline 
      \begin{tabular}{c}
   Shortening  \\ Conditions
\end{tabular} 
 &   \begin{tabular}{c}
    $\boldsymbol{\mathbf{\overline{{L}}}} $  \\ $ \Delta>2+\delta$
\end{tabular}  & 
\begin{tabular}{c}
    $\boldsymbol{\mathbf{\overline{{A}}}} $  \\ $ \Delta=2+\delta$
\end{tabular}  &  \ccy
\begin{tabular}{c}
    $\boldsymbol{\mathbf{\overline{{B}}}} $  \\ $ \Delta=\delta$ \\ Chiral 
\end{tabular}\\ \hline \hline
\begin{tabular}{c}
    $\mathbf{\boldsymbol L}$  \\ $\Delta>2+\overline{\delta}$
\end{tabular} & 
\begin{tabular}{c}
    $[j_L;j_R]_{\Delta}^{(\mathbf{R};r)}$ \\ \textcolor{blue}{$\mathcal{A}^{\Delta}_{\mathbf{R},r,(j_L;j_R)}$}
\end{tabular} & 
\begin{tabular}{c}
    $[j_L;j_R]_{\Delta}^{(\mathbf{R};r)}$ \\   $r > r_* $  \\  \textcolor{blue}{$\overline{\mathcal{C}}_{\mathbf{R},r,(j_L;j_R)}$}
\end{tabular} 
& \ccy
\begin{tabular}{c}
    $[j_L;0]_{\Delta}^{(\mathbf{R};r)}$  \\  $r > r_* + \frac{1}{\beta}$  \\ \textcolor{blue}{$\overline{\mathcal{B}}_{\mathbf{R},r,j_L}$}
\end{tabular} \\ \hline
\begin{tabular}{c}
    $\mathbf{\boldsymbol A}$  \\ $\Delta=2+\overline{\delta}$
\end{tabular} & 
\begin{tabular}{c}
    $[j_L;j_R]_{\Delta}^{(\mathbf{R};r)}$ \\$r < r_* $  \\ \textcolor{blue}{$\mathcal{C}_{\mathbf{R},r,(j_L;j_R)}$}
\end{tabular} & 
\begin{tabular}{c}
    $[j_L;j_R]_{\Delta}^{(\mathbf{R};r)}$ \\   $r = r_* $ \\ \textcolor{blue}{$\hat{\mathcal{C}}_{\mathbf{R},(j_L;j_R)}$}
\end{tabular} 
& \ccy
\begin{tabular}{c}
    $[j_L;0]_{\Delta}^{(\mathbf{R};r)}$  \\  $r = r_* + \frac{1}{\beta}$  \\ \textcolor{blue}{$\overline{\mathcal{D}}_{\mathbf{R},j_L}$}
\end{tabular} \\ \hline \ccg
\begin{tabular}{c}
    $\mathbf{\boldsymbol B}$  \\ $\Delta=\overline{\delta}$ \\ Antichiral 
\end{tabular} & \ccg
\begin{tabular}{c}
    $[0;j_R]_{\Delta}^{(\mathbf{R};r)}$ \\$r < r_* - \frac{1}{\beta} $ \\ \textcolor{blue}{$\mathcal{B}_{\mathbf{R},r,j_R}$}
\end{tabular} & \ccg
\begin{tabular}{c}
    $[0;j_R]_{\Delta}^{(\mathbf{R};r)}$ \\   $r = r_* - \frac{1}{\beta} $  \\ \textcolor{blue}{$\mathcal{D}_{\mathbf{R},j_R}$}
\end{tabular} 
& \ccr
\begin{tabular}{c}
    $[0;0]_{\Delta}^{(\mathbf{R};r)}$  \\  $r = r_* $ \\ Chiral-Antichiral  \\ \textcolor{blue}{$\hat{\mathcal{B}}_{\mathbf{R}}$}
\end{tabular} \\ \hline
     \end{tabular}
\caption{Consistent unitary multiplets in four-dimensional theories with any number of supersymmetries $\cN=1,2,3,4$ are represented in the first table. Boxes shaded in yellow contain chiral multiplets, those shaded in green contain anti-chiral multiplets, and the box shaded in red contains chiral-antichiral multiplets. In each box we indicate in blue the translation in the notation of \cite{Dolan:2002zh}. The constants $r_*$, $\d$, $\bar\d$, $\bal$, $\bar\bal$ and $\b$ are given in equations (\ref{del}), (\ref{delbar}), (\ref{rstar}) and (\ref{constantsAlphaBeta}).     }
     \label{tabMultiplets}
\end{table}

An obvious choice for these special chiral supermultiplet components are those which carry highest weight, i.e., for which $\bl=\bR$. Their operator products can only close on other such chiral fields in multiplets whose Dynkin labels are the sum of those of the original multiplets. These are then holomorphic fields $\vf_a$ satisfying \eqref{alg0}.  Similarly, their hermitian conjugates are primaries of anti-chiral multiplets of lowest weight $\bl=-\bR$, and so are fields $\bar\vf^a$ satisfying anti-chiral ring relations \eqref{ring2}.   There are other non-highest-weight supermultiplet components which also satisfy \eqref{alg0}, but this kind of enlargement of the chiral ring only occurs for $\cN\ge3$ supersymmetry. 
%since it is only in this case that the dimension of the $\SU(\cN)_R$ weight space is greater than one.

This does not exhaust all the possibilities for subsets of primary chiral fields whose products satisfy chiral ring relations.  In particular, a highest weight $\bR$ of an irrep of the $\suf(\cN)_R$ Lie algebra is highest with respect to an arbitrary choice of a basis of simple roots of $\suf(\cN)_R$.  So other subsets of fields which form a chiral ring are the highest weight primaries of chiral supermultiplets with respect to each different basis of simple roots.   The different bases of simple roots form an orbit of the action of the Weyl group, $\text{Weyl}[\suf(\cN)_R] = \cS_\cN$. $\cS_\cN$, the symmetric group whose action on the weight space is generated by reflections through hyperplanes orthogonal to the roots, acts by permuting the supercharges $\bar{Q}^{(I)}$, $I=1,...,\cN$, and thus the corresponding different choices of co-chiral fields. 

In addition, the choice of Cartan subalgebra of $\suf(\cN)_R$ is not unique, and can be rotated to other choices by conjugation by the $\suf(\cN)_R$ symmetry generators.  Since, as we will review below, a choice of a supercharge defines a complex structure on $\cM$ and the nonabelian R-symmetries do not act holomorphically on $\cM$ it follows that 
\begin{result}
Rotating the choice of Cartan subalgebra in general corresponds to rotating the splitting of the $\cM$ coordinate ring into its chiral and anti-chiral parts  \eqref{ring0} and \eqref{ring2}. This corresponds to choosing a different  complex structure on $\cM$.
\end{result}

The rest of this section will be dedicated to characterizing the possible chiral multiplets which can arise with various amount of supersymmetry and their corresponding chiral fields. Following \cite{Cordova:2016emh}, the shortening conditions on the various multiplets for four dimensional SCFTs are summarized in Table \ref{tabMultiplets}. To simplify the notation, we choose a slightly different naming scheme compared to \cite{Cordova:2016emh}. Specifically we have removed the subscript for both the $\mathbf{A}$ (and $\mathbf{\bar{A}}$) and $\mathbf{B}$ (and $\mathbf{\Bb}$) shortening condition, so $\mathbf{A}_\ell, \mathbf{B}_\ell\to \mathbf{A},\mathbf{B}$. This is because $(i)$ the distinction between $\mathbf{A}_1$ and $\mathbf{A}_2$ will make no difference in the analysis below and $(ii)$ in four dimension there is only one $\mathbf{B}$ shortening condition. 

In Table \ref{tabMultiplets} the following definitions are used:
\begin{align}\label{del}
\d &:= j_R + \bal\cdot\bR + \b r  \\\label{delbar}
\bar\d &:= j_L+ \bar\bal\cdot\bR - \b r \\ 
r_* &:=  \frac{1}{2\beta}(j_L-j_R+(\bar\bal-\bal) \cdot\bR)) \label{rstar}
\end{align}
where $\bal$, $\bar\bal$ and $\b$ are (vectors of) positive constants given by
\begin{equation}
\label{constantsAlphaBeta}
    \begin{array}{|c|c|ccc|} \hline 
        \mathrm{Supersymmetry} & \mathrm{Algebra} & \bal & \bar\bal & \b \\ \hline 
        \cN=1  & \uf(1)_R &   &  &  \frac{3}{2} \\
        \cN=2  &\ \suf(2)_R\oplus\uf(1)_R \ \ & (1) & (1) &\  \frac{1}{2}\ \  \\
        \cN=3  &\ \suf(3)_R\oplus\uf(1)_R\ \  & \left(\frac{2}{3},\frac{4}{3}\right)& \left(\frac{4}{3},\frac{2}{3}\right)&  \frac{1}{6} \\
        \cN=4  & \suf(4)_R & \left(\frac{1}{2},1,\frac{3}{2}\right)& \left(\frac{3}{2},1,\frac{1}{2}\right)&  \\[3pt] \hline 
    \end{array}
\end{equation}

\subsubsection{Non-supersymmetric 4d CFTs} 
\label{sectionN=0}

In this case only the Lorentz weights are available to play the role of the $\U(1)$ charge protecting the chiral ring relations.  Of the unitary representations listed in table 26 of \cite{Cordova:2016emh}, only the identity operator saturates a bound proportional to a linear combination of the Lorentz spins, so there is no chiral algebra enforced simply by unitarity and conformal invariance, and so presumably there is no spontaneous breaking of conformal invariance in genuinely $\cN=0$ CFTs if a chiral algebra is a necessary condition for the existence of a moduli space.  It is conceivable that other restrictions on the operator spectrum of a CFT, such as those following from crossing symmetry and some global internal symmetries, might lead to the ``accidental'' existence of a chiral algebra of local operators in specific theories.  But without the supersymmetric Ward identity that makes chiral correlators independent of the operator positions, it is not clear how or if the existence of such an algebra can be connected to the existence of a moduli space.

\subsubsection{$\cN=1$ SCFTs} 

These theories have a $\U(1)_R$ R-symmetry, and their chiral multiplets are the type $\mathbf{X\Bb}$ or $\mathbf{B \bar{X}}$ where $\mathbf{X}\in \{\mathbf{L}, \mathbf{A}, \mathbf{B}\}$. From Table \ref{tabMultiplets} and \eqref{del}-\eqref{rstar} these all have scaling dimension $\D=\frac32 |r|$. 
All such fields with positive $r$ then form a chiral ring as in \eqref{ring0} while their conjugates with negative $r$ form the anti-chiral ring as in \eqref{ring2}.  The $\mathbf{B \Bb}$ multiplet is the identity, and the $\mathbf{A\Bb}$ multiplets are all (perhaps higher-spin) free fields.  So only the $\mathbf{L\Bb}$ multiplets, which have $[j_L,j_R]=[j,0]$ and $r>\frac13(j+2)$, contribute chiral algebra operators in an interacting $\cN=1$ SCFT.

General restrictions on the types of chiral multiplets that can occur in $\cN=1$ SCFTs do not seem to be known.  In examples of $\cN=1$ SCFTs which come from relevant or marginal deformations of free gauge theories, $[1,0]$ chiral multiplets do occur,\footnote{In $\cN=1$ superfield language, they are the $\Tr(\Phi^k W_\a)$ multiplets of \cite{Seiberg:1994pq, Intriligator:1995id, Cachazo:2002ry}.} but the authors are not aware of any examples where higher-spin $[j\ge2,0]$ chiral multiplets are known to occur.  It is not known whether crossing symmetry forbids higher-spin bosonic chiral multiplets in interacting $\cN=1$ SCFTs, nor is it known whether it implies that the $j=0$ scalar $\mathbf{L\Bb}$ chiral multiplets must occur. In all examples we are aware of, however, such scalar chiral multiplets do occur, and a moduli space of vacua with spontaneously broken scale invariance exists.

\subsubsection{$\cN=2$ SCFTs}

\begin{figure}
\begin{center}
\begin{tikzpicture}
\newcommand\ew{3}
\newcommand\ma{4}
\draw[-latex] (\ma,\ma) -- (-\ma,-\ma) node[left] {$r$};
\draw[-latex] (\ma,-\ma) -- (-\ma,\ma) node[left] {$\mathbf{R}$};
% susy
\draw[-latex,thick,red!80!black] (0,0) -- (\ew,0) node[right] {$Q^{\mathbf{\lambda}=-1}$};
\draw[-latex,thick,red!80!black] (0,0) -- (0,\ew) node[above] {$Q^{\mathbf{\lambda}=+1}$};
\draw[-latex,thick,red!80!black] (0,0) -- (-\ew,0) node[left] {$\bar{Q}^{\mathbf{\lambda}=+1}$};
\draw[-latex,thick,red!80!black] (0,0) -- (0,-\ew) node[right] {$\bar{Q}^{\mathbf{\lambda}=-1}$};
% def
\tikzstyle{st2}=[circle,inner sep=1.5pt,fill=green!80!black];
\tikzstyle{st2big}=[circle,inner sep=2.5pt,fill=green!80!black];
\tikzstyle{st1}=[circle,inner sep=1.5pt,fill=blue!80!black];
\tikzstyle{st1big}=[circle,inner sep=2.5pt,fill=blue!80!black];
% hyper
\node[st1] at (.5*\ew,.5*\ew) {};
\node[st1big,label=left:$\varphi$] at (-.5*\ew,.5*\ew) {};
\node[st1big,label=below:$\bar{\varphi}$] at (.5*\ew,-.5*\ew) {};
\node[st1] at (-.5*\ew,-.5*\ew) {};
% vector 
\node[st2big,label=below:$\bar{a}$] at (1*\ew,1*\ew) {};
\node[st2] at (1*\ew,0*\ew) {};
\node[st2] at (0*\ew,1*\ew) {};
\node[st2,label=below:$\mathcal{F\bar{F}}$] at (0*\ew,0*\ew) {};
\node[st2big,label=left:$a$] at (-1*\ew,-1*\ew) {};
\node[st2] at (-1*\ew,0*\ew) {};
\node[st2] at (0*\ew,-1*\ew) {};
% Weyl group
\draw[latex-latex] (1.2*\ew,1.5*\ew) to[out=0,in=90] (1.5*\ew,1.2*\ew);  
\node at (1.5*\ew,-0.7*\ew) {Higgs branch operators};
\draw[-latex] (1.4*\ew,-0.8*\ew) -- (1.1*\ew,-1*\ew) node[left] {};
\node at (1.7*\ew,0.7*\ew) {Coulomb branch operators};
\draw[-latex] (1.4*\ew,0.8*\ew) -- (1.1*\ew,1*\ew) node[left] {};
\end{tikzpicture}
\end{center}
    \caption{Hypermultiplet (type $\mathbf{B\Bb}$ with $\mathbf{R}=1$, represented in blue) and vector multiplet (type $\mathbf{A\Bb}$ and $\mathbf{B\bar{A}}$, represented in green) in $\mathcal{N}=2$ theories. The big dots represent the scalar operators: $\varphi$, $\bar{\varphi}$ in the hypermultiplet (with $r=0$) and $a$,$\bar{a}$ in the vector multiplet (with $\mathbf{R}=0$). The small dots without label represent the fermions in the corresponding multiplets. The arrow in the upper right corner denotes the action of the Weyl group. }
    \label{figureN=2}
\end{figure}
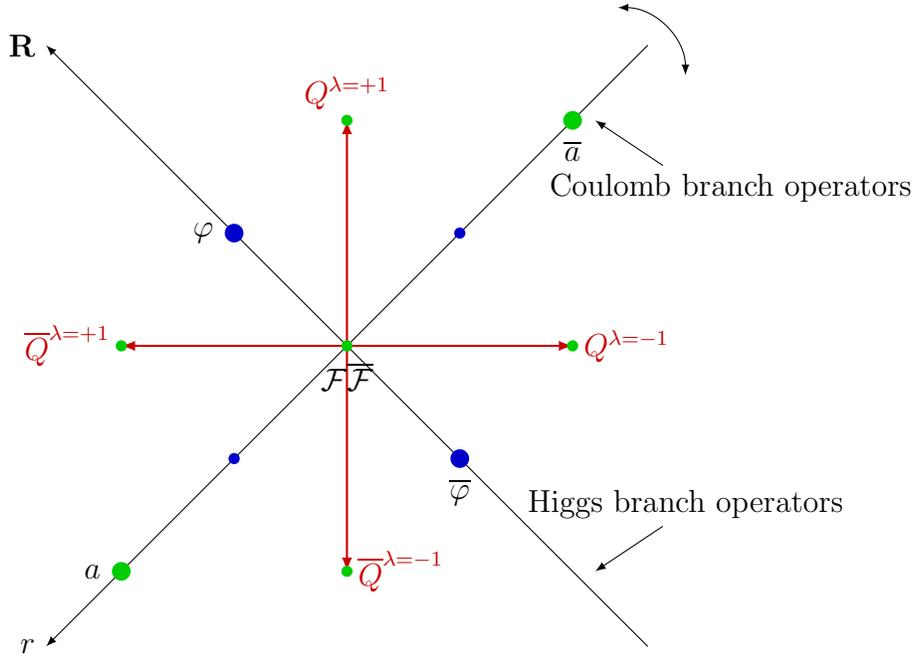

Now there are two independent $\uf(1)$ internal charges corresponding to the rank two $\uf(2)_R$ symmetry. 
The potential chiral multiplets have scaling dimension $\D=\bR + \frac{r}{2}$.
% In a given multiplet with $\D=\bR + \frac{r}{2}$, all scalar descendents which are annihilated by Qbar do not satisfy the bound. 
%\red{In each $\uf(2)_R$ multiplet with positive $r$, there is a single highest weight component of the $\suf(2)_R$ irrep. These fields all form a chiral ring as in \eqref{ring0}.  Likewise, the lowest-weight components of the conjugate multiplets form an anti-chiral ring as in \eqref{ring2}.} 
From the point of view of the $\cN=2$ moduli space, vevs of Lorentz scalar chiral primaries are complex coordinates on either a \emph{Coulomb branch} (if $\bR=0$), a \emph{Higgs branch} (if $r=0$), or a \emph{mixed branch} (if $\bR \, r \neq0$). The operators whose vevs parametrize each branch, form a corresponding chiral ring which are therefore called \emph{Coulomb}, \emph{Higgs} and \emph{mixed chiral rings}. See Figure \ref{figureN=2}.  

In detail, the $\mathbf{\mathbf{B\Bb}}$ multiplets with general $\bR$ are known as Higgs branch operators and their OPEs contain the Higgs branch chiral ring, and together with the $\bR\ge2$ $\mathbf{A \Bb}$ operators form the Hall-Littlewood chiral ring \cite{Gadde:2011uv}.  It is argued in \cite{Beem:2017ooy} that the $\bR\ge2$ $\mathbf{A \Bb}$ chiral operators are nilpotent, so the reduced Hall-Littlewood chiral ring is the Higgs branch chiral ring. The Coulomb branch chiral ring is generated by those scalar $\mathbf{L\Bb}$ chiral multiplet primaries $\vf_a^{[0,0]}$ with $\bR=0$. Finally the mixed branch chiral ring is generated by scalar primaries of the $\mathbf{L\Bb}$ chiral multiplet with $\bR\neq0$. Explicit  examples of chiral rings of theories containing mixed branches were worked out in \cite{Argyres:2016xmc}.

Some general restrictions on the types of chiral multiplets that can occur in $\cN=2$ SCFTs are known.  First, the $\mathbf{\mathbf{B\Bb}}$ multiplets (which have scalar primaries with $r=0$) are free fields for $\bR <2$, and the $\mathbf{A \Bb}$ multiplets (which have $[j_L, j_R] = [j, 0]$ and $r=j+2$) are free, or have extra supersymmetry currents, or have higher-spin currents for $\bR<2$, so do not occur in interacting, genuinely $\cN=2$ SCFTs.  The $\mathbf{\mathbf{B\Bb}}$ multiplet with $\bR=2$ contains a conserved flavor symmetry current, so any $\cN=2$ SCFT with a continuous flavor symmetry necessarily has a chiral algebra. Finally, it was shown in \cite{Manenti:2019jds} that $\cN=2$ SCFTs do not have ``exotic chiral multiplets'' --- non-scalar $\mathbf{L\Bb}$ chiral multiplets with $\bR=0$.

Extra constraints can be obtained by noticing that the choice of the Cartan generator $\bR$ in $\suf(2)_R$ is a matter of convention, and any given choice of Cartan subalgebra can be rotated to any other by conjugation by an element of $\SU(2)_R$.  Such a rotation, thus corresponds to a non-holomorphic isometry of the Higgs branch (or, more generally, the hyperk\"ahler factors) of $\cN=2$ moduli space, and so corresponds to rotating the $\C\P^1$ of hyperk\"ahler complex structures. In terms of the chiral ring, this rotation corresponds to different ways of splitting the moduli space coordinate ring \eqref{ring0} and \eqref{ring2} into its chiral and anti-chiral parts. The elements of a Lie group which leave any choice of Cartan subalgebra invariant (as a whole, not point-wise) defines its discrete Weyl subgroup.  In the case of $\SU(2)_R$, the Weyl group is $\Z_2$ whose nontrivial element acts to reflect weights as $\bl \mapsto -\bl$.  Since supermultiplets are irreps of $\SU(2)_R$, its Weyl subgroup maps supermultiplets to themselves. 

For the $\mathbf{B \Bb}$ chiral/anti-chiral multiplets, the ``Higgs branch operators'', this mapping takes the co-chiral set of highest weight $\bl=\bR$ operators to a different co-chiral set of lowest weight $\bl=-\bR$ operators.  The $\bl=\bR$ set are chiral with respect to (i.e., annihilated by) $\Qb^{\bl=1}$ while the $\bl=-\bR$ set are chiral with respect to the other supercharge $\Qb^{\bl=-1}$, see Figure \ref{figureN=2}. But for the $\mathbf{B \Bb}$ multiplets, components chiral with respect to $\Qb^{\bl=-1}$ are automatically also anti-chiral with respect to $Q^{\bl=-1} = [\Qb^{\bl=1}]^\dagger$.  Thus the $\Z_2$ Weyl group maps the Higgs branch chiral ring to its anti-chiral ring.  This implements the anti-holomorphic involution characteristic of any complex structure of a hyperk\"ahler manifold.

For the rest of the chiral multiplets --- those of types $\mathbf{L\Bb}$ or $\mathbf{A\Bb}$ --- the Weyl group maps the co-chiral operators with highest weights $\bl=\bR$ to operators with $\bl=-\bR$ but the same sign of $r$.  In the case $\bR=0$ --- the ``Coulomb branch operators'' --- the two sets coincide, but for $\bR>0$ --- the ``mixed branch operators'' --- they are distinct and the $\bl=-\bR$ set do not form a chiral ring.  Thus there is no Weyl group action on the coordinate ring of a mixed branch and it acts trivially on the Coulomb branch.

\section{Chiral rings for $\cN=3$ SCFTs}\label{sec2.3.3}

With the $\cN<3$ chiral rings behind us, we now turn to our main concern, the structure of $\cN=3$ chiral rings.  The discussion will be similar to that of the mixed branch rings of the $\cN=2$ case, but will be more constrained as there is now only a single type of complex geometry describing any component of the moduli space, the TSK structure which will be introduced in the next section. The constraints which arise by bringing together the TSK analysis of section \ref{sec2.1} and \ref{sec2-TSK}, with the chiral ring data discussed in this section, will be addressed in section \ref{TSKspace}.  The qualitatively new structure that arises in the chiral ring of $\cN=3$ CFTs compared to ones with $\cN<3$ is essentially due to the fact that the rank of  $\suf(3)_R$ is greater than 1 and that $\suf(3)_R$ has complex representations.

\begin{figure}
\centering
\begin{tikzpicture}
%defs
\tikzstyle{rc}=[circle,inner sep=2pt,fill=green!70!black];
\tikzstyle{bc}=[circle,inner sep=2pt,fill=blue!80!black!50];
\tikzstyle{gc}=[circle,inner sep=2pt,fill=red!80!black];
\tikzstyle{oc}=[circle,inner sep=2pt,fill=orange!80!black!50];
%% fig a
\begin{scope}[scale=1.0, xshift=-4cm]
\node at (0,-3) {\bf (a)};
\clip (-3.1,-3) rectangle (3.1cm,3cm); % clips the picture
% transformation matrix giving new unit vectors (hexagonal)
\pgftransformcm{.5}{-.866}{0.5}{0.866}{\pgfpoint{0cm}{0cm}}
% draw lattice dots
\foreach \x in {-7,-6,...,7}{\foreach \y in {-7,-6,...,7}{
\node[draw,circle,inner sep=0.5pt,fill] at (\x,\y) {};  }}
%chiral line
\draw[style=help lines,dashed] (0-4*2,1-4*1) -- (0+4*2,1+4*1);     
%draw and label some vectors
    \draw [ultra thick,-latex] (0,0) -- (0,1) 
   node [above] {\tiny $(0\, 1)$};
    \draw [ultra thick,-latex] (0,0) -- (1,1) 
   node [above] {\tiny $(1\,0)$};
%fund weights
   \node[draw,circle,inner sep=3pt] at (0,0) {};
   \node[rc] at (-1,-1) {};
   \node[rc] at (1,0) {};
   \node[rc] at (0,1) {};
%   \node[oc] at (0,0) {};
   \node[oc] at (1,1) {};
   \node[oc] at (-1,0) {};
   \node[oc] at (0,-1) {};
   \node[oc] at (-2,-2) {};
   \node[oc] at (2,0) {};
   \node[oc] at (0,2) {};
\end{scope}
%% fig b
\begin{scope}[scale=1.0, xshift=+4cm]
\node at (0,-3) {\bf (b)};
\clip (-3.1,-3) rectangle (3.1cm,3cm); % clips the picture
% transformation matrix giving new unit vectors (hexagonal)
\pgftransformcm{.5}{-.866}{0.5}{0.866}{\pgfpoint{0cm}{0cm}}
% weight lattice
\foreach \x in {-7,-6,...,7}{\foreach \y in {-7,-6,...,7}{
\node[draw,circle,inner sep=0.5pt,fill] at (\x,\y) {};  }}
%draw and label some vectors
    \draw [ultra thick,-latex] (0,0) -- (0,1) 
   node [above] {\tiny $(0\, 1)$};
    \draw [ultra thick,-latex] (0,0) -- (1,1) 
   node [above] {\tiny $(1\,0)$};
    \draw [ultra thick,-latex,blue!80!black!20] (0,0) -- (-1,1);
    \draw [ultra thick,-latex,blue!80!black!20] (0,0) -- (2,1);
% weights
   \node[draw,circle,inner sep=3pt] at (0,0) {}; %origin
   \node[gc] at (1,1) {}; %R=(1,0) weights
   \node[gc] at (-1,0) {}; %R=(1,0) weights
   \node[gc] at (0,-1) {}; %R=(1,0) weights
   \node[bc] at (0,0) {}; %R=(1,1) weights
   \node[bc] at (1,2) {}; %R=(1,1) weights
   \node[bc] at (-1,1) {}; %R=(1,1) weights
   \node[bc] at (2,1) {}; %R=(1,1) weights
   \node[bc] at (-2,-1) {}; %R=(1,1) weights
   \node[bc] at (-1,-2) {}; %R=(1,1) weights
   \node[bc] at (1,-1) {}; %R=(1,1) weights
%chiral line
\draw[style=help lines,dashed] (1-4*2,1-4*1) -- (1+4*2,1+4*1);     
\node at (0,-4) {\bf (b)};
\end{scope}
\end{tikzpicture}
\tdplotsetmaincoords{100}{20}
\begin{tikzpicture}[tdplot_main_coords,font=\sffamily]
\newcommand\ew{2}
\newcommand\ma{3}
\draw[-latex] (0,0,0) -- (\ma,0,0) node[left] {};
\draw[-latex] (0,0,0) -- (0,\ma,0) node[below] {};
\draw[-latex] (0,0,0) -- (0,0,\ma) node[left] {};
\draw[] (0,0,0) -- (-\ma,0,0) node[left] {};
\draw[] (0,0,0) -- (0,-\ma,0) node[left] {};
\draw[] (0,0,0) -- (0,0,-\ma) node[left] {};
\tdplotsetrotatedcoords{0}{0}{0}
\begin{scope}[tdplot_rotated_coords]
% susy
\draw[-latex,thick,red!80!black] (0,0,0) -- (\ew,0,0) node[above] {$Q$};
\draw[-latex,thick,red!80!black] (0,0,0) -- (0,\ew,0) node[below] {$Q$};
\draw[-latex,thick,red!80!black] (0,0,0) -- (0,0,\ew) node[right] {$Q$};
\draw[-latex,thick,green!80!black] (0,0,0) -- (-\ew,0,0) node[above] {$\bar{Q}$};
\draw[-latex,thick,green!80!black] (0,0,0) -- (0,-\ew,0) node[above] {$\bar{Q}$};
\draw[-latex,thick,green!80!black] (0,0,0) -- (0,0,-\ew) node[right] {$\bar{Q}$};
\end{scope}
\end{tikzpicture}
\caption{$\suf(3)_R$ weight lattice, with vectors showing a basis of fundamental weights.  (a) Green dots are the $\bar{\bf3}$, or $\bR=(0,1)$ weights, and the orange dots are the $\bR=(0,2)$ weights of the $\Qb^\bl$, $\bl\in\bR=(0,1)$, null states.  The only component of a chiral multiplet in the $\bar{\bf3}$ annihilated by $\Qb^{(0,1)}$ alone is the one with highest projection along the $\bl=(0,1)$ direction, shown as the one lying on the dashed line.  (b) Red dots are the ${\bf3}$, or $\bR=(1,0)$ weights, and the blue dots are the $\bR=(1,1)$ weights of the $\Qb^\bl$ null states.  The components of a chiral multiplet in the ${\bf3}$ annihilated by $\Qb^{(0,1)}$ alone are the two lying on the dashed line.  The light blue arrows show the choice of simple roots with respect to which our Dynkin labels are defined. Below in the three-dimensional $\uf(3)_R$ weight lattice, we represent the weights of $Q$ and $\bar{Q}$, which in addition to the $\suf(3)_R$ weights $(R_1,R_2)$ take into account the $\uf(1)_r$ charge $r$. }\label{su3fig1}
\end{figure}
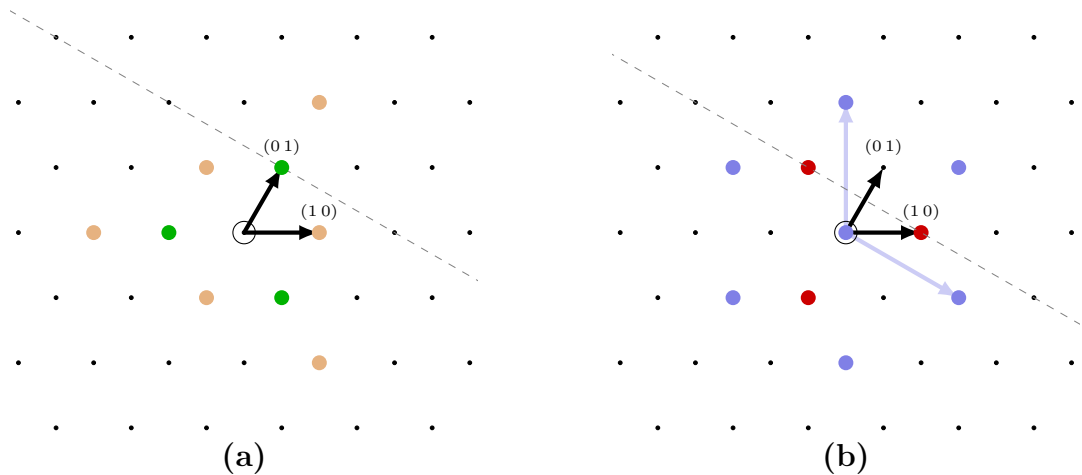

To identify a set of co-chiral primaries, we first choose a particular $\Qb^\bl$ supercharge with respect to which they are chiral (i.e., which annihilates them).  $\Qb$ transforms as $[0,1]^{(0,1),1}_{1/2}$, with $\suf(3)_R$ weights $\bl = (0,1),\, (1,-1),\, (-1,0)$. For concreteness, let's choose $\Qb^{\bl = (0,1)}$ to be our annihilating supercharge.  This is equivalent to choosing an $\cN=1$ subalgebra of the $\cN=3$ algebra and thus a particular complex structure on $\cM$.

Consider an $\mathbf{X\Bb}$ superconformal primary in the $\bR=(R_1,R_2)$ irrep of $\suf(3)_R$. Acting by $\Qb$ we obtain operators transforming in the $(R_1,R_2)\otimes (0,1)$ which is in general a reducible representation.  The null states lie in the $(R_1,R_2+1)$ \cite{Cordova:2016emh}.  A component of the $(R_1,R_2)$ irrep with weight $(\l_1,\l_2)$, is mapped by the top component $\Qb^{(0,1)}$ to a state with weight $(\l_1,\l_2+1)$.  The null representation always contains a component with such a weight, but that is not enough to assert that $(\l_1,\l_2)$ is annihilated by $\Qb^{(0,1)}$ since $(\l_1,\l_2+1)$ might also appear as a weight of a non-null representation in the decomposition of $(R_1,R_2)\otimes (0,1)$.  We conclude that $(\l_1,\l_2)$ is null if and only if $(\l_1,\l_2+1)$ does not appear in any non-null irreps of $(R_1,R_2)\otimes (0,1)$.  It follows that:
\begin{align}\label{N3crc}
    &\textit{The primary components of $\mathbf{X\Bb_{\bR}}$ with weights $\bl$ which have}\nn\\[-1.5mm]
    &\textit{the maximum value of $\bal\cdot\bl$ are the ones annihilated by $\Qb^{(0,1)}$.}
\end{align} 
This characterizes the chiral ring operators with respect to a given choice of $\cN=3$ supercharge.

\begin{figure}[ht]
\centering
\scalebox{.8}{
\begin{tikzpicture}
%defs
\tikzstyle{gc}=[circle,inner sep=2pt,fill=green!70!black];
\tikzstyle{bc}=[circle,inner sep=2pt,fill=blue!80!black];
\tikzstyle{rc}=[circle,inner sep=2pt,fill=red!80!black];
\tikzstyle{oc}=[circle,inner sep=2pt,fill=orange!80!black!50];
\clip (-5.1,-5) rectangle (5.1cm,5cm); % clips the picture
% transformation matrix giving new unit vectors (hexagonal)
\pgftransformcm{.5}{-.866}{0.5}{0.866}{\pgfpoint{0cm}{0cm}}
% weight lattice
\foreach \x in {-7,-6,...,7}{\foreach \y in {-7,-6,...,7}{
\node[draw,circle,inner sep=0.5pt,fill] at (\x,\y) {};  }}
%chiral lines
\draw[style=help lines,dashed,red] (0-4*2,4-4*1) -- (0+4*2,4+4*1);     
\draw[style=help lines,dashed,green!80!black] (0-4*2,2-4*1) -- (0+4*2,2+4*1);     
\draw[style=help lines,dashed,blue] (1-4*2,2-4*1) -- (1+4*2,2+4*1);     
\node[red!80!black] at (0+1.5*2,4+1.5*1) {$\chi$};
\node[green!60!black] at (0+2.3*2,2+2.3*1) {$\chi$};
\node[blue!80!black] at (1+2*2,2+2*1) {$\chi$};
\draw[style=help lines,dotted,red] (0-4*2,-2-4*1) -- (0+4*2,-2+4*1);     
\node[red!80!black] at (0-2.3*2,-2-2.3*1) {$\bar\chi$};
\draw[style=help lines,dotted,green!60!black] (0-4*2,-4-4*1) -- (0+4*2,-4+4*1);     
\node[green!60!black] at (0-1.5*2,-4-1.5*1) {$\bar\chi$};
\draw[style=help lines,dotted,blue] (1-4*2,-1-4*1) -- (1+4*2,-1+4*1); 
\node[blue!80!black] at (1-3*2,-1-3*1) {$\bar\chi$};
%fund weights
   \node[draw,circle,inner sep=3pt] at (0,0) {};
   \node[rc] at (0,4) {};
   \node[rc] at (-1,2) {};
   \node[rc] at (1,3) {};
   \node[rc] at (2,2) {};
   \node[rc] at (0,1) {};
   \node[rc] at (-2,0) {};
   \node[rc] at (3,1) {};
   \node[rc] at (1,0) {};
   \node[rc] at (-1,-1) {};
   \node[rc] at (-3,-2) {};
   \node[rc] at (-4,-4) {};
   \node[rc] at (-2,-3) {};
   \node[rc] at (0,-2) {};
   \node[rc] at (2,-1) {};
   \node[rc] at (4,0) {};
   \node[gc] at (0,0-4) {};
   \node[gc] at (-1,-1-2) {};
   \node[gc] at (1,1-3) {};
   \node[gc] at (2,2-2) {};
   \node[gc] at (0,0-1) {};
   \node[gc] at (-2,-2+0) {};
   \node[gc] at (3,3-1) {};
   \node[gc] at (1,1+0) {};
   \node[gc] at (-1,-1+1) {};
   \node[gc] at (-3,-3+2) {};
   \node[gc] at (-4,-4+4) {};
   \node[gc] at (-2,-2+3) {};
   \node[gc] at (0,+2) {};
   \node[gc] at (2,2+1) {};
   \node[gc] at (4,4+0) {};
   \node[bc] at (0,0) {}; %R=(1,1) weights
   \node[bc] at (1,2) {}; %R=(1,1) weights
   \node[bc] at (-1,1) {}; %R=(1,1) weights
   \node[bc] at (2,1) {}; %R=(1,1) weights
   \node[bc] at (-2,-1) {}; %R=(1,1) weights
   \node[bc] at (-1,-2) {}; %R=(1,1) weights
   \node[bc] at (1,-1) {}; %R=(1,1) weights
    \draw [ultra thick,-latex] (0,0) -- (0,1) 
   node [above] {\tiny $(0\, 1)$};
\end{tikzpicture}
}
\tdplotsetmaincoords{100}{20}
\begin{tikzpicture}[tdplot_main_coords,font=\sffamily]
\newcommand\ew{.8}
\newcommand\ma{3.6}
% axes
\draw[-latex] (0,0,0) -- (\ma,0,0) node[left] {};
\draw[-latex] (0,0,0) -- (0,\ma,0) node[below] {};
\draw[-latex] (0,0,0) -- (0,0,\ma) node[left] {};
\draw[] (0,0,0) -- (-\ma,0,0) node[left] {};
\draw[] (0,0,0) -- (0,-\ma,0) node[left] {};
\draw[] (0,0,0) -- (0,0,-\ma) node[left] {};
\tdplotsetrotatedcoords{0}{0}{0}
\begin{scope}[tdplot_rotated_coords]
\tikzstyle{st1}=[circle,inner sep=1.5pt,fill=green!80!black];
\tikzstyle{st2}=[circle,inner sep=1.5pt,fill=red!80!black];
\tikzstyle{st3}=[circle,inner sep=1.5pt,fill=blue!80!black];
% susy
\draw[-latex,thick,red!80!black] (0,0,0) -- (\ew,0,0) node[right] {$Q$};
\draw[-latex,thick,red!80!black] (0,0,0) -- (0,\ew,0) node[right] {$Q$};
\draw[-latex,thick,red!80!black] (0,0,0) -- (0,0,\ew) node[right] {$Q$};
% a 
\draw[fill=green,opacity=0.1] (0*\ew,-4*\ew,-4*\ew) -- (-4*\ew,-4*\ew,0*\ew) -- (-4*\ew,-0*\ew,-4*\ew) -- cycle;
\node[st1] at (0*\ew,-4*\ew,-4*\ew) {};
\node[st1] at (-1*\ew,-3*\ew,-4*\ew) {};
\node[st1] at (-2*\ew,-2*\ew,-4*\ew) {};
\node[st1] at (-1*\ew,-4*\ew,-3*\ew) {};
\node[st1] at (-3*\ew,-1*\ew,-4*\ew) {};
\node[st1] at (-2*\ew,-3*\ew,-3*\ew) {};
\node[st1] at (-4*\ew,0*\ew,-4*\ew) {};
\node[st1] at (-3*\ew,-2*\ew,-3*\ew) {};
\node[st1] at (-2*\ew,-4*\ew,-2*\ew) {};
\node[st1] at (-4*\ew,-1*\ew,-3*\ew) {};
\node[st1] at (-3*\ew,-3*\ew,-2*\ew) {};
\node[st1] at (-4*\ew,-2*\ew,-2*\ew) {};
\node[st1] at (-3*\ew,-4*\ew,-1*\ew) {};
\node[st1] at (-4*\ew,-3*\ew,-1*\ew) {};
\node[st1] at (-4*\ew,-4*\ew,0*\ew) {};
% red
\draw[fill=red,opacity=0.1] (0*\ew,4*\ew,4*\ew) -- (4*\ew,4*\ew,0*\ew) -- (4*\ew,-0*\ew,4*\ew) -- cycle;
\node[st2] at (0*\ew,4*\ew,4*\ew) {};
\node[st2] at (1*\ew,3*\ew,4*\ew) {};
\node[st2] at (2*\ew,2*\ew,4*\ew) {};
\node[st2] at (1*\ew,4*\ew,3*\ew) {};
\node[st2] at (3*\ew,1*\ew,4*\ew) {};
\node[st2] at (2*\ew,3*\ew,3*\ew) {};
\node[st2] at (4*\ew,0*\ew,4*\ew) {};
\node[st2] at (3*\ew,2*\ew,3*\ew) {};
\node[st2] at (2*\ew,4*\ew,2*\ew) {};
\node[st2] at (4*\ew,1*\ew,3*\ew) {};
\node[st2] at (3*\ew,3*\ew,2*\ew) {};
\node[st2] at (4*\ew,2*\ew,2*\ew) {};
\node[st2] at (3*\ew,4*\ew,1*\ew) {};
\node[st2] at (4*\ew,3*\ew,1*\ew) {};
\node[st2] at (4*\ew,4*\ew,0*\ew) {};
% blue 
\draw[fill=blue,opacity=0.1] (1*\ew,0*\ew,-1*\ew) -- (0*\ew,1*\ew,-1*\ew) -- (-1*\ew,1*\ew,0) -- (-1*\ew,0,1*\ew) -- (0,-1*\ew,1*\ew) -- (1*\ew,-1*\ew,0) -- cycle;
\node[st3] at (1*\ew,0*\ew,-1*\ew) {};
\node[st3] at (0*\ew,1*\ew,-1*\ew) {};
\node[st3] at (1*\ew,-1*\ew,0*\ew) {};
\node[st3] at (0*\ew,0*\ew,0*\ew) {};
\node[st3] at (0*\ew,0*\ew,0*\ew) {};
\node[st3] at (-1*\ew,1*\ew,0*\ew) {};
\node[st3] at (0*\ew,-1*\ew,1*\ew) {};
\node[st3] at (-1*\ew,0*\ew,1*\ew) {};
% cube 
\draw[] (-4*\ew,-4*\ew,0) -- (-4*\ew,-4*\ew,-4*\ew);
\draw[] (-4*\ew,0,-4*\ew) -- (-4*\ew,-4*\ew,-4*\ew);
\draw[] (0,-4*\ew,-4*\ew) -- (-4*\ew,-4*\ew,-4*\ew);
\draw[] (-4*\ew,-4*\ew,0) -- (-4*\ew,0,0);
\draw[] (-4*\ew,0,-4*\ew) -- (0,0,-4*\ew);
\draw[] (0,-4*\ew,-4*\ew) -- (0,0,-4*\ew);
\draw[] (-4*\ew,-4*\ew,0) -- (0,-4*\ew,0);
\draw[] (-4*\ew,0,-4*\ew) -- (-4*\ew,0,0);
\draw[] (0,-4*\ew,-4*\ew) -- (0,-4*\ew,0);
\draw[] (4*\ew,4*\ew,0) -- (4*\ew,4*\ew,4*\ew);
\draw[] (4*\ew,0,4*\ew) -- (4*\ew,4*\ew,4*\ew);
\draw[] (0,4*\ew,4*\ew) -- (4*\ew,4*\ew,4*\ew);
\draw[] (4*\ew,4*\ew,0) -- (4*\ew,0,0);
\draw[] (4*\ew,0,4*\ew) -- (0,0,4*\ew);
\draw[] (0,4*\ew,4*\ew) -- (0,0,4*\ew);
\draw[] (4*\ew,4*\ew,0) -- (0,4*\ew,0);
\draw[] (4*\ew,0,4*\ew) -- (4*\ew,0,0);
\draw[] (0,4*\ew,4*\ew) -- (0,4*\ew,0);
\draw[] (1*\ew,1*\ew,1*\ew) -- (-1*\ew,1*\ew,1*\ew) -- (-1*\ew,-1*\ew,1*\ew) -- (1*\ew,-1*\ew,1*\ew) -- (1*\ew,-1*\ew,-1*\ew) -- (1*\ew,1*\ew,-1*\ew) -- (-1*\ew,1*\ew,-1*\ew) -- (-1*\ew,-1*\ew,-1*\ew) -- (-1*\ew,-1*\ew,1*\ew);
\draw[] (-1*\ew,1*\ew,-1*\ew) -- (-1*\ew,1*\ew,1*\ew);
\draw[] (1*\ew,1*\ew,-1*\ew) -- (1*\ew,1*\ew,1*\ew);
\draw[] (1*\ew,-1*\ew,1*\ew) -- (1*\ew,1*\ew,1*\ew);
\draw[] (1*\ew,-1*\ew,-1*\ew) -- (-1*\ew,-1*\ew,-1*\ew);
\end{scope}
\end{tikzpicture}
\caption{Above: $\suf(3)_R$ weight lattice with weights of the $\bR=(0,4)$ in red, of the $\bR=(4,0)$ in green, and of the $\bR=(1,1)$ in blue.  The chiral components of these $\mathbf{\mathbf{B\Bb}}$ multiplets, i.e., those annihilated by $\Qb^{(0,1)}$, are the ones on the dashed ``$\chi$'' lines. The anti-chiral components are those on the dotted ``$\bar\chi$'' lines. Below: same objects in the full $\uf(3)_R$ weight lattice (see the Appendix for details on this representation). In the three-dimensional depiction, the chiral (respectively anti-chiral) components appear on the faces of the cubes. }\label{su3fig2}
\end{figure}
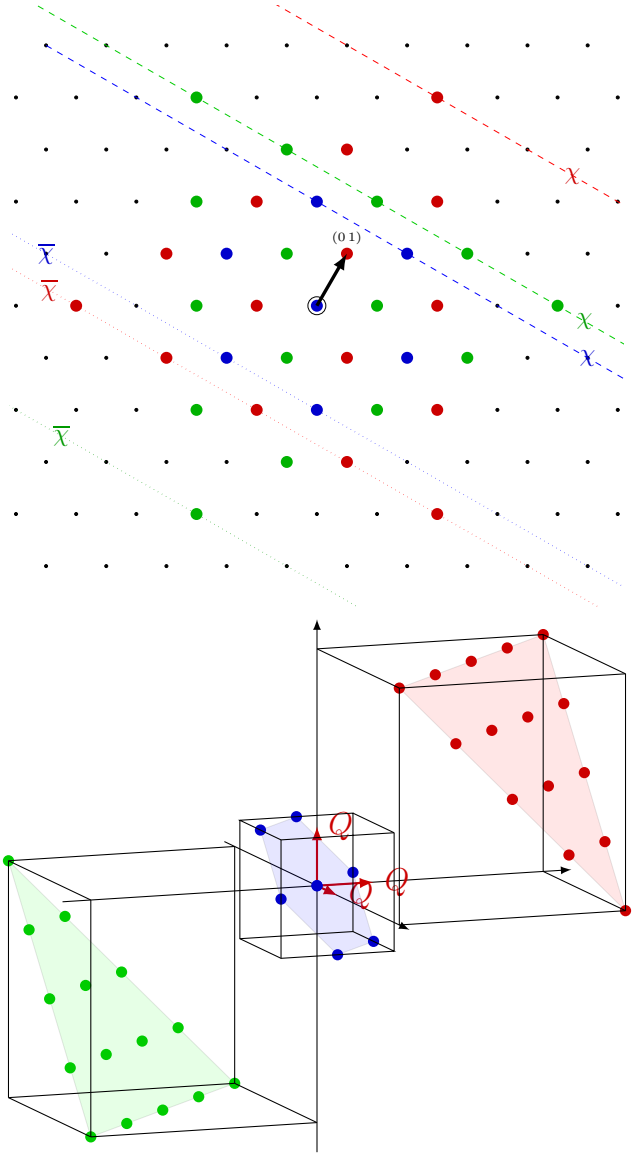

This can be understood very easily by drawing the weight diagrams.  For example, the highest weight component of a chiral multiplet in the $\bar{\bf3}$, or $\bR=(0,1)$, of $\SU(3)_R$ is annihilated $\Qb^{(0,1)}$, but the other two components are not.  This is because the $\bR=(0,2)$ null states are reached by a combination of $\Qb^{(0,1)}$ together with $\Qb^{(1,-1)}$ or $\Qb^{(-1,0)}$ acting on various components of the $\bar{\bf3}$ multiplet.  This is illustrated in figure \ref{su3fig1}(a).  On other hand, two weights of the ${\bf3}$, or $\bR=(1,0)$, irrep are annihilated by $\Qb^{(0,1)}$, as illustrated in figure \ref{su3fig1}(b). As described below the $\bR=(1,0)$ and $\bR=(0,1)$ are both identified with the $\cN=3$ free vector multiplet, thus the result depicted in fig. \ref{su3fig1} implies that under a given choice of the complex structure only a $U(2)\subset U(3)_R$ acts holomorphically on $\cM$. This we will be derived even more explicitly below. Note that although in these examples all the components chiral with respect to $\Qb^{(0,1)}$ are highest weights with respect to some choice of $\SU(3)_R$ simple roots, this need not always be the case; figure \ref{su3fig2} illustrates an example of the more general situation.

\paragraph{Type $\mathbf{\mathbf{B\Bb}}$ (chiral/anti-chiral) multiplets.}   

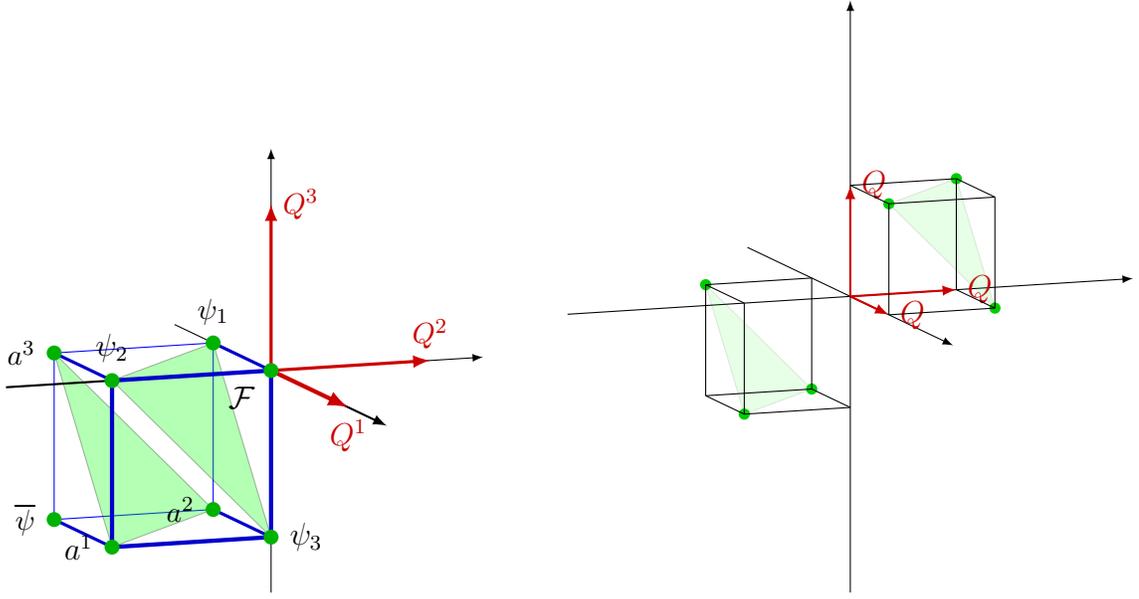
\begin{figure}
    \centering
    \begin{tabular}{cc}
        \tdplotsetmaincoords{100}{20}
\begin{tikzpicture}[tdplot_main_coords,font=\sffamily]
\tdplotsetrotatedcoords{0}{0}{0}
\begin{scope}[tdplot_rotated_coords, scale=1.5]
\tikzstyle{st}=[circle,inner sep=2pt,fill=green!70!black];
\newcommand\ew{1.5}
\newcommand\ma{2.5}
% +x, y, z coordinate axes
\draw[-latex] (0,0,0) -- (\ma-.5,0,0) node[left] {};
\draw[-latex,thick] (0,0,0) -- (0,\ma+.5,0) node[below] {};
\draw[-latex] (0,0,0) -- (0,0,\ma-.5) node[left] {};
\draw[] (0,0,0) -- (0,-\ma,0) node[left] {};
\draw[] (0,0,0) -- (0,0,-\ma+.5) node[left] {};
% Q
\draw[-latex,very thick,red!80!black] (0,0,0) -- (\ew,0,0) node[above] {\small $Q^2$};
\draw[-latex,ultra thick,red!80!black] (0,0,0) -- (0,\ew+.5,0) node[below] {$Q^1$};
\draw[-latex,very thick,red!80!black] (0,0,0) -- (0,0,\ew) node[right] {\small $Q^3$};
% back blue square
\draw[blue] (-\ew,-\ew,0) -- (0,-\ew,0);
\draw[blue] (0,-\ew,-\ew) -- (0,-\ew,0);
\draw[blue] (0,-\ew,-\ew) -- (-\ew,-\ew,-\ew);
\draw[blue] (-\ew,-\ew,0) -- (-\ew,-\ew,-\ew);
% blue edges
\draw[very thick, blue!80!black] (0,0,0) -- (0,-\ew,0);
\draw[very thick, blue!80!black] (0,-\ew,-\ew) -- (0,0,-\ew);
\draw[very thick, blue!80!black] (-\ew,-\ew,0) -- (-\ew,0,0);
\draw[very thick, blue!80!black] (-\ew,0,-\ew) -- (-\ew,-\ew,-\ew);
% green triangles
\draw[fill=green,opacity=0.3] (0,-\ew,-\ew) -- (-\ew,0,-\ew) -- (-\ew,-\ew,0) -- cycle;
\draw[fill=green,opacity=0.3] (0,0,-\ew) -- (-\ew,0,0) -- (0,-\ew,0) -- cycle;
% -x coordinate axis
\draw[thick] (0,0,0) -- (-\ma,0,0) node[left] {};
% front blue square (y=0 plane)
\draw[ultra thick, blue!80!black] (0,0,0) -- (0,0,-\ew);
\draw[ultra thick, blue!80!black] (0,0,0) -- (-\ew,0,0);
\draw[ultra thick, blue!80!black] (-\ew,0,-\ew) -- (-\ew,0,0);
\draw[ultra thick, blue!80!black] (-\ew,0,-\ew) -- (0,0,-\ew);
% a
\node[st,label=left:{\small $a^2$}] at (0,-\ew,-\ew) {};
\node[st,label=left:$a^1$] at (-\ew,0,-\ew) {};
\node[st,label=left:{\small $a^3$}] at (-\ew,-\ew,0) {};
% psibar
\node[st,label=left:{\small $\bar{\psi}$}] at (-\ew,-\ew,-\ew) {};
% psii 
\node[st,label=right:$\psi_3$] at (0,0,-\ew) {};
\node[st,label=$\psi_2$] at (-\ew,0,0) {};
\node[st,label={\small $\psi_1$}] at (0,-\ew,0) {};
% F
\node[st,label=below left:$\cF$] at (0,0,0) {};
\end{scope}
\end{tikzpicture} &
\tdplotsetmaincoords{100}{20}
\begin{tikzpicture}[tdplot_main_coords,font=\sffamily]
\newcommand\ew{1.5}
\newcommand\ma{4}
\draw[-latex] (0,0,0) -- (\ma,0,0) node[left] {};
\draw[-latex] (0,0,0) -- (0,\ma,0) node[below] {};
\draw[-latex] (0,0,0) -- (0,0,\ma) node[left] {};
\draw[] (0,0,0) -- (-\ma,0,0) node[left] {};
\draw[] (0,0,0) -- (0,-\ma,0) node[left] {};
\draw[] (0,0,0) -- (0,0,-\ma) node[left] {};
\tdplotsetrotatedcoords{0}{0}{0}
\begin{scope}[tdplot_rotated_coords]
\tikzstyle{st1}=[circle,inner sep=1.5pt,fill=green!80!black];
% susy
\draw[-latex,thick,red!80!black] (0,0,0) -- (\ew,0,0) node[right] {$Q$};
\draw[-latex,thick,red!80!black] (0,0,0) -- (0,\ew,0) node[right] {$Q$};
\draw[-latex,thick,red!80!black] (0,0,0) -- (0,0,\ew) node[right] {$Q$};
% a 
\draw[fill=green,opacity=0.1] (0*\ew,-1*\ew,-1*\ew) -- (-1*\ew,0*\ew,-1*\ew) -- (-1*\ew,-1*\ew,0*\ew) -- cycle;
\node[st1] at (0*\ew,-1*\ew,-1*\ew) {};
\node[st1] at (-1*\ew,0*\ew,-1*\ew) {};
\node[st1] at (-1*\ew,-1*\ew,0*\ew) {};
\draw[fill=green,opacity=0.1] (0*\ew,1*\ew,1*\ew) -- (1*\ew,0*\ew,1*\ew) -- (1*\ew,1*\ew,0*\ew) -- cycle;
\node[st1] at (0*\ew,1*\ew,1*\ew) {};
\node[st1] at (1*\ew,0*\ew,1*\ew) {};
\node[st1] at (1*\ew,1*\ew,0*\ew) {};
% cube 
\draw[] (-1*\ew,-1*\ew,0) -- (-1*\ew,-1*\ew,-1*\ew);
\draw[] (-1*\ew,0,-1*\ew) -- (-1*\ew,-1*\ew,-1*\ew);
\draw[] (0,-1*\ew,-1*\ew) -- (-1*\ew,-1*\ew,-1*\ew);
\draw[] (-1*\ew,-1*\ew,0) -- (-1*\ew,0,0);
\draw[] (-1*\ew,0,-1*\ew) -- (0,0,-1*\ew);
\draw[] (0,-1*\ew,-1*\ew) -- (0,0,-1*\ew);
\draw[] (-1*\ew,-1*\ew,0) -- (0,-1*\ew,0);
\draw[] (-1*\ew,0,-1*\ew) -- (-1*\ew,0,0);
\draw[] (0,-1*\ew,-1*\ew) -- (0,-1*\ew,0);
\draw[] (1*\ew,1*\ew,0) -- (1*\ew,1*\ew,1*\ew);
\draw[] (1*\ew,0,1*\ew) -- (1*\ew,1*\ew,1*\ew);
\draw[] (0,1*\ew,1*\ew) -- (1*\ew,1*\ew,1*\ew);
\draw[] (1*\ew,1*\ew,0) -- (1*\ew,0,0);
\draw[] (1*\ew,0,1*\ew) -- (0,0,1*\ew);
\draw[] (0,1*\ew,1*\ew) -- (0,0,1*\ew);
\draw[] (1*\ew,1*\ew,0) -- (0,1*\ew,0);
\draw[] (1*\ew,0,1*\ew) -- (1*\ew,0,0);
\draw[] (0,1*\ew,1*\ew) -- (0,1*\ew,0);
\end{scope}
\end{tikzpicture}
    \end{tabular}
   \caption{On the left is the weight diagram of $\mathfrak{u}(3)_R$, with the components of the free vector multiplet $\mathbf{B\bar{B}}_{(1,0)}$. Compare with the $\mathcal{N}=2$ case in Figure \ref{figureN=2}. On the right, this is the same diagram, with only the Lorentz scalars represented. Note that these are the same scalars as in the $\mathcal{N}=4$ vector multiplet. For clarity in the left diagram we have represented  only the $\mathbf{B\bar{B}}_{(1,0)}$ part of the full vector multiplet, the other half $\mathbf{B\bar{B}}_{(0,1)}$ having opposite weights.  }
    \label{vectorplet3d}
\end{figure}

\begin{figure}
    \centering
\tdplotsetmaincoords{100}{20}
\begin{tikzpicture}[tdplot_main_coords,font=\sffamily]
\newcommand\ew{1.5}
\newcommand\ma{4}
% axes
\draw[-latex] (0,0,0) -- (\ma,0,0) node[left] {};
\draw[-latex] (0,0,0) -- (0,\ma,0) node[below] {};
\draw[-latex] (0,0,0) -- (0,0,\ma) node[left] {};
\draw[] (0,0,0) -- (-\ma,0,0) node[left] {};
\draw[] (0,0,0) -- (0,-\ma,0) node[left] {};
\draw[] (0,0,0) -- (0,0,-\ma) node[left] {};
\tdplotsetrotatedcoords{0}{0}{0}
\begin{scope}[tdplot_rotated_coords]
\tikzstyle{st1}=[circle,inner sep=1.5pt,fill=green!80!black];
\tikzstyle{st2}=[circle,inner sep=1.5pt,fill=blue!80!black];
\tikzstyle{st3}=[circle,inner sep=1.5pt,fill=red!80!black];
\tikzstyle{st4}=[circle,inner sep=1.5pt,fill=yellow!80!black];
% susy
\draw[-latex,thick,red!80!black] (0,0,0) -- (\ew,0,0) node[right] {$Q$};
\draw[-latex,thick,red!80!black] (0,0,0) -- (0,\ew,0) node[right] {$Q$};
\draw[-latex,thick,red!80!black] (0,0,0) -- (0,0,\ew) node[right] {$Q$};
% a 
\draw[fill=green,opacity=0.1] (1*\ew,0*\ew,-1*\ew) -- (0*\ew,1*\ew,-1*\ew) -- (-1*\ew,1*\ew,0) -- (-1*\ew,0,1*\ew) -- (0,-1*\ew,1*\ew) -- (1*\ew,-1*\ew,0) -- cycle;
\node[st1] at (1*\ew,0*\ew,-1*\ew) {};
\node[st1] at (0*\ew,1*\ew,-1*\ew) {};
\node[st1] at (1*\ew,-1*\ew,0*\ew) {};
\node[st1] at (0*\ew,0*\ew,0*\ew) {};
\node[st1] at (0*\ew,0*\ew,0*\ew) {};
\node[st1] at (-1*\ew,1*\ew,0*\ew) {};
\node[st1] at (0*\ew,-1*\ew,1*\ew) {};
\node[st1] at (-1*\ew,0*\ew,1*\ew) {};
\draw[fill=blue,opacity=0.1] (1*\ew,1*\ew,0*\ew) -- (1*\ew,0*\ew,1*\ew) -- (0*\ew,1*\ew,1*\ew) -- cycle;
\node[st2] at (1*\ew,1*\ew,0*\ew) {};
\node[st2] at (1*\ew,0*\ew,1*\ew) {};
\node[st2] at (0*\ew,1*\ew,1*\ew) {};
\draw[fill=red,opacity=0.1] (-1*\ew,-1*\ew,0*\ew) -- (-1*\ew,0*\ew,-1*\ew) -- (0*\ew,-1*\ew,-1*\ew) -- cycle;
\node[st3] at (0*\ew,-1*\ew,-1*\ew) {};
\node[st3] at (-1*\ew,0*\ew,-1*\ew) {};
\node[st3] at (-1*\ew,-1*\ew,0*\ew) {};
% cube 
\draw[] (1*\ew,1*\ew,1*\ew) -- (-1*\ew,1*\ew,1*\ew) -- (-1*\ew,-1*\ew,1*\ew) -- (1*\ew,-1*\ew,1*\ew) -- (1*\ew,-1*\ew,-1*\ew) -- (1*\ew,1*\ew,-1*\ew) -- (-1*\ew,1*\ew,-1*\ew) -- (-1*\ew,-1*\ew,-1*\ew) -- (-1*\ew,-1*\ew,1*\ew);
\draw[] (-1*\ew,1*\ew,-1*\ew) -- (-1*\ew,1*\ew,1*\ew);
\draw[] (1*\ew,1*\ew,-1*\ew) -- (1*\ew,1*\ew,1*\ew);
\draw[] (1*\ew,-1*\ew,1*\ew) -- (1*\ew,1*\ew,1*\ew);
\draw[] (1*\ew,-1*\ew,-1*\ew) -- (-1*\ew,-1*\ew,-1*\ew);
\end{scope}
\end{tikzpicture}
    \caption{$\mathcal{N}=3$ energy-momentum multiplet, which is the $\mathrm{B\bar{B}}$ with $\mathbf{R}=(1,1)$, where only the Lorentz scalars are represented. The bottom component transforms in $[0,0]^{(1,1),0}_2$ and is represented in green. The components at $\Delta=3$ transform in $[0,0]^{(0,1),-2}_3$ (blue) and $[0,0]^{(1,0),2}_3$ (red). The colored planes correspond to $\mathfrak{su}(3)_R$ irreducible representations. }
    \label{figN=3EM}
\end{figure}
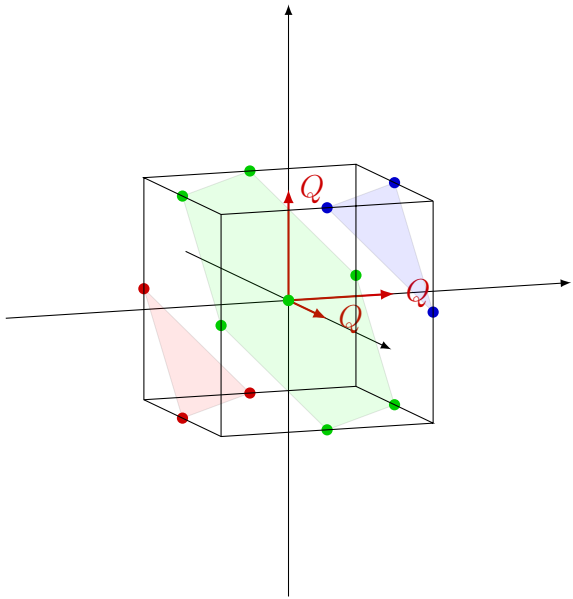

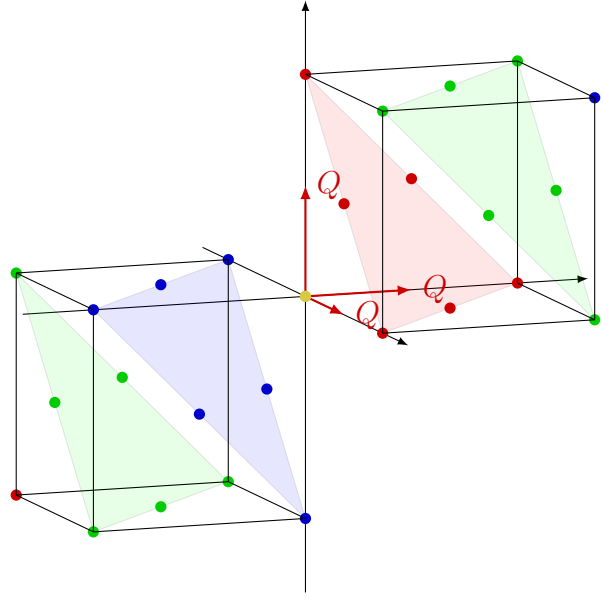
\begin{figure}
    \centering
\tdplotsetmaincoords{100}{20}
\begin{tikzpicture}[tdplot_main_coords,font=\sffamily]
\newcommand\ew{1.5}
\newcommand\ma{4}
\draw[-latex] (0,0,0) -- (\ma,0,0) node[left] {};
\draw[-latex] (0,0,0) -- (0,\ma,0) node[below] {};
\draw[-latex] (0,0,0) -- (0,0,\ma) node[left] {};
\draw[] (0,0,0) -- (-\ma,0,0) node[left] {};
\draw[] (0,0,0) -- (0,-\ma,0) node[left] {};
\draw[] (0,0,0) -- (0,0,-\ma) node[left] {};
\tdplotsetrotatedcoords{0}{0}{0}
\begin{scope}[tdplot_rotated_coords]
\tikzstyle{st1}=[circle,inner sep=1.5pt,fill=green!80!black];
\tikzstyle{st2}=[circle,inner sep=1.5pt,fill=blue!80!black];
\tikzstyle{st3}=[circle,inner sep=1.5pt,fill=red!80!black];
\tikzstyle{st4}=[circle,inner sep=1.5pt,fill=yellow!80!black];
% susy
\draw[-latex,thick,red!80!black] (0,0,0) -- (\ew,0,0) node[right] {$Q$};
\draw[-latex,thick,red!80!black] (0,0,0) -- (0,\ew,0) node[right] {$Q$};
\draw[-latex,thick,red!80!black] (0,0,0) -- (0,0,\ew) node[right] {$Q$};
% a 
\draw[fill=green,opacity=0.1] (0*\ew,-2*\ew,-2*\ew) -- (-2*\ew,0*\ew,-2*\ew) -- (-2*\ew,-2*\ew,0*\ew) -- cycle;
\node[st1] at (0*\ew,-2*\ew,-2*\ew) {};
\node[st1] at (-1*\ew,-1*\ew,-2*\ew) {};
\node[st1] at (-2*\ew,0*\ew,-2*\ew) {};
\node[st1] at (-1*\ew,-2*\ew,-1*\ew) {};
\node[st1] at (-2*\ew,-1*\ew,-1*\ew) {};
\node[st1] at (-2*\ew,-2*\ew,0*\ew) {};
\draw[fill=blue,opacity=0.1] (0,0,-2*\ew) -- (-2*\ew,0,0) -- (0,-2*\ew,0) -- cycle;
\node[st2] at (0*\ew,0*\ew,-2*\ew) {};
\node[st2] at (0*\ew,-1*\ew,-1*\ew) {};
\node[st2] at (-1*\ew,0*\ew,-1*\ew) {};
\node[st2] at (0*\ew,-2*\ew,0*\ew) {};
\node[st2] at (-1*\ew,-1*\ew,0*\ew) {};
\node[st2] at (-2*\ew,0*\ew,0*\ew) {};
\node[st3] at (-2*\ew,-2*\ew,-2*\ew) {};
\node[st4] at (0*\ew,0*\ew,0*\ew) {};
\draw[fill=green,opacity=0.1] (0*\ew,2*\ew,2*\ew) -- (2*\ew,0*\ew,2*\ew) -- (2*\ew,2*\ew,0*\ew) -- cycle;
\node[st1] at (2*\ew,2*\ew,0*\ew) {};
\node[st1] at (2*\ew,1*\ew,1*\ew) {};
\node[st1] at (1*\ew,2*\ew,1*\ew) {};
\node[st1] at (2*\ew,0*\ew,2*\ew) {};
\node[st1] at (1*\ew,1*\ew,2*\ew) {};
\node[st1] at (0*\ew,2*\ew,2*\ew) {};
\node[st2] at (2*\ew,2*\ew,2*\ew) {};
\draw[fill=red,opacity=0.1] (0,0,2*\ew) -- (2*\ew,0,0) -- (0,2*\ew,0) -- cycle;
\node[st3] at (2*\ew,0*\ew,0*\ew) {};
\node[st3] at (1*\ew,1*\ew,0*\ew) {};
\node[st3] at (0*\ew,2*\ew,0*\ew) {};
\node[st3] at (1*\ew,0*\ew,1*\ew) {};
\node[st3] at (0*\ew,1*\ew,1*\ew) {};
\node[st3] at (0*\ew,0*\ew,2*\ew) {};
\node[st4] at (0*\ew,0*\ew,0*\ew) {};
% cube 
\draw[] (-2*\ew,-2*\ew,0) -- (-2*\ew,-2*\ew,-2*\ew);
\draw[] (-2*\ew,0,-2*\ew) -- (-2*\ew,-2*\ew,-2*\ew);
\draw[] (0,-2*\ew,-2*\ew) -- (-2*\ew,-2*\ew,-2*\ew);
\draw[] (-2*\ew,-2*\ew,0) -- (-2*\ew,0,0);
\draw[] (-2*\ew,0,-2*\ew) -- (0,0,-2*\ew);
\draw[] (0,-2*\ew,-2*\ew) -- (0,0,-2*\ew);
\draw[] (-2*\ew,-2*\ew,0) -- (0,-2*\ew,0);
\draw[] (-2*\ew,0,-2*\ew) -- (-2*\ew,0,0);
\draw[] (0,-2*\ew,-2*\ew) -- (0,-2*\ew,0);
\draw[] (2*\ew,2*\ew,0) -- (2*\ew,2*\ew,2*\ew);
\draw[] (2*\ew,0,2*\ew) -- (2*\ew,2*\ew,2*\ew);
\draw[] (0,2*\ew,2*\ew) -- (2*\ew,2*\ew,2*\ew);
\draw[] (2*\ew,2*\ew,0) -- (2*\ew,0,0);
\draw[] (2*\ew,0,2*\ew) -- (0,0,2*\ew);
\draw[] (0,2*\ew,2*\ew) -- (0,0,2*\ew);
\draw[] (2*\ew,2*\ew,0) -- (0,2*\ew,0);
\draw[] (2*\ew,0,2*\ew) -- (2*\ew,0,0);
\draw[] (0,2*\ew,2*\ew) -- (0,2*\ew,0);
\end{scope}
\end{tikzpicture}
    \caption{$\mathcal{N}=3$ multiplets $\mathbf{B\bar{B}}$ with $\mathbf{R}=(2,0)$ and $\mathbf{R}=(0,2)$ where only the Lorentz scalars are represented. The bottom components transform in $[0,0]^{(2,0),4}_2$ and $[0,0]^{(0,2),-4}_2$ and are represented in green. Then we have components at conformal dimension $\Delta=3$, namely $[0,0]^{(0,2),2}_3$ (blue) and $[0,0]^{(0,0),6}_3$ (red) from the $\mathbf{R}=(2,0)$ multiplet, and similarly $[0,0]^{(2,0),-2}_3$ (red) and $[0,0]^{(0,0),-6}_3$ (blue) from the $\mathbf{R}=(0,2)$ multiplet. Finally there are two singlets at $\Delta=4$, $[0,0]^{(0,0),0}_4$ (yellow) which correspond to the exactly marginal deformations. The colored planes correspond to $\mathfrak{su}(3)_R$ irreducible representations. }
    \label{figN=3Other}
\end{figure}

The $\mathbf{\mathbf{B\Bb}}$ multiplets transform in representation (see Table \ref{tabMultiplets})
\begin{equation}
    [0,0]^{(R_1,R_2),r}_{\Delta} \qquad \textrm{with} \qquad r=2(R_1-R_2) \, \qquad  \Delta = R_1+R_2 \,. 
\end{equation}
They are entirely determined by $\bR=(R_1,R_2)$, so we denote them $\mathbf{B\Bb}_{\bR}$. 
They form the analog of the $\cN=2$ Higgs branch operators, though from the $\cN=2$ perspective they also contain Coulomb and Mixed branch operators. The $\mathbf{B\Bb}_{\bR}$ multiplets with $R_1+R_2 \le2$ are special: 
\begin{itemize}
    \item The $\bR=(0,0)$ is the identity operator. 
    \item The $\bR=(1,0)$ and $\bR=(0,1)$ are free vector multiplets. See Figure \ref{vectorplet3d}. 
        \item The $\bR=(1,1)$ is the $\cN=3$ stress-tensor multiplet, so must be present in a local CFT. See Figure \ref{figN=3EM}. Thus every local $\cN=3$ SCFT has a chiral ring, and, assuming this ring is not nilpotent, therefore a moduli space of vacua. The $\mathbf{B\Bb}_{\bR=(1,1)}$ multiplet contains an $\cN=2$ $\mathbf{B\Bb}_{\bR=1}$ Higgs branch multiplet and an $X\mathbf{\Bb}$ mixed branch multiplet both of dimension 2, but no $\cN=2$ Coulomb multiplets \cite{Lemos:2016xke}.  Since, as will be discussed in the next section, low energy $\cN=3$ supersymmetry implies that an $\cN=3$ moduli space will have both an $\cN=2$ Higgs branch and an $\cN=2$ Coulomb branch subspace, it follows that additional chiral multiplets beyond the $\mathbf{B\Bb}_{\bR=(1,1)}$ multiplet must occur in an $\cN=3$ SCFT. 
    \item The $\bR=(2,0)$ and $\bR=(0,2)$ contain additional conserved supercurrents, so their occurrence indicates an enhancement to $\cN=4$ supersymmetry. See figure \ref{figN=3Other}. 
\end{itemize}

The $\mathbf{B\Bb}_\bR=$ transforming in the $\bR=(R,0)$ contains CB operators \cite{Aharony:2015oyb,Lemos:2016xke}. The same argument for the anti-chiral ring implies that there must be a $\mathbf{B\Bb}$ multiplet with $\bR= (0,R)$ as well.  Symmetric products of the $\SU(3)_R$ $\bR=(1,1)$ (adjoint) irrep contain $(3n,0)$ and $(0,3n)$ irreps, so it is possible that the stress-energy OPEs can close on $\mathbf{B\Bb}$ multiplets of dimension $3n$ containing CB and mixed Higgs-Coulomb branch operators. Of course the $\cN=3$ operator spectrum can contain $\mathbf{B\Bb}$ in other $(R,0)$ and $(0,R)$ irreps, even though they can't appear in the stress-tensor multiplet OPEs with itself. In Figure \ref{su3fig2} we depict instead the chiral operators provided by three $\mathbf{B\Bb}$ in the $(4,0)$, $(0,4)$ and $(1,1)$ representation.\footnote{This example corresponds to the operator content of a physical rank-1 $\cN=3$ theory: the chiral ring generated by these operators is the coordinate ring of the $(k=4,\ell=4)$ theory in \cite{Aharony:2016kai}.}  The $\cN=3$ $\mathbf{B\Bb}_{(0,4)}$ multiplet has one $\cN=2$ $\mathbf{B\Bb}_4$ highest weight Higgs branch chiral ring field, while the $\cN=3$ $\mathbf{B\Bb}_{(4,0)}$ has as co-chiral ring fields: one $\cN=2$ $\mathbf{B\Bb}_4$ Higgs branch, three $\mathbf{X\Bb}_{3,2,1}$ mixed branch, and one $\mathbf{X\Bb}_0$ Coulomb branch, all of dimension 4.

Let's now repeat the analysis performed in the $\cN=2$ case and discuss the constraints arising from rotating the choice of Cartan subalgebra by conjugation by $\SU(3)_R$. These transformations act as a non-holomorphic isometries of $\cM$.  It translates the moduli space complex structure within the $\C\P^2$ of distinct TSK complex structures, as described in Section \ref{sec2.1}.  Such an $\SU(3)_R$ rotation thus generally gives a different way of splitting the moduli space coordinate ring into holomorphic and anti-holomorphic rings.

Again the Weyl group identifies those elements which preserve a given choice of the Cartan. In the $\SU(3)_R$, the Weyl group is the permutation group on 3 objects, $S_3 \simeq \Z_3 \rtimes \Z_2$, generated by elements $W_{(12)} \in \Z_2$ and $W_{(123)} \in \Z_3$  which we take to act on $\SU(3)$ weights as $W_{(12)}: (\l_1,\l_2) \mapsto ( -\l_1 , \l_2+\l_1 )$ and $W_{(123)}: (\l_1,\l_2) \mapsto (-\l_1-\l_2,\l_1)$.  Take the complex structure of $\cM$ induced by the $\Qb^{\bl=(0,1)}$ supercharge.  Then, since $W_{(12)}$ fixes $\bl=(0,1)$, a $\Z_2$ subgroup of the Weyl subgroup of $\SU(3)_R$ preserves the complex structure of $\cM$ and it maps the chiral ring fields to themselves, so its action gives a holomorphic involution of the chiral ring. This is apparent in the weight diagrams of Figures \ref{su3fig1} and \ref{su3fig2}, where $W_{(12)}$ acts by reflection through the line in the $(0,1)$ direction.  Note that from an $\cN=2$ perspective this involution mixes up the Higgs, mixed, and Coulomb branch operators.  For example, in the case of the $\cN=3$ $\mathbf{\mathbf{B\Bb}}_{(4,0)}$ multiplet, the $\cN=2$ $\mathbf{B\Bb}_4$ Higgs chiral ring field is exchanged with the $\cN=2$ $\mathbf{X\Bb}_4$ Coulomb field, and the $\cN=2$ $\mathbf{X\Bb}$ and $\mathbf{X\Bb}_3$ mixed branch chiral ring fields are also interchanged.  In the case of the $\cN=3$ $\mathbf{\mathbf{B\Bb}}_{(1,1)}$ multiplet, the $\cN=2$ $\mathbf{\mathbf{B\Bb}}_2$ Higgs chiral ring field is exchanged with the $\cN=2$ $X\mathbf{\Bb}$ mixed branch chiral ring field.

The other four non-trivial elements of the $\SU(3)_R$ Weyl group besides $W_{(12)}$ reflect or rotate $\Qb^{(0,1)}$ to the $\Qb^{(-1,0)}$ or $\Qb^{(1,-1)}$ supercharges so they change the  complex structure on $\cM$.  These elements thus map the chiral ring to a different co-chiral set of multiplet components forming an isomorphic chiral ring.

\subsection{Differences between $\cN=4$ and genuinely $\cN=3$ chiral rings}\label{sec2.4}

Because of the enlarged bosonic symmetry of the super-conformal algebra, $\cN=4$ superconformal representations are labelled by a $\SU(4)$ weights $\bR=(R_1,R_2,R_3)$. It can be shown that conserved currents can only resides in $\mathbf{A\bar{A}}$ in the singlet representation of $\SU(4)$ and $\mathbf{\mathbf{B\Bb}}$ multiplet with $R_1+R_2+R_3\leq2$ \cite{Cordova:2016emh}. The $\mathbf{B\Bb}$ have $\D=R_1+R_2+R_3$ as well as $R_1=R_3$. Of particular relevance are those with $R_1=R_3=0$ and in particular:

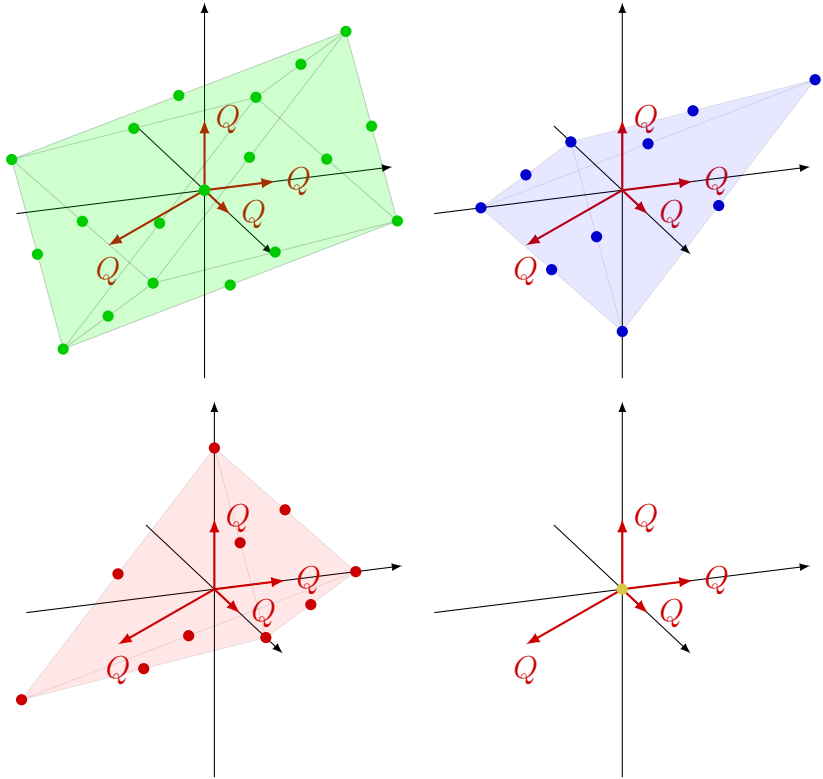
\begin{figure}
    \centering
\tdplotsetmaincoords{110}{20}
\begin{tikzpicture}[tdplot_main_coords,font=\sffamily]
\newcommand\ew{1}
\newcommand\ma{2.66}
\draw[-latex] (0,0,0) -- (\ma,0,0) node[left] {};
\draw[-latex] (0,0,0) -- (0,\ma,0) node[below] {};
\draw[-latex] (0,0,0) -- (0,0,\ma) node[left] {};
\draw[] (0,0,0) -- (-\ma,0,0) node[left] {};
\draw[] (0,0,0) -- (0,-\ma,0) node[left] {};
\draw[] (0,0,0) -- (0,0,-\ma) node[left] {};
\tdplotsetrotatedcoords{0}{0}{0}
\begin{scope}[tdplot_rotated_coords]
\tikzstyle{st1}=[circle,inner sep=1.5pt,fill=green!80!black];
% susy
\draw[-latex,thick,red!80!black] (0,0,0) -- (\ew,0,0) node[right] {$Q$};
\draw[-latex,thick,red!80!black] (0,0,0) -- (0,\ew,0) node[right] {$Q$};
\draw[-latex,thick,red!80!black] (0,0,0) -- (0,0,\ew) node[right] {$Q$};
\draw[-latex,thick,red!80!black] (0,0,0) -- (-\ew,-\ew,-\ew) node[below] {$Q$};
% cycles
\draw[fill=green,opacity=0.1] (0*\ew,2*\ew,2*\ew) -- (-2*\ew,-2*\ew,0*\ew) -- (2*\ew,0*\ew,2*\ew) -- cycle;
\draw[fill=green,opacity=0.1] (2*\ew,0*\ew,2*\ew) -- (0*\ew,-2*\ew,-2*\ew) -- (2*\ew,2*\ew,0*\ew) -- cycle;
\draw[fill=green,opacity=0.1] (2*\ew,2*\ew,0*\ew) -- (-2*\ew,0*\ew,-2*\ew) -- (0*\ew,2*\ew,2*\ew) -- cycle;
\draw[fill=green,opacity=0.1] (0*\ew,-2*\ew,-2*\ew) -- (2*\ew,2*\ew,0*\ew) -- (-2*\ew,0*\ew,-2*\ew) -- cycle;
\draw[fill=green,opacity=0.1] (-2*\ew,0*\ew,-2*\ew) -- (0*\ew,2*\ew,2*\ew) -- (-2*\ew,-2*\ew,0*\ew) -- cycle;
\draw[fill=green,opacity=0.1] (-2*\ew,-2*\ew,0*\ew) -- (2*\ew,0*\ew,2*\ew) -- (0*\ew,-2*\ew,-2*\ew) -- cycle;
\draw[fill=green,opacity=0.1] (0*\ew,2*\ew,2*\ew) -- (2*\ew,0*\ew,2*\ew) -- (2*\ew,2*\ew,0*\ew) -- cycle;
\draw[fill=green,opacity=0.1] (0*\ew,-2*\ew,-2*\ew) -- (-2*\ew,0*\ew,-2*\ew) -- (-2*\ew,-2*\ew,0*\ew) -- cycle;
% a 
\node[st1] at (1*\ew,0*\ew,-1*\ew) {};
\node[st1] at (0*\ew,1*\ew,-1*\ew) {};
\node[st1] at (1*\ew,-1*\ew,0*\ew) {};
\node[st1] at (0*\ew,0*\ew,0*\ew) {};
\node[st1] at (0*\ew,0*\ew,0*\ew) {};
\node[st1] at (-1*\ew,1*\ew,0*\ew) {};
\node[st1] at (0*\ew,-1*\ew,1*\ew) {};
\node[st1] at (-1*\ew,0*\ew,1*\ew) {};
\node[st1] at (0*\ew,-2*\ew,-2*\ew) {};
\node[st1] at (-1*\ew,-1*\ew,-2*\ew) {};
\node[st1] at (-2*\ew,0*\ew,-2*\ew) {};
\node[st1] at (-1*\ew,-2*\ew,-1*\ew) {};
\node[st1] at (-2*\ew,-1*\ew,-1*\ew) {};
\node[st1] at (-2*\ew,-2*\ew,0*\ew) {};
\node[st1] at (2*\ew,2*\ew,0*\ew) {};
\node[st1] at (2*\ew,1*\ew,1*\ew) {};
\node[st1] at (1*\ew,2*\ew,1*\ew) {};
\node[st1] at (2*\ew,0*\ew,2*\ew) {};
\node[st1] at (1*\ew,1*\ew,2*\ew) {};
\node[st1] at (0*\ew,2*\ew,2*\ew) {};
\end{scope}
\end{tikzpicture}
\begin{tikzpicture}[tdplot_main_coords,font=\sffamily]
\newcommand\ew{1}
\newcommand\ma{2.66}
\draw[-latex] (0,0,0) -- (\ma,0,0) node[left] {};
\draw[-latex] (0,0,0) -- (0,\ma,0) node[below] {};
\draw[-latex] (0,0,0) -- (0,0,\ma) node[left] {};
\draw[] (0,0,0) -- (-\ma,0,0) node[left] {};
\draw[] (0,0,0) -- (0,-\ma,0) node[left] {};
\draw[] (0,0,0) -- (0,0,-\ma) node[left] {};
\tdplotsetrotatedcoords{0}{0}{0}
\begin{scope}[tdplot_rotated_coords]
\tikzstyle{st2}=[circle,inner sep=1.5pt,fill=blue!80!black];
% susy
\draw[-latex,thick,red!80!black] (0,0,0) -- (\ew,0,0) node[right] {$Q$};
\draw[-latex,thick,red!80!black] (0,0,0) -- (0,\ew,0) node[right] {$Q$};
\draw[-latex,thick,red!80!black] (0,0,0) -- (0,0,\ew) node[right] {$Q$};
\draw[-latex,thick,red!80!black] (0,0,0) -- (-\ew,-\ew,-\ew) node[below] {$Q$};
% cycles
\draw[fill=blue,opacity=0.05] (-2*\ew,0,0) -- (0,-2*\ew,0) -- (2*\ew,2*\ew,2*\ew) -- cycle;
\draw[fill=blue,opacity=0.05] (0,0,-2*\ew) -- (0,-2*\ew,0) -- (2*\ew,2*\ew,2*\ew) -- cycle;
\draw[fill=blue,opacity=0.05] (0,0,-2*\ew) -- (-2*\ew,0,0) -- (2*\ew,2*\ew,2*\ew) -- cycle;
\draw[fill=blue,opacity=0.05] (0,0,-2*\ew) -- (-2*\ew,0,0) -- (0,-2*\ew,0) -- cycle;
% a 
\node[st2] at (1*\ew,1*\ew,0*\ew) {};
\node[st2] at (1*\ew,0*\ew,1*\ew) {};
\node[st2] at (0*\ew,1*\ew,1*\ew) {};
\node[st2] at (0*\ew,0*\ew,-2*\ew) {};
\node[st2] at (0*\ew,-1*\ew,-1*\ew) {};
\node[st2] at (-1*\ew,0*\ew,-1*\ew) {};
\node[st2] at (0*\ew,-2*\ew,0*\ew) {};
\node[st2] at (-1*\ew,-1*\ew,0*\ew) {};
\node[st2] at (-2*\ew,0*\ew,0*\ew) {};
\node[st2] at (2*\ew,2*\ew,2*\ew) {};
\end{scope}
\end{tikzpicture}
\begin{tikzpicture}[tdplot_main_coords,font=\sffamily]
\newcommand\ew{1}
\newcommand\ma{2.66}
\draw[-latex] (0,0,0) -- (\ma,0,0) node[left] {};
\draw[-latex] (0,0,0) -- (0,\ma,0) node[below] {};
\draw[-latex] (0,0,0) -- (0,0,\ma) node[left] {};
\draw[] (0,0,0) -- (-\ma,0,0) node[left] {};
\draw[] (0,0,0) -- (0,-\ma,0) node[left] {};
\draw[] (0,0,0) -- (0,0,-\ma) node[left] {};
\tdplotsetrotatedcoords{0}{0}{0}
\begin{scope}[tdplot_rotated_coords]
\tikzstyle{st3}=[circle,inner sep=1.5pt,fill=red!80!black];
% susy
\draw[-latex,thick,red!80!black] (0,0,0) -- (\ew,0,0) node[right] {$Q$};
\draw[-latex,thick,red!80!black] (0,0,0) -- (0,\ew,0) node[right] {$Q$};
\draw[-latex,thick,red!80!black] (0,0,0) -- (0,0,\ew) node[right] {$Q$};
\draw[-latex,thick,red!80!black] (0,0,0) -- (-\ew,-\ew,-\ew) node[below] {$Q$};
% cycles
\draw[fill=red,opacity=0.05] (2*\ew,0,0) -- (0,2*\ew,0) -- (-2*\ew,-2*\ew,-2*\ew) -- cycle;
\draw[fill=red,opacity=0.05] (0,0,2*\ew) -- (0,2*\ew,0) -- (-2*\ew,-2*\ew,-2*\ew) -- cycle;
\draw[fill=red,opacity=0.05] (0,0,2*\ew) -- (2*\ew,0,0) -- (-2*\ew,-2*\ew,-2*\ew) -- cycle;
\draw[fill=red,opacity=0.05] (0,0,2*\ew) -- (2*\ew,0,0) -- (0,2*\ew,0) -- cycle;
% a 
\node[st3] at (0*\ew,-1*\ew,-1*\ew) {};
\node[st3] at (-1*\ew,0*\ew,-1*\ew) {};
\node[st3] at (-1*\ew,-1*\ew,0*\ew) {};
\node[st3] at (-2*\ew,-2*\ew,-2*\ew) {};
\node[st3] at (2*\ew,0*\ew,0*\ew) {};
\node[st3] at (1*\ew,1*\ew,0*\ew) {};
\node[st3] at (0*\ew,2*\ew,0*\ew) {};
\node[st3] at (1*\ew,0*\ew,1*\ew) {};
\node[st3] at (0*\ew,1*\ew,1*\ew) {};
\node[st3] at (0*\ew,0*\ew,2*\ew) {};
\end{scope}
\end{tikzpicture}
\begin{tikzpicture}[tdplot_main_coords,font=\sffamily]
\newcommand\ew{1}
\newcommand\ma{2.66}
\draw[-latex] (0,0,0) -- (\ma,0,0) node[left] {};
\draw[-latex] (0,0,0) -- (0,\ma,0) node[below] {};
\draw[-latex] (0,0,0) -- (0,0,\ma) node[left] {};
\draw[] (0,0,0) -- (-\ma,0,0) node[left] {};
\draw[] (0,0,0) -- (0,-\ma,0) node[left] {};
\draw[] (0,0,0) -- (0,0,-\ma) node[left] {};
\tdplotsetrotatedcoords{0}{0}{0}
\begin{scope}[tdplot_rotated_coords]
\tikzstyle{st4}=[circle,inner sep=1.5pt,fill=yellow!80!black];
% susy
\draw[-latex,thick,red!80!black] (0,0,0) -- (\ew,0,0) node[right] {$Q$};
\draw[-latex,thick,red!80!black] (0,0,0) -- (0,\ew,0) node[right] {$Q$};
\draw[-latex,thick,red!80!black] (0,0,0) -- (0,0,\ew) node[right] {$Q$};
\draw[-latex,thick,red!80!black] (0,0,0) -- (-\ew,-\ew,-\ew) node[below] {$Q$};
% a 
\node[st4] at (0*\ew,0*\ew,0*\ew) {};
\end{scope}
\end{tikzpicture}
    \caption{$\mathcal{N}=4$ stress-tensor multiplet, which is the $\mathbf{B\bar{B}}$ with $\mathbf{R}=(0,2,0)$, where only the Lorentz scalars components are represented. Each dot represents an $\suf(4)_R$ weight; the $\suf(4)_R$ weight lattice has been drawn in such a way that the $\uf(3)_R$ lattice of the previous figures is a sublattice. The bottom component transforms in $[0,0]^{(0,2,0)}_2$ and is represented in green. The components at $\Delta=3$ transform in $[0,0]^{(0,0,2)}_3$ (blue) and $[0,0]^{(2,0,0)}_3$ (red). Finally we have two singlet scalars in $[0,0]^{(0,0,0)}_4$ which correspond to the exactly marginal deformations (yellow).  }
    \label{figN=4EM}
\end{figure}

\begin{enumerate}[leftmargin=2.7cm]
    \item[$\boldsymbol{\mathbf{B\Bb}}_{(0,1,0)}$] This is the $\cN=4$ free vector multiplet that coincides with the $\cN=3$ $\mathbf{B\Bb}_{(1,0)}\oplus\mathbf{B\Bb}_{(0,1)}$ free vector multiplet, see figure \ref{vectorplet3d}.
    \item[$\boldsymbol{\mathbf{B\Bb}}_{(0,2,0)}$] This is the $\cN=4$ stress-tensor multiplet, see figure \ref{figN=4EM}. In $\cN=3$ language this decomposes as the sum of three multiplet, namely the $\mathbf{B\Bb}_{(2,0)}$, $\mathbf{B\Bb}_{(0,2)}$ and $\mathbf{B\Bb}_{(1,1)}$. The latter is the $\cN=3$ stress-tensor multiplet while the first two contain extra supercurrents, an exactly marginal deformation as well as a chiral $\cN=2$ CB operator.
    \item[$\boldsymbol{\mathbf{B\Bb}}_{(0,p,0)}$] This multiplet contains $\cN=2$ Coulomb branch of dimensions $p$. It decomposes into $p+1$ $\cN=3$ multiplets, namely 
    \beq
    \mathbf{B\Bb}_{(0,p,0)}\ \quad \xrightarrow[\text{$\cN{=}4$ to $\cN{=}3$}]\quad \bigoplus_{R_1+R_2=p}\mathbf{B\Bb}_{(R_1,R_2)}
    \eeq
\end{enumerate}

\noindent From the decomposition above it follows, even more straightforwardly than in the $\cN=3$ case, that $\cN=4$ theories always have a CB and even more specifically a dimension 2 globally defined CB coordinate as well as an exactly marginal operator which is identified with the holomorphic gauge coupling of the theory. A genuine $\cN=3$ theory has, on the contrary, no exactly marginal deformation and no dimension 2 CB coordinate.

Since the structure of the free $\cN=3$ superconformal multiplet, $\mathbf{B\Bb}_{(1,0)}$ and $\mathbf{B\Bb}_{(0,1)}$ coincides with the free $\cN=4$ one, $\mathbf{B\Bb}_{(0,1,0)}$, the structure of the moduli space of $\cN=4$ theories is largely similar to what we have thus far described. The main difference is that in the $\cN=4$ case the full space of $\C\P^3\cong \SU(4)/\SU(3)\times \U(1)$ of possible complex structures, is physical. This will be described in more details below but can already be understood by noticing that two components of the $\mathbf{B\Bb}_{(0,1,0)}$ primary are chiral/anti-chiral along with two more that are each separately chiral and anti-chiral. Thus the complex structure of the $\cN=3$ moduli space where all the special coordinates are holomorphic, is now compatible with the structure of the SCFT.

Let's quickly discuss some possible restriction on $\cN=4$ multiplets. In \cite{Bonetti:2018fqz} is noted that the multiplets parametrizing the moduli space of a rank-$r$ $\cN=4$ theory, can be accounted by the chiral component of $r$ $\mathbf{B\Bb}_{0,\D_i,0}$, where $\D_i$, $i=1,...,r$, are the CB scaling of the theory under the $\cN=2$ decomposition. This is true for all known cases but those with $D_n$ gauge Lie Algebra with $n\geq4$. In this case extra multiplet are present. For $n<6$ those are all of the form $\mathbf{B\Bb}_{(3,n-4,3)}$ but more operators are present to account for the moduli space structure as $n$ increases. 

%Any $\mathbf{X\Bb}$, with $\mathbf{X}\in\{\mathbf{A},\mathbf{L}\}$, that has extra scalar chiral components is therefore not allowed. It is likely that this condition is very restrictive. 

\section{Conditions from low energy $\cN=3$ supersymmetry}\label{sec2.1}

In the sections above, we have analyzed necessary conditions for the existence of an $\cN=3$ moduli space of vacua and some of the contraints implied by the structure of the interacting superconformal operator algebra. We now revise this discussion assuming the existence of vacua which preserve $\cN=3$ supersymmetry and anlyze in detail the implications on the low-energy effective theory that follow. 

First, if the CFT has a vacuum spontaneously breaking conformal invariance but preserving $\cN=3$ supersymmetry, then it necessarily has a dilaton, and so a whole moduli space of vacua,\footnote{We assume that the different vacua of the moduli space are distinguished by the vevs of a finite set of real scalar fields, and $\cM$ inherits the topology from open sets in the space of these vevs.  This makes $\cM$ regular Hausdorff, second-countable, and thus metrizable, as a topological space.} $\cM$.  In the IR, the dilaton, being a Nambu-Goldstone boson, is free.  The only free massless $\cN{=}3$ multiplets respecting CPT symmetry with scalars are the vector multiplet, or $\mathbf{B\Bb}_{(1,0)}\oplus\mathbf{B\Bb}_{(0,1)}$, (with field content 3 complex scalars, 4 Weyl fermions, and a vector boson, see Figure \ref{vectorplet3d}) and a gravitino multiplet (1 complex scalar, 3 Weyl fermions, 3 vector bosons, and a spin-3/2 fermion) \cite{Wess:1992cp}.  Assuming the existence of a local energy-momentum tensor, the massless particles must have spins $\le 1$ \cite{Weinberg:1980kq}, so the dilaton and any other IR free massless scalars on $\cM$ must belong to $\cN=3$ vector multiplets.

The vector multiplet contains a single $\U(1)$ gauge field, so spontaneously broken $\cN=3$ superconformal invariance implies $\cM$ is a ``Coulomb branch''. Furthermore the bottom components of an $\cN=3$ vector multiplet are 3 complex scalars, $a^I$, $I=1,2,3$, which transform in the ${\bf 3}_2$ representation of an IR effective $\U(3)_*$ R-symmetry, where the subscript refers to the charge of the overall $\U(1)$ factor.\footnote{By choosing the vector multiplet scalars to have overall $\U(1)$ R-charge $+2$ we are using the convention of \cite{Cordova:2016emh,Lemos:2016xke}. This differs from that of others \cite{Aharony:2015oyb} which assigns them charge $-2$. } The $*$ subscript is to emphasize that this $\U(3)_*$ is an ``accidental'' symmetry in the IR at any given vacuum $*\in\cM$.  Thus, at a vacuum where the vector multiplet scalars have vevs $a^I_*$, the $\U(3)_*$ IR symmetry acts on the scalar fields as
\begin{align}\label{u3*}
    (a^I - a^I_*) &\mapsto R^I_J (a^J - a^J_*), &
    R &\in \U(3)_* .
\end{align}

$\U(3)_R$ is the automorphism group of the unbroken $\cN=3$ supersymmetry algebra, but a priori need not be a symmetry of the full theory. It is a symmetry of an $\cN=3$ SCFT.  We will work out below how the $\U(3)_*$ action on the moduli space is related to that of the $\U(3)_R$ symmetry of an $\cN=3$ SCFT.\footnote{The low energy theory of a free $\cN{=}3$ vector multiplet actually has an accidentally enhanced $\cN{=}4$ supersymmetry as well as an accidental $\SU(4)_*$ R-symmetry. We will focus here on the $\U(3)_*$ subgroup under which the microscopic $\cN{=}3$ supercharges transform as a ${\bf 3}_{-1}$.  The additional constraints on the moduli space geometry that occur when there is a microscopic $\cN{=}4$ supersymmetry were discussed in section \ref{sec2.4}.}

The points of $\cM_\text{smooth} \subset \cM$ where there are only free massless vector multiplets --- i.e., where there is a finite mass gap to any other states in the theory --- are thus the target space of an $\cN=3$ sigma model.  $\cM_\text{smooth}$ inherits a metric from the kinetic terms of the massless scalar fields.  This means that $\cM$ is a metric space in the sense that it admits a distance function.  $\cM$ must be metrically complete:  any ``missing'' points at finite distance can be probed by an arbitrarily low energy fluctuation from a neighboring vacuum, and so must be included in the vacuum moduli space.

But $\cM$ need not be smooth as a metric space: its distance function can have non-analyticities along subspaces where additional electrically and magnetically charged states become massless.  Equivalently, its Riemannian metric tensor is ill-defined at these subspaces.  We will call the locus of metric non-analyticities of $\cM$ the \emph{metric singularities} of $\cM$, $\cM_\text{metric} := \cM \setminus \cM_\text{smooth}$.  We will see that this locus is the source of (generalized) deficit angles or, equivalently, delta function supported curvature contributions.  The metric singularities can also contain the locus of singularities in the complex structures, $\cM_{\rm cplx}$, of $\cM$ and places where $\cM$ fails to be a topological manifold. This will be discussed in detail in section \ref{TSKspace}. 

If a vacuum of the SCFT has $n$ massless vector multiplets at low energies with complex scalar components $a^I_i$, $i=1,\ldots,n$, then low energy $\cN=3$ supersymmetry prohibits any potential terms in the effective action for the $a^I_i$.  Then the component of $\cM_\text{smooth}$ containing that vacuum is a metrically smooth $3n$-complex-dimensional manifold.  $n$ is called the rank of (that component of) $\cM$.  Also, we will call the vevs of the $a^I_i$ \emph{special coordinates} on $\cM_\text{smooth}$, since they are closely analogous to the special coordinates of $\cN=2$ Coulomb branch special K\"ahler geometry.

At any vacuum in $\cM_\text{smooth}$ the $\cN=3$ supertranslation algebra is
\begin{align}\label{N3susy}
\{Q^I,Q^J\} & = %\e_{\a\b}
\e^{IJK}\bar Z_K ,&
\{Q^I,\bar Q_J\} & = \d^I_J P,
\end{align}
where the $Z^K$ are a triplet of complex scalar central charges.  Lorentz indices, Pauli matrices and the like have been suppressed, as they are easily and uniquely determined from the Lorentz irreps of the generators which are given below in \eqref{vectorcharges}.

The on-shell transformation rules of the massless $\cN=3$ vector multiplet component fields are easily worked out to be
\begin{align}\label{vectortransf}
Q^I a_j^J\ \,&= \e^{IJK}\, \psi_{Kj} &
\bar Q_I a_j^J\ \,&= \d^J_I\, \bar\psi_j \nn\\
Q^I \psi_{Jj} &= \d^I_J\, \cF_j &
\bar Q_I \psi_{Jj} &= \e_{IJK}\, P\, a_j^K \nn\\
Q^I \psi_j \ \,&= 0 &
\bar Q_I \psi_j \ &= P\, \bar a_{Ij} \\
Q^I \cF_j \ &= P\, \bar\psi_j^I &
\bar Q_I \cF_j \ &= P\, \psi_{Ij} \nn
\end{align}
where the bars denote complex conjugation, and we have used the Lorentz and $R$-charges of the generators and fields,
\begin{align}\label{vectorcharges}
\begin{array}{c|cc|cccc}
\text{operator} & Q^I & P & a^I_j & \bar{\psi}_j & \psi_{Ij} & \cF_j \\
\hline
[j_L,j_R]_\D^{\bR \, , r} &
[1,0]^{(1,0),-1}_{1/2} &
[1,1]^{(0,0),\,0}_{1} &
[0,0]^{(1,0),+2}_{1} &
[0,1]^{(0,0),+3}_{3/2} &
[1,0]^{(0,1),+1}_{3/2} &
[2,0]^{(0,0),\,0}_{2}
\end{array}
\end{align}
where $j=1,\ldots,n$ labels the rank $n$ different vector multiplets. We have used the notation (\ref{eqA2}), and the $\U(3)_R$ charges of the $\cN=3$ vector multiplet fields are illustrated in figure \ref{vectorplet3d}.

The supersymmetry algebra \eqref{N3susy} implies the BPS bound on masses 
\begin{equation}\label{BPSbound}
    M\ge (Z^K \bar Z_K)^{1/2}:=|\bZ| \, . 
\end{equation}
Since the triplet of charges $Z^K$ are central, they are linear combinations of conserved (non-R) $\uf(1)$ charges in the theory.  Since there are no $\cN=3$ global flavor symmetries in a local unitary $\cN=3$ SCFT \cite{Aharony:2015oyb}, on $\cM_\text{smooth}$ the central charge is a linear combination of the $n$ electric and $n$ magnetic charges $(q^i,p_i)$ of a state. By computing the central charge from the low energy $\U(1)^n$ $\cN=3$ sigma model, just as for $\cN=2$ supersymmetry \cite{Seiberg:1994rs}, their coefficients are, up to possible constant shifts, the special coordinates, $a^I_i$, and the dual special coordinates $a^{Ii}_D$ (defined below).  Since the constant parts of the special and dual special coordinates are not determined by the sigma model, we are free to choose them in each coordinate chart on $\cM_\text{smooth}$ so that the central charges are given by
\begin{align}\label{N3Z}
Z^K(q^i,p_i) = q^i a_i^K + p_i a_D^{Ki}
\end{align}
for $(q^i, p_i) \in \Z^{2n}$.  The BPS bound implies that metric singularities can occur when $Z^K(q^i, p_i) = 0$ for some charges $(q^i, p_i)$ and for all $K \in \{1,2,3\}$. 

The set of metric singularities $\cM_\mathrm{metric} \subset \cM$ is given by the locus of points where charged states become massless on $\cM$, which can only happen when the BPS mass bound \eqref{BPSbound} is zero.  This means that $\cM_\mathrm{metric}$ is given by the locus of points where the $\U(3)$ vector of central charges, $Z^I$ for $I=1,2,3$, for charges $Q := (p_i,q^i)$ of states which saturate the BPS bound, satisfy
\begin{equation}\label{vanishZ}
    Z^I(Q) = 0, \quad I=1,2,3 . 
\end{equation}
Note that, unlike the $\cN=2$ case where the BPS mass bound on the Coulomb branch, $M \ge |Z(Q)|$, vanishes when a single complex function vanishes, in the $\cN=3$ case the bound is $M \ge |\bZ(Q)|$ where we define $|\bZ(Q)| := (Z^K(Q) \bar Z_K(Q))^{1/2}$.  So $M$ only vanishes when all three components of $Z^K(Q)$ vanish.

This implies not only that metric singularities occur in complex codimension a multiple of 3\footnote{As discussed in an analogous situation in section 2.2 of \cite{Argyres:2018zay}, and in more detail in section \ref{secBPS}, an additional assumption that the common zeros of the central charges for BPS states in the spectrum are nowhere dense in $\cM$ may be necessary to arrive at this conclusion.}, but also that possible ``walls'' of marginal stability of BPS states occur in real codimension at least 5, so are not walls at all.  To see this, consider two BPS states with charges $Q$ and $Q'$.  Then a point of marginal stability for these two states is one where $|\bZ(Q) + \bZ(Q')| = |\bZ(Q)| + |\bZ(Q')|$, which is equivalent to 
% $\Re (\bZ_Q \cdot \bar\bZ_{Q'}) = |\bZ_Q|\, |\bZ_{Q'}|$.  Since $\Re (\bZ_Q \cdot \bar\bZ_{Q'}) \le |\bZ_Q \cdot \bar\bZ_{Q'}| \le |\bZ_Q|\, |\bZ_{Q'}|$ by the Cauchy-Schwarz inequality, points of marginal stability only occur if 
$\bZ(Q') = \a \bZ(Q)$ for $\a$ a positive real number.  Thus a locus of marginal stability is defined by 5 real conditions. So the BPS spectrum of an $\cN=3$ SCFT is constant on each connected component of $\cM_\text{smooth}$. Metric singularities will be discussed in much more detail in section \ref{secBPS}.

$\cM_\text{smooth}$ is flat, since, following an argument of \cite{Cordova:2016xhm}, there are no appropriate irrelevant $\cN{=}3$ preserving deformations that can give rise to a metric curvature term.  In fact, $\cN=3$ supersymmetry implies more: not only is $\cM_\text{smooth}$ metrically flat, but also flat coordinates on $\cM_\text{smooth}$ are the vevs of the $a^I_i$ fields:  the special coordinates are flat coordinates.   This follows immediately from familiar facts about $\cN=2$ Coulomb branch effective actions.  The $\cN=3$ vector multiplet decomposes as a neutral $\cN=2$ hypermultiplet and an $\cN=2$ vector multiplet.  The $\cN = 3$ IR effective action has the same form as the $\cN = 2$ enhanced Coulomb branch effective action, but now since the $\cN = 2$ hypermultiplet and vector multiplet scalars are related by the $\U(3)_*$ IR effective global symmetry, they must have the same kinetic terms, 
\begin{align}\label{LIR}
\cL_\text{bosonic} = \Im \left[ \t^{ij}(a)
\left(\del a_i^I \cdot \del \bar a_{Ij} + \cF_i \cdot \cF_j\right)
\right] .
\end{align}
This form and unitarity of the low energy effective action imply
\begin{align}\label{TSK0}
\t^{ij} &= \t^{ji} & &\text{and}& \Im \, \t  > 0 .
\end{align}
The second equation says that the matrix $\Im \, \t $ is positive definite. 
Since an $\cN = 2$ selection rule \cite{Argyres:1996eh} forbids the vector multiplet metric and the hypermultiplet metric from depending on the same fields, we must have
\begin{align}\label{TSK1}
\t^{ij} = \text{constant} .
\end{align}
This has the immediate consequence that $\cM_\text{smooth}$ is not only flat, but also that the $a_i^I$ are its flat coordinates.  

Indeed, the moduli space effective action \eqref{LIR} implies the metric on $\cM_\text{smooth}$ is
\begin{align}\label{TSKmetric}
    g &= \Im\t^{ij} \, \big(
     \mathrm{d} a^I_i \mathrm{d}\ab_{Ij} + \mathrm{d}\ab_{Ii} \mathrm{d}a^I_j \big) . 
\end{align}
As noted earlier, this extends to a metric structure (distance function) on all of $\cM$.

Since they are flat coordinates, on overlaps of special coordinate charts the two sets of special coordinates must be linearly related, perhaps up to a constant shift.  Since BPS masses are well-defined functions on $\cM$, it follows from \eqref{N3Z} that on overlaps of special coordinate charts the two sets of special coordinates are strictly linearly related, i.e., without constant shifts.

While the effective action \eqref{LIR} is invariant under $\U(3)_*$ transformations \eqref{u3*} for any choice of $a^J_{j*}$, this extends to covariance of the central charges \eqref{N3Z} only for the choice $a^J_{j*}=0$ (whether or not this value of the coordinates occurs in any given special coordinate chart).  We will denote this action of the automorphism group of the $\cN=3$ supersymmetry algebra on the low energy theory and BPS masses by $\U(3)_R$.  We will see below that it coincides with that of the $\U(3)_R$ global symmetry in the case of $\cN=3$ SCFTs.  The $\U(3)_R$ acts such that the $a^J_j$ special coordinates transform in the ${\bf 3}_2 \otimes \I_n := {\bf 3}_2 \oplus {\bf 3}_2 \oplus \cdots \oplus {\bf 3}_2$ representation.  This follows from the $\cN=3$ supersymmetry algebra \eqref{N3susy} which implies that the central charges $Z^K$ transform in the ${\bf 3}_2$.  For this to be true for all $(q^i,p_i)$ EM charge sectors, \eqref{N3Z} then implies the claimed $\U(3)_R$ action. We have identified the IR effective $\cN=3$ supercharges in \eqref{N3susy} with the microscopic supercharges in the $\cN=3$ supersymmetry algebra since there is no other triplet of supercharges with which they could mix. 

The assumed existence of the $\cN=3$ supersymmetry charges means that relative to a choice of basis of the supercharges, $\{Q^J, J=1,2,3\}$, there is a corresponding choice of equivalence class, $[a^J_j]$, of the special coordinates in every coordinate chart of the atlas, determined by \eqref{vectortransf}. The equivalence relation is $a^J_j \sim \r^i_j a^J_i$ with $\r\in\GL(n,\C)$.  In other words, the choice of supercharge basis induces a preferred choice of $\U(3)_R$ orientation of the special coordinates everywhere on $\cM$.  

Since $\cM$ is a Coulomb branch, it carries an $\cN=3$ analog of a special K\"ahler structure which we now describe.  The low energy $\U(1)^n$ EM duality group is $\spdnz$, the subgroup of $\GL(2n,\Z)$ which preserves the non-degenerate integral skew Dirac pairing $\DD$ on the EM charge lattice,
\begin{align}\label{sp2rZact1}
    M & \in \spdnz &  &\Longleftrightarrow& 
    M \DD M^T &= \DD & &\text{and}& 
    M&\in \GL(2n,\Z).
\end{align}
The Dirac pairing endows the charge lattice with a symplectic structure, and a splitting of the lattice into magnetic and electric sublattices is polarization of the charge lattice (a splitting into complementary lagrangian sublattices).  A polarized basis can always be chosen so that the Dirac pairing takes the \emph{canonical form}
\begin{align}\label{Dpairing}
    \DD &= \e_2 \otimes \d_n, &
    \d_n &= \text{diag}\{d_1, d_2, \ldots, d_n\}, &
    d_i &\in \Z_+, &
    & d_i|d_{i+1} ,
\end{align}
and where $\e_2$ is the $2\times 2$ unit antisymmetric matrix.  The skew eigenvalues $d_i$ invariantly characterize $\DD$.  When all the $d_i=1$, we say that the polarization and the Dirac pairing are \emph{principal}, and the EM duality group is the familiar $\Sp(2n,\Z)$.

The low energy effective action \eqref{LIR} on $\cM$ is written in terms of ``electric'' variables --- i.e., the vector potential of the vector multiplet couples to electric charges --- so presupposes a choice of polarization of the charge lattice.  In analogy to special K\"ahler geometry, we call the vevs of the complex scalar fields $a^I_i$ in the vector multiplet the \emph{special coordinates}, and define the ``magnetic'' \emph{dual special coordinates} by
\begin{align}\label{TSK2}
a^{Ii}_D &:= d_i \t^{ij} a^I_j &
& \text{(no sum on $i$)},
\end{align}
where the $d_i$ are the skew eigenvalues of the Dirac pairing \eqref{Dpairing}.  It is convenient to introduce a matrix notation where we treat $a^I$ as an $n$-component column vector, in which \eqref{TSK2} becomes
\begin{align}\label{TSK2a}
a^I_D := \d_n \t a^I.
\end{align}
Then \eqref{N3Z} and \eqref{LIR} imply that together $(a^{Ii}_D, a^I_i)$ transform in the $6n$-dimensional representation of $\U(3)_R\times \spdnz$ which is the tensor product of the defining representations of the two factors.  This, together with the fact that special coordinates are linearly related and that there is a preferred choice of $\U(3)_R$ orientation of the special coordinates everywhere on $\cM$, means that on chart overlaps the two sets of special coordinates are related by
\begin{align}\label{TSKu3}
\bpm a'_D \\ a' \epm =
\I_3 \otimes M \bpm a_D \\ a \epm, \qquad M \in \spdnz.
\end{align}

Write $M$ in terms of $n\times n$ matrices as
\begin{align} 
M &:= \bpm A&B\\C&D\epm \in \spdnz.
\end{align}
Together with the relation \eqref{TSK2} between $a$ and $a_D$, \eqref{TSKu3} implies that the action of $M$ on the special coordinates is given by
\begin{align} \label{TSKr3}
M:  a^I &\mapsto {a'}^I = (\r_\t) a^I , &
&\text{with}&
\r_\t &= C \d_n \t + D \in \GL(n,\C).
\end{align}
\eqref{TSKu3} also implies the EM duality group action on the $n\times n$ matrix $\t^{ij}$ of low energy $\U(1)^n$ gauge couplings on $\cM$ is given by the action of $\spdnz$ on the Seigel upper half space,\footnote{Our definition \eqref{TSK2a} leads to an action on $\t$ which differs from the one usually defined --- see e.g., \cite{hulek1998geometry} --- by the placement of factors of the $\d_n$ matrix. Our conventions are chosen to preserve the standard forms \eqref{N3Z} and \eqref{LIR} of the central charges and IR effective action appearing in the physics literature.} which we denote by $\circ$,
\begin{align} \label{TSKs3}
M:  \t \mapsto \t' = M\circ\t 
:= \d_n^{-1} (A \d_n \t+B) (C \d_n \t+D)^{-1} .
\end{align}

% \red{\paragraph{Correction?} The group $\spdnz$ acts naturally on the space of $2n \times n$ matrices of rank $n$ divided by the equivalence relation 
% \begin{equation} 
%     \begin{pmatrix} R_1 \\ R_2 \end{pmatrix} \sim  \begin{pmatrix} R_1 \Lambda \\ R_2 \Lambda \end{pmatrix} 
% \end{equation}
% for $\Lambda \in \mathrm{GL}(n,\mathbb{C})$, via left matrix multiplication by $M$. It is natural to embed the Siegel upper-half space spanned by $\tau$ via \cite{hulek1998geometry}
% \begin{equation}
%     \tau \rightarrow \begin{pmatrix} \tau  \\ \delta_n \end{pmatrix} \, . 
% \end{equation}
% Then the action of $M$ on $\tau$ is computed as follows: 
% \begin{equation}
%   M   \begin{pmatrix} \tau  \\ \delta_n \end{pmatrix} = \begin{pmatrix} A \tau + B \delta_n  \\ C \tau + D \delta_n \end{pmatrix} \sim \begin{pmatrix} \delta_n (A \tau + B \delta_n )( C \tau + D \delta_n )^{-1} \\ \delta_n \end{pmatrix} \, , 
% \end{equation}
% which gives 
% \begin{equation}
%     M \circ \tau = \delta_n (A \tau + B \delta_n )( C \tau + D \delta_n )^{-1} \, . 
% \end{equation}}

\section{Triple special K\"ahler (TSK) manifolds}
\label{sec2-TSK}

We will call a $3n$-complex-dimensional manifold with this EM duality structure a \emph{triple special K\"ahler} (TSK) manifold. In this section we assemble the properties of $\cM_\text{smooth}$ described above into a formal definition of a TSK manifold and then discuss some properties which follow from the definition. 

Note that the manifolds $\cM_\text{smooth}$ described here are not metrically complete: they are missing the singular points where the interesting physics occurs.  We will propose a definition of a metrically complete, but singular, triple special K\"ahler \emph{space}, $\cM$, in section \ref{TSKspace} by using the constraints coming from the assumed unitarity, locality, and superconformal invariance of the quantum field theory on the kinds of singular behavior that $\cM$ can and should have.

\begin{definition}\label{defn1}
A TSK manifold is given by $(\cM_\textrm{smooth}, \DD, \{a,\t\})$ where:
\begin{description}
    \item[(a)] $\cM_\textrm{smooth}$ is a $3n$-complex-dimensional differentiable manifold with an atlas of \emph{special coordinate} charts, $\{U_\a\}$. 
    \item[(b)] $\DD$ is an integer non-degenerate skew-symmetric $2n \times 2n$ matrix in canonical form. 
    \item[(c)] Denote the complex special coordinates in a given chart by $a^J_j$, $J\in\{1,2,3\}$ and $j\in\{1,\ldots,n\}$, and define the \emph{dual special coordinates} in that chart by $a_D^{Jj}= d_j \t^{jk} a^J_k$, where in each chart $\t$ is a fixed complex symmetric $n\times n$ matrix with positive definite imaginary part.
    \item[(d)] Special coordinates are linearly related on a chart overlap, $U_\a \cap U_\b$, by ${a_D \choose a}^{(\a)} = \I_3 \otimes M^{(\a\b)} {a_D \choose a}^{(\b)}$ where $M^{(\a\b)} \in \spdnz$ is an integer $2n\times 2n$ matrix satisfying $M^{(\a\b)} \DD M^{(\a\b)\,T} = \DD$.  
\end{description}
Two $\cM_\textrm{smooth}$ with \emph{special structures}, $ \{a,\t\}$ and $\{a',\t'\}$, are considered equivalent if they are related in every chart by $\t' = M\circ \t$ and ${a'_D \choose a'} = F \otimes M {a_D \choose a}$, where $F \in \GL(3,\C)$ and $M \in \spdnz$ are independent of the chart.
\end{definition}

There are additional structures on $\cM_\text{smooth}$ which follow naturally from the definition. 
The first is a fiber bundle $\L \to \cM_\text{smooth}$ whose fibers are rank-$2n$ lattices $\simeq \Z^{2n}$ and whose transition functions are the inverse transposes of the same $M^{(\a\b)}\in\spdnz$ as in the definition of $\cM_\text{smooth}$. Physically, $\L$ is the charge lattice and $\DD$ gives the Dirac pairing.  With $\L$ one can then construct the central charges \eqref{N3Z}. Two central charges, however, are considered equivalent only for a subgroup $\U(3)\subset \GL(3,\C)$ of the equivalence of special structures. 

The second structure is a flat positive-definite metric given in each chart by $g = \Im \big( d\ab_{Jj} \otimes d a_D^{Jj} + d a_D^{Jj} \otimes d\ab_{Jj} \big)$.  Note that this metric is also only invariant under  $\U(3) \subset \GL(3,\C)$ of the equivalence.  Thus specification of the central charges or metric determines an overall normalization of length scales which is absent in the definition of a TSK manifold given above.

A third structure which follows from the definition is a $\C\P^2 \coprod \C\P^0$ of inequivalent K\"ahler structures on $\cM_\text{smooth}$. We will describe these complex structures in more detail in the next subsection. The $\C\P^2$ of complex structures of $\cM_\text{smooth}$ are the \emph{physical complex structures} and are related to the action of the $\cN=3$ supersymmetry charges.  They form an orbit under a non-holomorphic action of the $\U(3)$ isometry group of $\cM_\text{smooth}$. The remaining ``isolated'' \emph{special complex structure} is not related to the $\cN=3$ supercharges; the $\U(3)$ isometry acts holomorphically with respect to the special complex structure.  

As we will see in section \ref{TSKspace} when we discuss TSK \emph{spaces} (as opposed to manifolds), these three structures will play a central role and so will be considered as part of the TSK space definition in the physical application to $\cN=3$ moduli spaces.

In the above definition we have restricted the chart transitions to $M^{(\a\b)}\in\spdnz$.  This is in contrast to alternative definitions of special K\"ahler geometry, e.g., \cite{Freed:1997dp}, in which the transition functions are allowed to be affine transformations with linear part in $\Sp(2n,\R)$ plus constant shifts. The extra rigidity from the $\spdnz$ condition is an important feature of the low energy physics of $\cN=3$ theories and plays a central role in our discussion in \cite{PAMN3II} of the global structure of $\cN=3$ moduli spaces. However, it may be useful to develop an analog for TSK geometry of the intrinsic definitions of the $\Sp(2n,\R)$ special K\"ahler geometry in \cite{Freed:1997dp} or in terms of Hessian structures \cite{LopesCardoso:2019mlj}.

Our definition can mostly be rephrased in a way which does not mention coordinates by generalizing the description \cite{Donagi:1995cf} of special K\"ahler geometry in terms of an algebraic integrable system as follows.
\begin{definition}\label{defn2}
A TSK manifold $\cM_\textrm{smooth}$ is given by $(A,\DD,E,e,\Omega)$ where:
\begin{description}
	\item[(a)] $\cM_\textrm{smooth}$ is a $3n$-dimensional complex manifold.
	\item[(b)] $A \to \cM_\textrm{smooth}$ is a holomorphic fiber bundle whose fibers are complex abelian varieties of dimension $n$ with a fixed Hodge form $\DD$ and fixed complex structure.
	\item[(c)] $E \to \cM_\textrm{smooth}$ is a rank-3 holomorphic vector bundle with a fixed euclidean bundle metric $e$ and fixed compatible complex structure.  
	\item[(d)] $\DD$, $e$, and the complex structure are constant across fibers, so $\DD\otimes e$ extends to a real closed $(1,1)$-form $\hat \DD$ on $A\otimes E$.  (It is the unique form which restricts to $\DD\otimes e$ on fibers and which has rank $6n$, i.e., half the real dimension of $A\otimes E$.) 
	\item[(e)] $\Omega$ is a non-degenerate holomorphic symplectic form on $A\otimes E$ such that the fibers are Lagrangian.  
\end{description}
\end{definition}

This is related to our previous definition as follows.  The complex modulus of the abelian fibers of $A$ is $\t$, and its symmetry and the  positive definiteness of its imaginary part follow from the Riemann conditions for an abelian variety.  A K\"ahler form on $\cM$ is given by $\int_\text{fibers} \Omega \wedge \bar\Omega \wedge \hat \DD^{3n-1}$, and its associated Riemannian metric is $g$.  The first homology classes of the fibers of $A$ form the EM charge lattice, $\L$, and $\int_{\a\ast\b} \DD$ is the Dirac pairing on those charges, where $\a\ast\b$ is the Pontryagin product of two homology cycles on the abelian variety.  Finally, the differential of the central charge is given by the map $dZ: \L \to T^*E$ which assigns to a charge $\a\in\L$ the $\C^3$-valued 1-form $\int_\a \Omega$.  

One drawback of this definition is that although it automatically restricts the linear part of the chart transition functions to be in $\spdnz$, because it only defines the differential of the central charges, it only specifies the special coordinates up to constant shifts. The requirement that we can set all the constant shifts in the transition functions to zero is equivalent to the additional condition that there is no global obstruction to integrating $dZ$ to $Z$ on $\cM_\text{smooth}$. For this reason, definition \ref{defn2} is weaker than definition \ref{defn1}.  (Note that this drawback of the integrable system definition also applies to the analogous integrable system definition of special K\"ahler geometry.)

A second drawback is that this definition picks a complex structure on $\cM$.  In fact it is the ``unphysical'' one in which the $\U(3)$ isometry acts holomorphically.  It would be interesting to improve this definition to one which does not single out any complex structure.  For instance, a definition of hyperk\"ahler geometry, $\cH$, can be given in terms of the complex geometry of a complex twistor space $T\to\C\P^1$ with $\cH$ fiber without singling-out any one of the $\C\P^1$ of complex structures of $\cH$ (see, e.g., \cite{joyce2000compact}).  Perhaps, analogously, there is a definition of TSK geometry in terms of the complex geometry of a fiber bundle over $\C\P^2$ with fibers given by the abelian variety bundle $A\to\cM$.

%%%%%%%
%%%%%%%

\subsection{Examples} 
\label{5.examples}

We now illustrate the above definitions on two sets of simple examples, which will accompany us throughout this section. 

\subsubsection*{Example A: Rank 1 $\Z_k$ orbifolds}

Let $\cM_\text{smooth} = (\C^3-\{0\})/\Z_k$ for $k=1,2,3,4,6$, where the $\Z_k$ identifications are $\C^3 \ni (a^1,a^2,a^3) \sim \xi (a^1,a^2,a^3)$ for $\xi$ any $k$-th root of unity. Only the origin is fixed by these identifications, which guarantees that $\cM_\text{smooth}$ is smooth indeed.

To specify the TSK structure we choose a Dirac pairing to be principal, $\DD = \e_2$, the $a^I$, $I=1,2,3$, to be special coordinates, and the EM couplings to be
\begin{align}
    \t &\in \mathscr{H}_1 & &\text{for}& k &=1 , \nn\\
    \t &\in \mathscr{H}_1 & &\text{for}& k &=2 , \nn\\
    \t &= e^{2 \pi i /3} & &\text{for}& k &=3 , \nn\\
    \t &= i & &\text{for}& k &=4 , \nn\\
    \t &= e^{2 \pi i /3} & &\text{for}& k &=6 , 
\end{align}
where $\mathscr{H}_1$ is the complex upper half-plane.  Then the dual coordinates are $a_D^I = \t a^I$, for $I=1,2,3$.

In these examples the duality group is $\spdnz = \SL(2,\Z)$. So the special structures specified above are equivalent to ones with $\t$ transformed by any element of $\SL(2,\Z)$ under its fractional linear action on $\mathscr{H}_1$.  Thus for $k=1,2$, one can restrict to $\t$ in any fundamental domain of the $\SL(2,\Z)$ action. 

In the language of definition \ref{defn2}, the complex abelian varieties fibered over $\cM_\text{smooth}$ are elliptic curves with complex structure given by $\t$. 

Note that the orbifolds studied in this paragraph are the full list of rank 1 $\mathcal{N} \geq 3$ theories. For reference we gather them in Table \ref{tabCBrank1}. 

%%%%%%%%

\subsubsection*{Example B: Rank 2 $S_3$ orbifold}

Define the orbifold $\cM = \C^6/S_3$ where the symmetric group $S_3$ acts as follows.  Let $a^I_i$ for $I=1,2,3$, $i=1,2$ be complex coordinates on the covering space $\til\cM = \C^6 = \C^3 \otimes \C^2$. $S_3$ is generated by an order two and an order three element which we take to act on $\C^3 \otimes \C^2$ by multiplication of the vector of $a^I_i$ coordinates by the matrices (the second matrix acts on the right)
\begin{align}\label{S3B}
    \I_3 \otimes \bpm 0 & 1 \\ 1 & 0 \epm \, , \qquad     
    \I_3 \otimes \bpm 0 & -1 \\ 1 & -1 \epm \, . 
\end{align}
Note that this action is purely real. 

For the TSK structure, we choose the $a^I_i$ to be the special coordinates, a principal polarization $\DD = \I_3\otimes\e_2$, and an EM coupling matrix
\begin{align}\label{tauB}
    \t &= \bpm \varsigma & \frac12\varsigma \\ 
    \frac12\varsigma & \varsigma + 1 \epm 
\end{align}
for any complex $\varsigma$ with $\Im \, \varsigma>0$. 

It is then straightforward to check that the orbifold identifications on the 12-component vector of dual special coordinates and special coordinates are of the form $\left(\bsm a_D\\ a\esm\right) \sim \I_3\otimes M \left(\bsm a_D\\ a\esm\right)$ with $M \in \spdnz = \Sp(4,\Z)$.  $(\cM_\text{smooth}, \DD, \{a,\t\})$ thus satisfies the requirements of definition \ref{defn1} of a TSK manifold.

%%%%%%%%%%
%%%%%%%%%%

\subsection{Some properties of $\cM_\text{smooth}$}
\label{5.properties}

Some properties of TSK manifolds which follow easily from  definition \ref{defn1} are:
\begin{itemize}
    \item special coordinate monodromies give a homomorphism of the fundamental group into the EM duality group whose image is the EM monodromy group (see section \ref{sec5.2.1});
    \item the matrix $\t$ of low energy $\U(1)^n$ gauge couplings is fixed by the EM monodromy group in each connected component of $\cM_\text{smooth}$ (see section \ref{sec5.2.2});
    \item there is a maximum finite order of all elements of the EM monodromy group (see section \ref{sec5.2.3});
    \item $\cM_\text{smooth}$ is flat and has a $\U(3)$ group of isometries (see section \ref{sec5.2.4});
    \item there is a $\C\P^2$ of inequivalent metric-compatible ``physical'' complex structures on $\cM_\text{smooth}$ which are induced by the $\cN=3$ supercharges (see section \ref{sec5.2.5-kahler}); and
    \item there is one additional ``special'' metric-compatible complex structure --- not induced by any supercharge --- in which the special coordinates are holomorphic coordinates (see section \ref{sec5.2.5-kahler}).
\end{itemize}
We now derive and discuss these properties.

%%%%%%

\subsubsection{Fundamental group and EM monodromy group}\label{sec5.2.1}

Let $\{ U_\a \}$ be a good open cover of $\cM_\text{smooth}$ by special coordinate charts. To each overlap $U_\a \cap U_\b$ corresponds a special coordinate transformation $M^{(\a\b)} \in \spdnz$ as in definition \ref{defn1}, which can thus be associated to the corresponding $(6n-1)$-dimensional face of the nerve of the open cover. Upon continuing along a closed path $\g$ in $\cM$ the special coordinates thus experience a monodromy of the form 
\begin{align}\label{TSK4}
\left( \bsm a_D \\ a \esm\right) & \overset\g\longrightarrow
\I_3 \otimes M_\g \left( \bsm a_D \\ a \esm\right), &
M_\g & := M^{(\a_1\a_m)} M^{(\a_m \a_{m-1})} \cdots 
M^{(\a_2\a_1)} 
\in \spdnz, 
\end{align}
i.e., the product (in order, from some chosen starting point) of the overlap transformations of each face that $\g$ intersects.  The consistency condition on triple overlaps, $M^{(\a_1\a_3)} M^{(\a_3\a_2)} M^{(\a_2\a_1)} =1$, means that $M_\g$ only depends on the homotopy class, $[\g]$, of $\g$ in $\cM_\text{smooth}$. Therefore $M_{[\g]}$ gives a representation of $\pi_1( \cM_\text{smooth})$ in $\spdnz$.  Note that since $\cM_\text{sing}$ is of real codimension 6 in $\cM$, and so cannot link a path $\g$, it does not contribute directly to non-trivial elements of $\pi_1( \cM_\text{smooth})$.  However, as we will explore in more detail in section \ref{TSKspace} there is a sense in which nontrivial elements of $\pi_1(\cM_\text{smooth})$ are associated with specific components of $\cM_\text{sing}$. Fom now on we will just write $\pi_1(\cM)$ in place of the more precise $\pi_1( \cM_\text{smooth} )$.  

Denote by $\G(\cM) \subset \spdnz$ the subgroup of the EM duality group generated by the monodromies $M_\g$ for all $[\g] \in \pi_1(\cM)$. We will call $\G(\cM)$ the \emph{EM monodromy group} of $\cM$. 

\paragraph{Examples.}

In the examples given in section \ref{5.examples}, the groups $\pi_1(\cM)$ are respectively $\mathbb{Z}_k$ and $S_3$, by definition of the orbifolds.

\paragraph{A-} 

In the rank one cases, the monodromy groups are 
\begin{align}
k=1 & \qquad & \Gamma(\mathcal{M}) = \{1\} \nn\\
k=2 & \qquad & \Gamma(\mathcal{M}) = \{1,S^2\} \nn\\
k=3 & \qquad & \Gamma(\mathcal{M}) = \{1,U^2 ,U^4 \} \nn\\
k=4 & \qquad & \Gamma(\mathcal{M}) = \{1,S,S^2,S^3\} \nn\\
k=6 & \qquad & \Gamma(\mathcal{M}) = \{1,U,U^2,U^3,U^4,U^5\}
\end{align}
where $S = \left(\bsm 0 & -1 \\ 1 & 0 \esm\right)$ and $U = \left(\bsm 0 & -1 \\ 1 & 1 \esm\right)$ are generators of $\mathrm{SL}(2,\mathbb{Z})$ which satisfy $S^4 = U^6 = 1$.

\paragraph{B-} 

In the rank two case, we have (see \cite{Argyres:2019ngz} for the details of the construction)
\begin{equation}
    \Gamma(\mathcal{M}) = \{1,s_1,s_2,s_1s_2,s_2s_1,s_1s_2s_1\} \cong S_3 \, , 
\end{equation}
with 
\begin{align}\label{eqs1s2}
    s_1 &:=\left(\bsm -1 & 0 & 0 & 0 \\
    -1 & 1 & 0 & 0 \\
    0 & 0 & -1 & -1 \\
    0 & 0 & 0 & 1 \esm\right) ,& 
    s_2 &:= \left(\bsm 0 & 1 & 0 & -1 \\
    1 & 0 & 1 & 0 \\
    0 & 0 & 0 & 1 \\
    0 & 0 & 1 & 0 \esm\right) ,&
    s_1, s_2 &\in \Sp_\DD(4,\Z),
\end{align}
satisfying the presentation of $S_3$, namely $s_1^2 = s_2^2 = (s_1 s_2)^3=1$. 

%%%%%%

\subsubsection{Global constancy of $\t$}\label{sec5.2.2}

Since $\t$ is constant in each chart it must be fixed by all EM monodromies.  One way of seeing this is by noting that one can ``straighten out'' the special structure along any path $\g$ by transforming the special coordinates in the $\a_j$-th chart in \eqref{TSK4} by acting with $M^{(\a_1\a_2)} M^{(\a_2 \a_3)} \cdots M^{(\a_{j-2} \a_{j-1})} M^{(\a_{j-1}\a_j)}$.  In these new straightened coordinates the induced transition transformations along $\g$ are simply the identity, $\til M^{(\a_j\a_{j-1})} = 1$, for $j\in\{2,\ldots,m\}$, and the monodromy is given by the ``final'' transition element $M_\g = \til M^{(\a_1\a_m)}$.  Furthermore, it then follows from the relation between the special coordinates and the dual special coordinates given in definition \ref{defn1} that $\t^{(\a_j)} = \t^{(\a_1)}$ for all $j\in\{1,\ldots,m\}$ in the straightened coordinates.

By picking base points in each connected component of $\cM_\text{smooth}$ and doing this straightening for all paths starting at these base points, we see that we can choose the special structure so that $\t$ is constant in each component.  Furthermore, by going to the simply connected universal cover of each component of $\cM_\text{smooth}$ we can choose all their transition transformations to be the identity.  In this way it is always possible to cover each connected component of $\cM_\text{smooth}$ by a single contractible closed chart, $\bar\cF$, which is a fundamental domain of the action of $\pi_1(\cM)$ on the universal covering space of that component of $\cM_\text{smooth}$.  The special structure at the boundaries of $\bar\cF$ are then identified by $\G(\cM)$ which gives a representation of the action of $\pi_1(\cM)$.  This is the way we presented the special structures of the orbifold examples given in section \ref{5.examples}.

If we write $M_\g$ in terms of $n\times n$ matrices as $M_\g = \Big(\bsm A_\g&B_\g\\C_\g&D_\g\esm\Big)$, then, together with the relation between $a$ and $a_D$, \eqref{TSK4} implies that the action of $M_\g$ on the straightened special coordinates is given by
\begin{align}\label{sp2rZact2}
M_\g:  a_j^I &\mapsto (\r_{\t,\g})_j^i a_i^I , &
&\text{with}&
\r_{\t,\g} &= C_\g \d_n \t+D_\g \in \GL(n,\C).
\end{align}
The constancy of $\t$ in these coordinates implies that $\t$ must be fixed by all $M_\g\in\G(\cM)$,
\begin{align}\label{eqtau}
\d_n \t &= (A_\g \d_n \t+B_\g) (C_\g \d_n \t+D_\g)^{-1} &
&\text{for all}& \g\in \pi_1(\cM),
\end{align}
for each connected component of $\cM_\text{smooth}$.

By virtue of the condition \eqref{eqtau} that $\t$ is fixed by $\G(\cM)$, the set
\begin{equation}
\label{GLnrep}
    \r_\t(\cM) := \{ \r_{\t,\g} \,|\, \g \in \pi_1(\cM) \} 
\end{equation}
forms a group, i.e., a representation of $\G(\cM)$ in $\C^n$.

In the case of an $\cN=4$ theory, the monodromy group is restricted to lie in a subgroup of $\GL(n,\R) \subset \GL(n,\C)$ with respect to some choice of real structure on $\C^n$.  This will be discussed in section \ref{sec5.4-N=4}.

\subsubsection*{Examples}

\paragraph{A-} In the rank one cases, evaluating (\ref{GLnrep}) gives that $\r_\t(\cM)$ is the multiplicative group of $k$-th complex roots of unity.  Note that for $k=1,2$ this is a real group action. 

\paragraph{B-} In the rank two case, one checks that \eqref{tauB} is the set of matrix $\t$ which satisfy (\ref{eqtau}) for all the elements of $S_3$ as generated by (\ref{eqs1s2}) and furthermore have positive definite imaginary part. One then computes the $\GL(2,\C)$ representation (\ref{GLnrep}). Since the matrices (\ref{eqs1s2}) have been chosen upper-block-diagonal, $\r_\t(\cM)$ does not depend on $\varsigma$, and we find the real group action
\begin{align}\label{rhoMrank2}
     \r_\t(\cM) &= \left\{ 
     \left( \bsm -1 & -1 \\ 0 & 1 \esm \right),
     \left( \bsm 0 & 1 \\ -1 & -1 \esm \right),
     \left( \bsm -1 & -1 \\ 1 & 0 \esm\right),
     \left( \bsm 0 & 1 \\ 1 & 0 \esm\right),
     \left( \bsm 1 & 0 \\ -1 & -1 \esm\right),
     \left( \bsm 1 & 0 \\ 0 & 1 \esm\right) 
     \right\} .
\end{align}

%%%%%%

\subsubsection{Constraints on the EM monodromy group}\label{sec5.2.3}

$\spdnz$ elements which fix a $\t$ with $\Im\t>0$ are diagonalizable over $\C$ with eigenvalues which are roots of unity.
%\footnote{It follows from very general considerations --- see, e.g., lemma 4.5 in \cite{Schmid:1973} --- that all $M_\g$ are quasi-unipotent, i.e., that all their eigenvalues are roots of unity. The additional statement here is that the $M_\g$ are also diagonalizable over $\C$.}  
One way to see this is to note that the $\GL(n,\C)$ action of $\r_{\t,\g}$ preserves the metric on $\cM$ mentioned in the paragraph after definition \ref{defn1} (and discussed below).  Since the metric is positive definite, $\r_{\t,\g}$ is conjugate to a unitary matrix and so is semisimple over $\C$ and has all unit norm eigenvalues. Since $\r_{\t,\g}$ is also represented by $M_\g \in \GL(2n,\Z)$ its eigenvalues are roots of unity. In particular, each $M_\g \in \G(\cM)$ is of finite order. Furthermore, the set of roots of unity that can appear as eigenvalues of elements of $\GL(2n,\Z)$ for a given value of $n$ is finite \cite{Argyres:2018zay, Caorsi:2018zsq, Argyres:2018urp}.  Thus there is a maximum finite order of any element of $\G(\cM)$ if we assume that all connected components of $\cM_\text{smooth}$ have bounded dimension. 

\paragraph{Examples.}

In the examples given in section \ref{5.examples} the maximum orders of $\G(\cM)$ is $k$ for the rank one examples, and the maximum order is 3 for the rank two example. 

%%%%%%

\subsubsection{Flat metric, isometries, and quadratic forms}\label{sec5.2.4}

The special structure induces a natural Riemannian metric on $\cM_\text{smooth}$ given by 
\begin{align}\label{TSKmetric2}
    g &= \Im\t^{ij} \, \big(
    da^I_i d\ab_{Ij} + d\ab_{Ii} da^I_j \big) , 
\end{align}
which is the physical metric on the moduli space induced by the fluctuations of the massless scalar fields.  It is a flat positive definite metric by virtue of the constancy of $\t$ and the positivity of $\Im\t$, and the special coordinates are flat coordinates.  

It is easy to check that the defining condition \eqref{sp2rZact1} that EM duality transformations preserve the Dirac pairing \eqref{Dpairing} implies, together with the definition \eqref{TSK2a} of $a_D$ and the action, \eqref{sp2rZact2} and \eqref{eqtau}, of the monodromy group that $\r_\t(\cM)$ preserves the metric:
\begin{align}\label{TSKmetric3}
    \r^\dag (\Im\t) \r &= \Im\t & &\text{for all }& \r &\in\r_\t(\cM) .
\end{align}
Thus $\r_\t(\cM)$ is a discrete isometry group of $\cM_\text{smooth}$.

The identifications \eqref{sp2rZact2} on the special coordinates upon traversing loops $\g\in\pi_1(\cM)$ break the local $\SO(6n)$ isometry group of a flat metric to the subgroup commuting with the linear action of $\r_\t(\cM) \subset \GL(n,\C) \subset \GL(3n,\C)$ on the special coordinates.  The commutant of $\GL(n,\C)$ in $\GL(3n,\C)$ is $\GL(3,\C)$, whose intersection with $\SO(6n,\R)$ is $\U(3)$.  This is the expected $\U(3)_R$ continuous isometry group coming from the broken $R$-symmetry on $\cM$.

Decompose the isometry group of $\cM$ as $\U(3)_R = \SU(3)_R \ltimes \Uot_r$.  We write the overall $\Uot_r$ factor with a tilde to emphasize that a priori it not need be compact, so write it as its universal cover.  This just means that, until we learn otherwise, $\Uot_r$ charges of fields need not be integral or even rational. Even though the special coordinates have $\Uot_r$ charges $r=2$ it does not immediately follow that the $r$-charges of coordinates on $\cM$ are even, because the special coordinates are generally not good (globally defined) coordinates on $\cM$.

There is always a set of 9 independent globally defined functions on $\cM$ quadratic in the special coordinates.  Because the metric is flat it induces a monodromy-invariant hermitean quadratic form on the special coordinates given by $a^\dag_I\, \Im\t\, a^I := \ab_{Ii}\, \Im\t^{ij}\, a^I_j$ which therefore extends to a well-defined quadratic form on all of $\cM_\text{smooth}$. Moreover, since the monodromies commute with the $\U(3)_R$ isometries, $\cM_\text{smooth}$ inherits a whole set of quadratic forms, 
\begin{align}\label{TSKmetric4}
    a^\dag_I\, \Im\t\, a^J, \qquad I,J\in\{1,2,3\},
\end{align}
which transforms in the ${\bf 8}_0 \oplus {\bf 1}_0$ of $\U(3)_R$.  In section \ref{subspaces} we will examine the circumstances under which some of these coordinates are holomorphic with respect to various complex structures.

If $\r_\t(\cM)$ is restricted, the continuous isometry group can be larger. This is what happens for the moduli spaces of $\cN=4$ theories, where the continuous isometry group is enhanced to $\SO(6)$; see section \ref{sec5.4-N=4}.

%%%%%%

\subsubsection*{Examples.}

\paragraph{A-} For our rank-1 examples, the groups $\r_\t(\cM)$ are the complex $k$-th roots of unity. For $k=3,4,6$, we obtain immediately that the continuous isometry group is $\U(3)$. In the case $k=1,2$, since the roots are real, we get instead a continuous isometry group $\SO(6)$. Thus the $k=1,2$ orbifold moduli space is consistent with the SCFT having $\cN=4$ supersymmetry, but the $k=3,4,6$ cases are not.

\paragraph{B-} Since the matrices in (\ref{rhoMrank2}) are real the isometry group is $\SO(6)$, thus compatible with $\cN=4$ supersymmetry. 

%%%%%%%%
%%%%%%%%

\subsubsection{K\"ahler structures}
\label{sec5.2.5-kahler}

Any TSK manifold $\cM_\text{smooth}$ has a $\C\P^2 \coprod \C\P^0$ of inequivalent metric-compatible complex structures.  Since they are all compatible with a flat metric, they are, in fact, K\"ahler structures on $\cM_\text{smooth}$.  These complex structures are constructed as follows.

The real and imaginary parts of the special coordinates in a given chart form a euclidean coordinate system --- i.e., flat with respect to the metric \eqref{TSKmetric2} --- on an open set in $\R^{6n}$.  A metric-compatible almost complex structure, $J$, on this chart can then be represented in this coordinate system as a $6n \times 6n$ real matrix such that 
\begin{align}\label{cplx-1}
    J^2=-\I_{6n}.
\end{align}
$J$ is compatible with the metric \eqref{TSKmetric2} if $g(J\cdot,J\cdot) = g(\cdot,\cdot)$, which is equivalent to the conditions that $J$ is a constant matrix satisfying
\begin{align}\label{cplx0}
    J^T (\I_6\otimes \Im\t) J = \I_6 \otimes \Im\t,
\end{align}
where we are decomposing $\R^{6n} = \R^6 \otimes \R^{n}$.

The transition map \eqref{TSKr3} induces a unique almost complex structure which coincides with $J$ on the chart overlap.  By straightening the coordinate charts along any closed path in $\cM_\text{smooth}$, as described in section \ref{sec5.2.2}, it follows that the condition that $J$ extends to a complex structure of $\cM_\text{smooth}$ is that
\begin{align}\label{cplx1}
    R_\g J &= J R_\g  \quad
    \text{for all } \g \in \pi_1(\cM) , &
    &\text{where}&
    R_\g &:= \I_3 \otimes \r_{\t,\g}^\R.
\end{align}
Here we are using the decomposition $\R^{6n} = \R^3 \otimes \R^{2n}$, and $\r_{\t,\g}^\R$ is the representative of the monodromy group element $\r_{\t,\g} \in \GL(n,\C)$ of \eqref{sp2rZact2} considered as an element of $\GL(2n,\R)$.  Concretely, if $\r = r+ i s$ is the decomposition of $\r\in\GL(n,\C)$ into its real and imaginary parts, then $\r^\R = \I_2 \otimes r + \e_2 \otimes s$ is its representation in $\GL(2n,\R)$ where $\e_2 := \left(\bsm 0&-1 \\ 1&0 \esm \right)$.  Thus a monodromy group element has the form 
\begin{align}\label{cplx1.5}
  R_\g = \I_6 \otimes r_\g + \J \otimes s_\g ,  
\end{align}
where
\begin{align}\label{cplx2}
    \J := \I_3 \otimes \e_2
\end{align}
obeys $\J^2=-\I_6$, so can be thought of as a complex structure on $\R^6$.  Then the requirement \eqref{cplx1} for general $R_\g$ (i.e., without assuming any extra constraints on the monodromy group) together with the conditions \eqref{cplx-1} and \eqref{cplx0} that $J$ be a metric-compatible complex structure implies that
\begin{align}\label{cplx3}
    J &= j \otimes \I_n , &
    j^2 &= -\I_6 , &
    j^Tj &= \I_6 , &
    &\text{and}&
    [j,\J] &= 0.
\end{align}

The last two conditions in \eqref{cplx3} imply that $j \in \U(3)$ when the $\R^6$ is interpreted as $\C^3$ with complex structure $\J$.  This is because $[j,\J]=0$ means that $j\in\GL(3,\C)$ and $j^Tj=\I_6$ means that $j\in\O(6,\R)$, and the intersection of these two groups is $U(3)$.  The remaining condition on $j$, namely, $j^2=-\I_6$ implies that its eigenvalues are $\pm i$.  As an element of $\U(3)$ there are thus two possibilities, eigenvalues $\{+i,+i,+i\}$, or eigenvalues $\{-i, +i, +i\}$ (and the negatives of these, which are just the conjugate complex structures).  

In the first case $j$ is the unique element $+i \I_3 \in \U(3)$, which in the real basis is just $j = \J$.  We call this the \emph{special complex structure} on $\cM$.  In this complex structure holomorphic coordinates $\z^I_i$ on $\cM_\text{smooth}$ in any chart can simply be taken to be the complex special coordinates themselves,
\begin{align}\label{cplx4}
    \z^I_i &= a^I_i &
    &\text{(special complex structure)}.
\end{align}
The $\U(3)_R$ isometry thus acts holomorphically in the special complex structure.

In the second case $j$ can be any element of $\U(3)$ of the form 
\begin{align}\label{cplx5}
    j &= U \left( \bsm-i&&\\&+i&\\&&+i\esm \right) U^{-1} , &
    &\text{for }\  U \in \U(3) &
    &\text{(physical complex structures)}.
\end{align}
Since the $U$'s which commute with the diagonal matrix form a $\U(2) \times \U(1)$ subgroup of $\U(3)$, the inequivalent such complex structures form the coset space $\U(3)/[\U(2) \times \U(1)] \simeq \C\P^2$.  We call this $\C\P^2$ of complex structures the \emph{physical complex structures} on $\cM$. One can think of this $\C\P^2$ of K\"ahler structures of a TSK space as an analog of the $\C\P^1$ of metric-compatible complex structures that form a hyperk\"ahler structure.

This $\C\P^2$ of complex structures are called physical because they are the ones induced by the $\cN=3$ supersymmetry charges.  To see this, recall that any choice of $\cN=1$ subalgebra of the $\cN=3$ supersymmetry algebra \eqref{N3susy} defines a K\"ahler structure on $\cM$.  For instance, if the $\cN=1$ subalgebra is generated by the $Q^1$ and $\Qb_1$ supercharges, then the holomorphic complex coordinates on $\cM$ are the vevs of those complex scalars, $\vf$, for whom $Q^1\vf$ is a left-handed Weyl spinor.  The effective $\cN=1$ sigma model for these scalars induces a K\"ahler structure on $\cM$ in the usual way. And since there is a $\U(3)/[\U(2)\times\U(1)] \simeq \C\P^2$ of inequivalent ways of embedding an $\cN=1$ subalgebra in the $\cN=3$ algebra, $\cM$ admits a $\C\P^2$ of inequivalent complex structures.

Explicitly, the transformation rules \eqref{vectortransf} of the massless $\cN=3$ vector multiplet component fields show that with respect to the complex structure induced by any given supercharge, and for each value of $i\in \{1,\ldots,n\}$, two of the three $a_i^I$ are holomorphic on $\cM$ and the other is antiholomorphic.  For example, with respect to the complex structure induced by $Q^1$, the holomorphic coordinates on $\cM$ in any chart are
\begin{align}\label{TSKcplx}
\z_{1j} &= \bar a_{1j}, & 
\z^2_j &= a^2_j, &
\z^3_j & = a^3_j.
\end{align}
Note that these do not transform as a triplet under $\SU(3)_R$.  In these coordinates, the metric \eqref{TSKmetric2} on $\cM$ is $g = \frac12 \Im\t^{ij} \big(
    d\z_{1i} d\bar\z^1_j 
    + d\z^2_i d\bar\z_{2j} 
    + d\z^3_i d\bar\z_{3j} +\text{c.c.} \big)$,
whose associated K\"ahler form is $\w = \frac{i}2 \Im\t^{ij} \big(
    d\z_{1i} \^ d\bar\z^1_j 
    + d\z^2_i \^ d\bar\z_{2j} 
    + d\z^3_i \^ d\bar\z_{3j} \big)$.

The $\U(3)_R$ automorphism group of the $\cN=3$ supersymmetry algebra changes the complex and K\"ahler structures to
\begin{align}\label{TSKcplx2}
    \z^R_{1j} &= \bar{R^1_J a^J_j} = \ab_{Jj} (R^{-1})^J_1, 
    \qquad
    \z^{R\,2}_j = R^2_J a^J_j,
    \qquad
    \z^{R\,3}_j = R^3_J a^J_j,
    \\
    \w^R &= \frac{i}2 \Im\t^{ij} \Big(
    d\z^R_{1i} \^ d\bar\z^{R\,1}_j 
    + d\z^{R\,2}_i \^ d\bar\z^R_{2j} 
    + d\z^{R\,3}_i \^ d\bar\z^R_{3j} \Big) ,
    \nn 
\end{align}
where $R \in \U(3)$.  These are the complex structures induced by the $R^1_J Q^J$ supercharge.  \eqref{TSKcplx2} is the $\C\P^2$ of inequivalent complex structures and K\"ahler forms on $\cM_\text{smooth}$ constructed above in \eqref{cplx5}.

The relation of the holomorphic coordinates on $\cM$ to the special coordinates given in \eqref{TSKcplx2} shows that the $\U(3)_R$ isometry does not act holomorphically on $\cM$ with respect to any of the physical complex structures since their holomorphic coordinates do not form among themselves a representation of $\U(3)_R$.  Indeed, the action of the $\U(3)_R$ isometry of $\cM$ can be interpreted as a change in the complex structure of a fixed $\cM$, inducing an action on the $\C\P^2$ of physical complex structures of $\cM$.   From \eqref{TSKcplx2} it follows that a $\U(2) \times \U(1) \subset \U(3)_R$ acts holomorphically on $\cM$, so does not change its physical complex structure.  Thus the orbit of the $\U(3)_R$ action on $\C\P^2$ is all of $\C\P^2$, since $\C\P^2 \simeq \U(3)_R/[\U(2) \times \U(1)]$. This is again analogous to the way the non-holomorphic $\SU(2)$ isometry of a hyperk\"ahler space acts on the $\C\P^1$ of its complex structures.

The physical complex structures and the $\U(3)_R$ isometry action on $\cM_\text{smooth}$ are related by the properties of the special coordinates through \eqref{TSKcplx2}.  This relationship is characterized in a coordinate invariant way by the condition, noted above, that there is a $\U(2) \times \U(1) \subset\SU(3)_R$ holomorphic isometry with respect to any choice of physical complex structure on $\cM_\text{smooth}$.

If the monodromy group, $\G(\cM)$, is such that its elements all have a restricted form, as occurs for theories with $\cN=4$ supersymmetry, it can happen that some of the conditions on the compatible complex structure $J$ are relaxed.  In particular, $\cN=4$ SCFTs have a $\C\P^3$ of metric-compatible complex structures which form a single orbit under the action on $\cM_\text{smooth}$ of the $\SU(4)_R$ symmetry of an $\cN=4$ SCFT; see section \ref{sec5.4-N=4} below.

%%%%%%

\subsubsection*{Examples.}

\paragraph{A-} For our rank-1 examples, the holomorphic coordinates, $\z^I$, in the special complex structure are simply the special coordinates, $\z^I = a^I$.  In these coordinates the $\Z_k$ orbifold action multiplies all the $\z^I$ by a common root of unity.  Holomorphic coordinates in one of the physical complex structures can be taken to be, e.g., $(\z^1,\z^2,\z^3) = (\ab_1, a^2,a^3)$. In these coordinates the orbifold action is now $(\z^1,\z^2,\z^3) \mapsto (\bar\xi \z^1, \xi \z^2, \xi \z^3)$ for $\xi$ an appropriate root of unity.  The orbifold action is still holomorphic in these coordinates, so $\C^3/\Z_k$ is a complex space with this choice of almost complex structure.  Since the orbifold action for $k=2$ is purely real, $\C^3/\Z_2$ has additional complex structures beyond the special and physical ones, which form a $\C\P^3$ of inequivalent complex structures as described in section \ref{sec5.4-N=4} below.

\paragraph{B-} The $S_3$ orbifold action was defined to be holomorphic for the special coordinates, so the special almost complex structure extends to a complex structure of the orbifold space. Since the $S_3$ action \eqref{S3B} commutes with the $\U(3)_R$ rotations, it is easy to see that the $S_3$ action is holomorphic with respect to the physical almost complex structures as well.  In fact, since the $S_3$ action is real, $\C^6/S_3$ in fact has the larger $\C\P^3$ space of complex structures.

%%%%%%%%%%
%%%%%%%%%%
%%%%%%%%%%

\subsection{Special and hyperk\"ahler slices}
\label{subspaces}

There are natural complex submanifolds of a TSK manifold which are $n$-dimensional special K\"ahler and $2n$-dimensional hyperk\"ahler spaces.  We call them the special and hyperk\"ahler \emph{slices} of $\cM$, respectively.  We will see that there is actually a family of such slices parameterized by points in $\C\P^2$.

We define these slices as follows. Choose any $\U(2)\times\U(1) \subset \U(3)_R$ subgroup\footnote{Here we really mean an appropriate cover of $\U(2)\times \U(1)$ which is a subgroup of $\U(3)_R = \SU(3)_R \ltimes \Uot_r$.  See the discussion in section \ref{sec5.2.4}.} of the isometry group of $\cM$. Then a \emph{special splice} is any submanifold of $\cM$ which is fixed pointwise by the action of the $\U(2)$ subgroup, and a \emph{hyperk\"ahler slice} is any submanifold which is fixed pointwise by the $\U(1)$ subgroup.

We will now show that these slices have the following properties:
\begin{itemize}
    \item A special slice is an $n$-complex-dimensional flat special K\"ahler manifold.  It has a unique complex structure and a holomorphic $\U(1)$ isometry group.
    \item There is a $\C\P^1$ of complex structures in which a hyperk\"ahler slice is a $2n$-complex-dimensional flat hyperk\"ahler manifold with a triholomorphic $\U(1)$ isometry.  It also has a complex structure in which it has a holomorphic $\U(2)$ isometry group, but is not hyperk\"ahler.
    \item If any special or hyperk\"ahler slice exists, then a whole $\C\P^2$ of inequivalent slices also exist.
\end{itemize}
From the point of view of a choice of $\cN=2$ subalgebra of the $\cN=3$ superconformal symmetry, a special slice is ``just'' the Coulomb branch and the hyperk\"ahler slice the Higgs branch.  This makes many of the above properties appear natural. But we will see that there are subtleties in making this identification which, in principle, could prevent the slices from existing.

In any chart of $\cM_\text{smooth}$, under almost\footnote{The special cases of measure zero in which the $\U(2)$ factor is a subgroup of the $\SU(3)_R$ factor of the isometry group, then the statement does not hold.  But in these cases it is easy to see that the only subspaces fixed pointwise under either the $\U(1)$ or $\U(2)$ action are just a single point: the common origin of the special coordinate systems.  Furthermore, as we will see in the next section, the origin is only a point in $\cM_\text{smooth}$ when $\cM_\text{smooth}$ is flat $\C^{3n}$, i.e., the moduli space of $n$ free massless $\cN=3$ vector multiplets.} any $\U(1)\times\U(2)$ subgroup of the isometry group, the special coordinates transform under the $\SU(2)$ factor as $n$ copies of a singlet and $n$ copies of a doublet representation.  By an overall $\U(3)_R$ rotation of the special coordinates we can thus describe the slices by the coordinate conditions
\begin{align}\label{slice1}
    \text{special slice:}& & \cC &= \{ a^1_i=a^2_i=0,\ i=1,\ldots,n \}, \nn\\
    \text{hyperk\"ahler slice:}& & \cH &= \{ a^3_i=0,\ i=1,\ldots,n \},
\end{align}
in each chart.  The $\U(3)_R$ global action ensures that both are also globally defined as nowhere dense $n$- and $2n$-complex-dimensional submanifolds of $\cM_\text{smooth}$, respectively, which both go through the origin of $\cM$.  If the slices in \eqref{slice1} exist, then clearly they can be rotated into other such slices by the action of the isometry group, giving $\U(3)_R/[\U(2)\times\U(1)] \simeq \C\P^2$ inequivalent slices, since the equations in \eqref{slice1} are invariant under the $\U(2)\times\U(1)$ subgroup.

Under any of the complex structures of $\cM$, the special slice $\cC$ is a holomorphic submanifold, and it inherits the same complex structure form all of them.  Furthermore, by taking $a_i = a^3_i$ to be special coordinates on $\cC$, its K\"ahler form is $\w_\cC =  \frac i2 \Im\t^{ij} \, da_i \^ d\ab_j$, the $\U(3)_R$ isometry of $\cM$ acts as a holomorphic $\U(1)_R$ isometry on $\cC$, and the $\spdnz$ EM duality structure carries over unchanged.  Thus $\cC$ is a rank-$n$ flat special K\"ahler manifold.

There is a $\C\P^1$ of inequivalent physical complex structures of $\cM$ under which the hyperk\"ahler slice $\cH$ is a hyperk\"ahler manifold.  These are the complex structures of the form \eqref{TSKcplx2} but where $R$ is chosen to be in the $\U(2)$ subgroup. For example, choose the physical complex structure on $\cM$ with holomorphic coordinates $\z^a_i$ for $a=1,2$ given by $\z_{1i} = \ab_{1i}$ and $\z^2_i = a^2_i$.  Then the K\"ahler form of $\cH$ with respect to this complex structure is $\w_\cH = \frac{i}2 \Im\t^{ij} ( d\z_{1i} \^ d\bar\z^1_j + d\z^2_i \^ d\bar\z_{2j} )$ and the holomorphic 2-form made from the K\"ahler forms with respect to the other two orthogonal complex structures of $\cH$ is $\w^{(2,0)}_\cH = 2i\Im\t^{ij} d\z_{1i} \^ d\z^2_j$.  The $\U(3)_R$ isometry of $\cM$ acts on $\cH$ as a $\U(2) = \SU(2)_R \times \U(1)_F$ isometry.  The $(\z_{1i},\bar\z_{2i})$ transform as a doublet of the $\SU(2)_R$ non-holomorphic isometry, while the $\U(1)_F$ action under which $\z_{1i}$ and $\bar\z_{2i}$ all have the same charge is a tri-holomorphic isometry.  Thus the $\U(1)_F$ is the part of the $\cN=3$ R-symmetry which acts as a flavor symmetry from the $\cN=2$ perspective.

Note that $\cH$ inherits from the TSK structure of $\cM$ some additional structure beyond its hyperk\"ahler structure.  Firstly, it is flat.  Secondly, it inherits from the special complex structure of $\cM$ an additional complex structure beyond the $\C\P^1$ of hyperk\"ahler complex structures described above.  In this new complex structure the $a^1_i$ and $a^2_i$ special coordinates are holomorphic coordinates, and the $\U(2)$ isometry group acts holomorphically.  Thirdly, the $\spdnz$ EM duality special structure carries over to $\cH$.  

In summary, the hyperk\"ahler and special slices of a TSK manifold are linear subspaces of complex dimension $2n$ and $n$, respectively, which are reductions of the TSK structure to analogs with $\U(2)$ and $\U(1)$ isometry groups, respectively.

The main caveat to this whole discussion, however, is that the hyperk\"ahler and special slices might fail to exist for rank $n\ge 2$. The reason is easy to see: there may be no solutions to the equations \eqref{slice1} in a given special coordinate chart.  When $n=1$, solutions are always assured since the special coordinates $(a^1,a^2,a^3)$ of any given point can be rotated to $(\a,0,0)$ by the $\U(3)_R$ action, and so preserves some $\U(2)\subset \U(3)_R$.  Thus it lies on a special slice, and, by picking a $\U(1)$ subgroup of the $\U(2)$, also on a hyperk\"ahler slice.  But for $n=2$ this argument fails, for the $2\times 3$ matrix $a^I_i$ of special coordinates can generically only be rotated to the form $\left(\bsm \a & 0 & 0\\ \b & \g & 0\esm\right)$, and thus only preserves a $\U(1)$, so only a hyperk\"ahler slice is assured to go through the point.  The general condition for a hyperk\"ahler or special slice to go through a point is that its $n\times3$ matrix of special coordinates $a^I_i$ have rank 2 or 1, respectively.  Clearly, for $n\ge 3$ a generic point has rank 3, so no subgroup of $\U(3)_R$ is preserved and neither a special nor a hyperk\"ahler slice goes through it.  

Since special and hyperk\"ahler slices do not go through generic points, it is possible that there are no points in a given special coordinate chart lying on these slices.  It then becomes a global question whether upon patching together all the charts, any have slices going through them. In particular, as described in section \ref{sec5.2.2}, we can straighten out the special coordinates to form a single ``chart'' covering all of $\cM_\text{smooth}$ with its boundaries identified by monodromy group transformations $\r_{\t,\g}$ of the form \eqref{sp2rZact2}. The question then becomes whether these identifications always force the existence of points on the chart whose matrix of special coordinates have reduced rank. We do not know the answer to this question.

Since the special and hyperk\"ahler sections can be thought of as the Coulomb and Higgs branches of the $\cN=3$ theory viewed as an $\cN=2$ theory with respect to a choice of an $\cN=2$ subalgebra, it may seem surprising that we claim that neither a Higgs branch, $\cH$, nor a Coulomb branch, $\cC$, might exist in an $\cN=3$ theory with a moduli space, $\cM$.  $\cM$ is a mixed branch from the $\cN=2$ point view, \emph{locally} the Riemannian product $\til\cC\times\til\cH$, where $\til\cC$ is some open set of a special K\"ahler space and $\til\cH$ is some hyperk\"ahler cone.  But globally $\cM$ may fail to be a product, and even fail to be a bundle of, say, $\til\cH$ fibers over a $\cC$ base, but be ``twisted'' in both the $\til\cH$ and $\til\cC$ directions. Then both a $\cH$ and $\cC$ section might fail to exist.  

This should be contrasted to the usual situation \cite{Argyres:1996eh} in a genuinely $\cN=2$ theory, where the Higgs branch is typically fibered over the Coulomb branch along a singular subvariety, $\cC'$, of $\cC$, and $\cC'$ and $\til\cH$ meet, moreover, at a singularity of the $\til\cH$ fiber. In such a case both the Higgs and Coulomb sections necessarily exist. But there are a special class of mixed branches, called \emph{enhanced Coulomb branches} (ECBs) in \cite{Argyres:2016xmc}, in which a \emph{regular} $\til\cH \simeq \H^h$ is fibered over every point --- in particular all the regular points --- of $\cC$.  In this case we are saying that it is not a priori evident that the $\cC$ section need exist.\footnote{Note that such a section was implicitly assumed to exist in the discussion of ECBs in \cite{Argyres:2016xmc}.} The moduli space $\cM$ of an $\cN=3$ theory is an example of such an ECB.

%%%%%%%
%%%%%%%

\subsection{Enhancement to $\cN=4$}
\label{sec5.4-N=4}

The $\cN=4$ superconformal symmetry group includes an $\SU(4)_R$ $R$-symmetry factor which is spontaneously broken on the moduli space of vacua.  So what distinguishes the moduli space of an $\cN=4$ SCFT from that of a genuinely $\cN=3$ SCFT is that its continuous isometry group is enlarged to $\SO(6) \subset \SU(4)_R$. The reason only the index-2 $\SO(6)$ subgroup acts as an isometry on $\cM$ is that the scalar fields of the free massless $\cN=4$ vector multiplet transforms in the $\bf 6$ of $\SU(4)_R$. This implies also that the $\SO(6)$ isometry group actions is such that the special coordinates transform in $n$ copies of the $\bf 6$.

This enlargement of the continuous isometry group is the only further condition that $\cN=4$ supersymmetry imposes on the geometry of smooth points of an $\cN=4$ moduli space.  This is because the field content of the massless $\cN=3$ and $\cN=4$ vector multiplets and the leading (2-derivative) terms of their IR effective actions coincide.  

This enlargement of the continuous isometry group relative to the $\cN=3$ case implies a restriction on the monodromy group $\G(\cM)$ such that its representation $\r_\t(\cM) \subset \GL(n,\C)$ \eqref{GLnrep} on $\C^n$ can be realized as a subgroup of a real group $\GL(n,V)$ where $V\simeq\R^n$ is some real subspace of $\C^n$. Since the monodromy group is now real, there is a quadratic form which is left invariant by it and this in turn implies that the coordinate ring always has a holomorphic invariant of degree 2. This matches with a constraint discussed in \ref{sec2.4} on the chiral ring of $\cN=4$ SCFTs.  It also implies that there is an enlargement of the set of inequivalent metric-compatible complex structures on $\cM$ to a whole $\C\P^3$. It is a single orbit of the action of the $\SU(4)_R$ isometry group on any given complex structure.  

We now derive each of these consequences of the existence of an $\SO(6)$ isometry group on a TSK manifold.

\paragraph{Restriction of the monodromy group.} 

The requirement that the isometry group be enhanced from $\U(3)$ to $\SO(6)$ implies that the identifications \eqref{sp2rZact2} on the special coordinates upon traversing loops $\g\in\pi_1(\cM)$ must be restricted: the commutant of $\r_\t(\cM) \subset \GL(n,\C) \subset \GL(3n,\C)$ in $\SO(6n)$ must contain an $\SO(6)$ which acts on the first factor of the decomposition $\R^{6n} = \R^6 \otimes V$, where $V\simeq \R^n \subset \C^n$ is a real subspace.  Since the commutant of such an $\SO(6)$ in $\SO(6n)$ is $\GL(n,\R)$ it follows that $\r_\t(\cM) \subset \GL(V) \subset \GL(n,\C)$.  

%This is equivalent to saying that there is a choice of real structure on $\C^n$ --- that is, an antilinear involution $\s:\C^n\to\C^n$ --- with respect to which all elements of $\r_\t(\cM)$ are real, i.e., $[\s,\r_\t(\cM)]=0$.  So if $V\simeq\R^n\subset\C^n$ is the real subspace with respect to $\s$ (i.e., its $+1$ eigenspace) then $\r_\t(\cM): V \to V$.

The real and imaginary decomposition $\C^n = V \oplus iV$ gives $\C^n \simeq \R^2 \otimes V$ as a real vector space, and the monodromy group elements have the form $\r_{\t,\g}^\R = \I_2 \otimes \til r_\g$ with respect to this decomposition.  Then in $\R^{6n} = \R^6 \otimes V$ the monodromies all have the form
\begin{align}\label{N4monods}
    R_\g = \I_6 \otimes \til r_\g
\end{align}
for $\til r_\g \in \GL(V)$.  

An isometry, $S$, must commute with all the monodromies: $[S,R_\g]=0$ for all $\g$. This implies that in the $\R^6\otimes V$ decomposition, $S = s \otimes \I_n$ with $s \in \GL(6,\R)$.

An isometry $S$ must also preserve the metric. The metric components in the $\R^6\otimes V$ decomposition are $g_{6n} = \I_3 \otimes g'_{2n}$ where $g'_{2n}$ is some positive symmetric matrix on $\R^2\otimes V$. Recall that with respect to the ``standard'' real decomposition of $\C^n \simeq \R^2\otimes \R^n$, $g'_{2n} = \I_2 \otimes \Im \t$, but this factorized form need not persist in the $\C^n \simeq \R^2 \otimes V$ real decomposition. The condition that $S=s \otimes \I_n$ in the $\R^6 \otimes V$ decomposition implies that for such $S$ to span $\SO(6)$ and leave the metric invariant,  $\Im\t$ and $V$ must be such that $g_{6n}$ factorizes as
\begin{align}\label{N4monods2}
    g_{6n} = \I_6 \otimes g''_n
\end{align}
for $g''_n$ some positive symmetric matrix on $V$.  Then $g_{6n} = S^T g_{6n} S$ implies $s^Ts=\I_6$, so $s \in \O(6) \simeq \SO(6) \ltimes \Z_2$.

The two conditions \eqref{N4monods} and \eqref{N4monods2} require that there is a real decomposition $\C^n = \R^2\otimes V$ with respect to which the monodromies are real and which at the same time preserves the factorization of the metric derived from $\Im\t$.  In addition the monodromy group and the form of $\t$ are related by the fact \eqref{eqtau} that $\t$ is fixed by the monodromy group action.  We do not know what the most general form of the solution of these interlocking constraints are, but it is clear that they tightly restrict the possible monodromy groups that an $\cN=4$ theory can have.  In all the $\cN=4$ cases we know, the constraints are satisfied by monodromy groups which are real in the ``standard'' real decomposition $\C^n \simeq \R^2 \otimes \R^n$, for which $g''_n = \Im\t$. Since $\Im\t$ is positive definite, the only way this can happen is if the lower left $C_\g$ $n\times n$ block of each $M_\g \in \spdnz$ monodromy vanishes.  Then all monodromies will have the form 
\begin{align}
    M_\g &= \bpm A_\g & S_\g A_\g^{-T} \d_n^{-1} \\ 
    0 & \d_n A_\g^{-T} \d_n^{-1} \epm & &\text{with}&
    A_\g \d_n\t\d_n A_\g^T + S_\g &= \d_n\t\d_n
\end{align}
for all $\g \in \pi_1(\cM)$, where $S_\g$ is a symmetric integer matrix and $A_\g\in\GL(n,\Z)$. Example B described in section \ref{5.examples} is of this form.

Note that we have shown that the $\cN=4$ isometry group actually contains $\O(6) \simeq \SO(6) \ltimes \Z_2$, a slight enlargement over the initially assumed $\SO(6)$ isometry group.

\paragraph{Invariant holomorphic quadratic form on special slices.}

The set of nine independent globally defined quadratic forms, $a^\dag_I\, \Im\t\, a^J$, on $\cM_\text{smooth}$ found in section \ref{sec5.2.4} for a general TSK manifold are enlarged to a set of 21 such forms on an $\cN=4$ TSK manifold. This follows by decomposing $\C^n = V \oplus i V$ into real and imaginary subspaces as in the last paragraph and writing $a^I = \a^I + i \a^{I+3}$ with respect to this decomposition.  Then the $\a^a$, $a=1,\ldots,6$, transform in the $\bf 6$ of the $\O(6)$ isometry group, and the metric gives rise to the set of monodromy-invariant quadratic forms, $\a^a g''_n \a^b$, which transform in the ${\bf 21}$ of $\O(6)$. In particular, defining $S := \I_3 \otimes \left(\bsm 1 & i \\ i & -1 \esm\right)$, a globally defined complex quadratic form is
\begin{align}\label{N4quad1}
    S_{ab} \, \a^a g''_n \a^b 
    = \a^I g''_n \a^I - \a^{I+3} g''_n \a^{I+3}
    + 2 i \a^I g''_n \a^{I+3} 
    = a^I g''_n a^I ,
\end{align}
which is holomorphic in the special coordinates. Therefore it gives a holomorphic quadratic invariant on any special slice of $\cM$.  

%\red{Converse?}

\paragraph{Complex structures.}

Because of the restricted form \eqref{N4monods} of the $\cN=4$ monodromy group, the set of inequivalent metric-compatible complex structures, $J$, is enlarged. The condition that $[J,R_\g]=0$ for all $\g$ gives $J = j \otimes \I_n$ and $j^2 = - \I_6$, just as in \eqref{cplx3} except now in the $\R^{6n} = \R^6 \otimes V$ decomposition. The condition that $J$ is metric compatible then implies $j^Tj=\I_6$. This means simply that $j$ is an orthogonal complex structure on $\R^6$. The inequivalent such complex structures form the coset space $\SO(6)/\U(3) \simeq \C\P^3$.

They form a single orbit under the action on $\cM_\text{smooth}$ of the $\SU(4)_R$ symmetry of an $\cN=4$ SCFT.  This follows because each complex structure is left invariant by a $\U(3) \subset \SU(4)_R$ which acts holomorphically with respect to that complex structure.  Indeed, with respect to the $\cN=3$ subalgebra for which that $\U(3)$ is the R-symmetry, the complex structure is the $\cN=3$ special complex structure introduced in section \ref{sec5.2.5-kahler}.  So the R-symmetry orbit of this complex structure is all of $\C\P^3$ since $\C\P^3 \simeq \SU(4)_R / \U(3)$. 

%%%%%%
%%%%%%
%%%%%%

\section{TSK spaces}\label{TSKspace}

$\cN=3$ moduli spaces are not TSK \emph{manifolds}, since interacting physics implies metric singularities where additional charged states become massless. Beyond metric singularities, singularities in the complex structure of the space are also physically important. The latter are largely captured by the chiral ring structure extensively described in section \ref{sec2.3} and \ref{sec2.3.3}.  Furthermore, we are interested in the moduli spaces of $\cN=3$ SCFTs, as a result of which their TSK geometries will be cones with (at least) a metric singularity at its tip for any interacting theory.  The question then arises as to what singularities are physically allowable.  We do not have a definitive answer to this question, but  suggest here what we think are a reasonable minimum set of properties that a general TSK space should have.

Singularities of the moduli space of vacua can be of two types, with different physical interpretation. 
\begin{itemize}
    \item The \emph{metric singularities} are points where the metric becomes non-analytic. These are places where $\cM$ is no longer a smooth Riemannian manifold, but may still have a regular complex structure. We call the set of these points $\cM_\text{metric}$. They are interpreted as the loci where states which are charged under the low energy $\U(1)^n$ gauge group, and are massive on a generic point of $\cM$, becomes massless. Since the central charge \eqref{N3Z} provides a lower bound for the mass of a state given its EM charges, the locus of metric singularities $\cM_{\mathrm{metric}}$ is a subvariety of the set of zeros of $Z^K(q,p)$, for all $K=1,2,3$ and a set of occupied EM charges $(q,p)$. It is thus complex-analytic with respect to any of the complex structures of $\cM$. 
    \item The set of \emph{complex singularities}, which we will indicate as $\cM_\mathrm{cplx}$, occur instead where $\cM$ is singular as a complex variety. If the chiral ring associated to the operators whose vev parameterizes the moduli space is not freely generated \cite{Argyres:2018wxu}, this generically implies that complex singularities arise. In particular, given the explicit expression of $\C\{\cM\}$, the holomorphic coordinate ring of $\cM$ \eqref{holoring}, it is straightforward to compute $\cM_{\mathrm{cplx}}$. 
\end{itemize}
Generically, $\cM_{\mathrm{cplx}}$ is a proper subvariety of $\cM_{\mathrm{metric}}$, for explicit examples see \cite{Argyres:2019ngz,PAMN3II}. For this reason in the following we will use $\rm{sing}(\cM)$, the singular locus of a given TSK space, and $\cM_{\textrm{metric}}$ interchangeably. 

\subsubsection*{Examples.}

It can be helpful to see how these structures work in practice in the two examples that we have been analyzing in detail.

\paragraph{A-} In the rank one examples, the case $k=1$ is particular because it has no singularity. For $k>1$, the origin is obviously a metric non-analyticity as it is the locus where the scale invariant theory lives. Metrically $\cM$ is a flat cone with a curvature singularity at its tip. Furthermore, $\cM\equiv\C^3/\Z_k$ as a complex space. Writing explicitly the corresponding holomorphic coordinate ring it is easy to show that the origin is also a complex singularity. Thus in this case $\cM_{\textrm{cplx}}\cong\cM_{\textrm{metric}}\cong\rm{sing}(\cM)\equiv\{0\}$.

\paragraph{B-} In $\til\cM\equiv\C^6$ we can determine the varieties which are fixed by non-trivial elements of $S_3$. The order-two elements each fix a codimension 3 variety, 
\begin{align}\label{metsing1}
\til\cV_{(12)} &= \left\{ \ \, a_1^I - a_2^I = 0 \, , I=1,2,3 \right\}, \\\label{metsing2}
\til\cV_{(23)} &= \left\{ a_1^I +2 a_2^I = 0 \, , I=1,2,3 \right\}, \\\label{metsing3}
\til\cV_{(13)} &= \left\{ 2a_1^I - a_2^I = 0 \, , I=1,2,3 \right\}, 
\end{align}
and the order-three elements fix only the origin and permute the $\til\cV_{(ij)}$. 

Now we descend to the orbifold $\cM=\til\cM/S_3$. On that space we can pick coordinates 
\begin{align}\label{invS3}
    u^I &:=\frac{1}{3} [(a^I_1)^2 + a^I_1 a^I_2 + (a^I_2)^2 ] &
    &\text{and}&
    v^I &:= \frac{1}{2} a^I_1 a^I_2 (a^I_1+a^I_2) \, ,
\end{align}
which are six $S_3$-invariant polynomials. The three submanifolds $\til\cV_{(12)}$, $\til\cV_{(23)}$, $\til\cV_{(13)}$ all descend to
\begin{equation}\label{metsing}
   \cM_\text{metric} = \left\{ (u^I)^3 = (v^I)^2  \, , I=1,2,3 \right\} \,. 
\end{equation}
This is indeed the whole of $\cM_\text{metric}$, as the other non-trivial elements of the orbifold group fix only the origin. It has complex dimension 3 and complex codimension 3, and it is a complete intersection. We then define $\cM_\text{smooth} = \cM \setminus \cM_\text{metric}$.

The study of complex singularities is more intricate and explicit computations can become cumbersome given the large number of both generators of the holomorphic coordinate ring and relations among them. To illustrate how things work, let's focus on its special slice, $\cC\cong \C^2/A_2$, identified as the $a^2_i=a^3_i=0$ sub-locus. Since here the orbifold group, that is the Weyl group of $A_2$, acts irreducibly, the Chevalley-Shephard-Todd theorem \cite{Shephard:1954,Chevalley:1955} simplifies things considerably and it ensures that the holomorphic coordinate ring is a polynomial ring:
\beq\label{PolRing}
\C\{\cC\}:=\C[u^1,v^1],
\eeq
where $u^1$ and $v^1$ are the $A_2$-invariant homogeneous polynomials in the special coordinates $(a_1^1,a_2^1)\in\C^2$ of degree two and three respectively, defined in \eqref{invS3}. \eqref{PolRing} would in particular imply that no complex singularity arises in this case, though this is an artifact of considering only $\cC$. From the analysis of $\C\{\cM\}$, it is in fact possible to show that the origin of $\cC$ is a complex singularity, so $\cM_{\rm cplx}=\{0\}$. In this case $\cM_{\rm cplx}\subset\cM_{\rm metric}$.

\vspace{1em}

In the rest of this section, we will uncover constraints on the allowed singular behavior $\cN=3$ moduli spaces by adding in the requirement of superconformal invariance of the underlying $\cN=3$ field theory.  
 
\subsection{Conditions from spontaneously broken scale and R-symmetries}\label{sec2.2}

Using the spontaneously broken dilatation symmetry we show that $\cM$ is a bouquet of flat cones (see figure \ref{cones}), and the $\U(3)_R$ symmetry together with dilatations acts linearly on the special coordinates.

To see this, note that the spontaneously broken dilatation symmetry implies that $\cM$ has a real homothety ($\R^+$ scaling action), under which we take the line element to scale with (mass) weight one.  Thus the differentials of the special coordinates scale linearly, and so the special coordinates themselves scale as $a_j^I \to \m\ a_j^I $, where $\m\in\R^+$ is the scale factor.  By scaling $\m\to0$ we see that the chart where the special coordinates form a non-degenerate coordinate system on $\cM$ approaches the point with coordinates $a_j^I  = 0$ arbitrarily closely. This point is at finite distance and is on the boundary of that chart.  The fixed point is not necessarily in the chart, but may instead only be on its boundary, because the special coordinate system may (and generally does) degenerate there.  

Broken scale invariance implies that if there are more than one fixed points of dilatations, they must be at infinite metric distance from one another, for otherwise their distance would be a scale in the theory.  Thus there can only be a single finite-distance fixed point of dilatations on $\cM$.  This is the scale-invariant vacuum which we have seen is the origin of every special coordinate chart.   $\cM$ thus inherits the aforementioned structure of a bouquet of cones with the origin as their shared tip; see figure \ref{cones}.  We will now concentrate on a single metrically complete component of such a bouquet.

So, together with the spontaneously broken dilatations, the $\R^+\times\U(3)_R$ symmetry acts linearly in terms of special coordinates.  Furthermore, since the $\R^+$ charge (i.e., scaling dimension) of $a^J_j$ is equal to half its $\U(1)_r$ charge, they combine to form a complex $\C^*$ action.  So the smooth part of an $\cN=3$ CB, $\cM_\text{smooth}$, is a flat $3n$-complex-dimensional manifold with a $\C^*$ complex homothety and an $\SU(3)_R$ isometry fixing one point of the metric completion of $\cM_\text{smooth}$.  Because we do not assume $\Uot_r \subset \U(3)_R$ to be compact,  the complex homothety of $\cM$ is not strictly given by a $\C^*$ action, but rather by some $\til\C^*$ action where $\til \C^*$ is the infinitely-sheeted cover of $\C^*$ with a logarithmic branch point at the (excluded) origin. 

In a special coordinate chart, which we have argued extends from the tip of the cone to infinity (though it need not cover the whole cone in the ``angular'' directions), an element $U_3\in\SU(3)_R$ acts on the $\C^{3n}$ special coordinates as multiplication by $U_3\otimes  \I_n$, while $\m\in\C^*$ acts as multiplication by $\m \I_3\otimes \I_n$.

The TSK monodromy condition \eqref{TSK4} implies that upon continuing along a path $\g$ nontrivial in $\pi_1(\cM)$, the special coordinates experience a monodromy in $\GL(3n,\C)$ of the form \eqref{sp2rZact2}.  From its form it is apparent that it commutes with the $\U(3)_R$ action on the special coordinates, and so is compatible with the $\U(3)_R$ action being an isometry of $\cM$. Since metric singularities are in complex co-dimension three, $\g$ does not link $\cM_{\mathrm{metric}}$. The situation is thus superficially different from the analysis of CBs of $\cN=2$ SCFTs where the special K\"ahler structure imposes that $\cM_{\mathrm{metric}}$ is in complex co-dimension one. In this case non-trivial loops link $\cM_{\mathrm{metric}}$ and thus the consistency between the monodromy group and EM duality in the IR sets strong constraints on what singularities can appear and, consequentially, on the spectrum of CB operators at the conformal point \cite{Caorsi:2018zsq, Argyres:2018urp}.  Though non-trivial loops in $\cM_\text{smooth}$ do not link $\cM_\text{metric}$ in TSK spaces, nevertheless we will see below that there are elements of $\pi_1(\cM)$ which are associated to components in $\cM_\text{metric}$ in the sense that they can only be shrunk to a point by passing them through $\cM_\text{metric}$. Thus the existence of $\cM_\text{metric}$ is the obstruction to these paths being homotopically trivial as effectively as if they linked $\cM_\text{metric}$. From this, it follows that for TSK spaces the analysis of the monodromy group gives similarly strong constraints as in the special K\"ahler case on the structure of possible metric non-analyticities. 

\subsection{Conditions from BPS spectrum: TSK stratification}\label{secBPS}

We now connect the geometry in a neighborhood of a point in $\cM_\mathrm{metric}$ to the charges of BPS states becoming massless there and to the EM duality monodromies they induce. They are connected to one another through the physical condition that the central charges associated to the vector of electric and magnetic charges of the BPS states vanish at $\cM_\mathrm{metric}$. 

It is convenient to first set up a streamlined notation for the EM charges, the central charges, and the special coordinates.  We define bold symbols to be complex 3-vectors in the $\bf 3$ of $\U(3)_R$, and define $\bA := (\ba_D^i,\ba_i)$ be the $2n$-component vector of such 3-vectors which transforms linearly under the $\spdnz$ group.  Let $Q := (q^i,p_i)$ be the $2n$-component vector of integers in the charge lattice $\L \simeq \Z^{2n}$.  Finally, let $\bZ$ be the 3-vector of central charges, so $\bZ_Q(\bA) := Q^T \bA$. The central charge defines a dual relationship between the charges and the special coordinates.  Specifically, introduce a complex ``charge space" $V:=\C \otimes_\Z \L \simeq \C^{2n}$.  The special coordinates $\bA$ then take values in $\C^3 \otimes V^*$ where $V^*$ is the linear dual of the charge space. Furthermore, $V$ and $\C^3\otimes V^*$ are symplectic spaces whose symplectic bilinear forms are induced by the Dirac pairing.  Denote the Dirac pairing on the charge lattice by $\vev{Q,Q'} = Q^T \DD Q'$, and extend it to $V$ by linearity.  Then $V^*$ inherits an $\spdnz$ action from that on $V$, and a Dirac pairing, $\vev{\bar\bA,\bA} = \bar\bA^T \cdot \DD^{-1} \bA$ from the Dirac pairing on $V$ and the $\U(3)$-invariant inner product on $\C^3$.  

If a state in the theory with charge $Q\neq0$ becomes massless at a point $P\in\cM$ where $\bZ_Q (P)=0$, then there will be charged massless states in the spectrum of the effective theory everywhere on the locus $\cM_\text{metric}^Q := \{P\in\cM\, |\, \bZ_Q(P)=0\}$. This follows since, as we discussed in section \ref{sec2.1}, the BPS spectrum is constant in any connected component of $\cM_{\rm smooth}$. Then $\cM_\text{metric}  = \bigcup_{Q\in \text{BPS}} \cM_\text{metric}^Q$, for $Q$ running over the set of charges of BPS states in the spectrum. In this notation, the metric on $\cM_\text{smooth}$ is
\begin{align}\label{metric-ci}
g = i \vev{d\bar\bA\, ,\, d\bA}\, .
\end{align}

Since $\bZ$ is a complex 3-vector, it seems reasonable that $\cM_\text{metric}$ is a complex codimension $3$ subspace of $\cM$.  However this conclusion is difficult to justify or even make sense of without some knowledge of the kinds of spaces $\cM$ and $\cM_\text{metric}$ are.\footnote{As an example of the kind of wild behavior that we are trying to avoid, if the cardinality of the set of BPS states is infinite, then the union of the $\cM_\text{metric}^Q$ could conceivably accumulate densely in $\cM$.} To make further progress, we make the following assumption whose physical justification is unclear to us:
%Since the equation defining $\cM_\text{metric}^Q$ is linear in $Q$, all $Q\in \Phi$ can be taken to be primitive vectors in $\L$.  However $\Phi$ need not be a sublattice of $\L$, since if there are BPS states with charges $Q$ and $Q'$ in the BPS spectrum, there need not be a BPS state with charge $Q+Q'$ in the BPS spectrum.
\begin{quote}
	{\bf Regularity assumption:}  
	\emph{$\cM$ is a complex analytic space,
	and $\cM_\text{metric} = \bigcup_{Q\in \text{\rm BPS}} 
	\cM_\text{metric}^Q$ is an analytic subset 
	of $\cM$ of complex codimension 3.}
\end{quote}

It is important to notice that the central charges $\bZ_Q$, a priori only locally defined on charts, do not extend to single-valued functions on $\cM_\text{smooth}$ --- i.e., they are branched over $\cM_\text{metric}^Q$. However, with the regularity assumption, $\bZ_Q$ does extend analytically to $\cM_\text{metric}^Q$ where it vanishes. This follows since $\cM$ is metrically complete, so points of $\cM_\text{metric}^Q$ are at finite distance in $\cM_\text{smooth}$, and so will be on the boundary of special coordinate charts on $\cM_\text{smooth}$ at finite values of the special coordinates. Because the central charges are linear in the special coordinates, the restriction to $\cM_\text{metric}^Q$ describes a complex codimension 3 linear subspace in the closure of a special coordinate chart at least at a generic point.

In particular, $\cM_\text{metric}^Q$ inherits a flat metric almost everywhere, and so can be decomposed into its smooth and singular parts just as we did for the ambient $\cM$ space. Furthermore, $\cM_\text{metric}^Q$ carries a rank-$n$ TSK structure, since $\DD$, $\t$, and the $\spdnz$ overlap transformations are all inherited from the ambient charts and their overlaps. 

In the remainder of this section we argue that upon restriction to $\cM_\text{metric}^Q$, its inherited rank-$n$ TSK structure consistently restricts to a rank-$(n-1)$ TSK structure.  Thus $\cM_\text{metric}^Q$ will itself be a TSK space.

Consider a point $P \in \cM_\text{metric}^Q$ where the induced metric on $\cM_\text{metric}^Q$ is flat (non-singular).  With the above assumption, a small neighborhood of $P$ exists which is the product of a flat $6(n-1)$-real-dimensional ball in $\cM_\text{metric}^Q$ with a $6$-real-dimensional flat cone over a base, $L^Q$, which is a connected $5$-real-dimensional closed manifold (the link of $\cM_\text{metric}^Q$ in $\cM$).  We will call this cone over $L^Q$ the \emph{transverse slice} to $\cM_\text{metric}^Q$ in $\cM$ through $P$.

Denote by $\ba$ the $n$-component vector of $\U(3)$ triplets of complex special coordinates on $\cM$.  A $\GL(n,\C)$ transformation can be chosen to bring it to the form $\ba = (\ba^\parallel, \ba^\perp)$ such that $\ba^\perp \propto \bZ_Q(\bA)$ are a triplet of coordinates vanishing along $\cM_\text{metric}^Q$, and where $\ba^\parallel$ are the $n-1$ triplets of complex coordinates such that $\del/\del\ba^\parallel$ are tangent to $\cM_\text{metric}^Q$.  The $\ba^\parallel$ and their complex conjugates are thus good (independent, single-valued) coordinates on the flat $6(n-1)$-real-dimensional ball in $\cM_\text{metric}^Q$ around $P$. The components of the metric along $\cM_\text{metric}$ must be non-degenerate. The metric on $\cM_\text{smooth}$ is $g = i \vev{d\bar\bA\, ,\, d\bA}$, where $d$ is the exterior derivative on $\cM_\text{smooth}$, so the induced metric along $\cM_\text{metric}$ is $g_\parallel = i \vev{d_\parallel\bar\bA\, ,\, d_\parallel\bA}$. Its non-degeneracy implies that $\del_\parallel\bA$ and $\delb_\parallel \bar\bA$ span a $6(n-1)$-real-dimensional symplectic subspace of $\C^3\otimes V^*$. But since the conditions determining this subspace are $\U(3)_R$-invariant, and the $\U(3)_R$ acts irreducibly on the $\C^3$ factor, this subspace is of the form $\C^3 \otimes S_\parallel$ with $S_\parallel \subset V^*$ a $2(n-1)$-real-dimensional symplectic subspace of $V^*$. This defines a decomposition $V^* = S_\parallel \oplus (S_\parallel)^\perp$ into symplectic subspaces, where $(S_\parallel)^\perp$ is the $2$-real-dimensional symplectic complement\footnote{The symplectic complement of $S$ is defined by $S^\perp := \{ v\in V^* \, |\, \vev{v,w}=0\ \text{for all}\ w\in S\}$.} of $S_\parallel$.

Recall that the analytic continuation along any closed path $\g$ in $\cM_\text{smooth}$ will give a monodromy, $\bA \xrightarrow[]{\,\, \g \,\,\,} M_\g \bA$, with $M_\g \in \spdnz$. Call EM duality monodromies along closed paths in an arbitrarily small  neighborhood of $P$ the \emph{local monodromies of $P$}.  The fact that the $\ba^\parallel$ are good coordinates around $P$ implies that they suffer no nontrivial local monodromies.  This means, in particular, that the $S_\parallel$ symplectic subspace is left invariant by the local monodromies of $P$.

This puts strong constraints on the possible local monodromies around $P\in\cM_\text{metric}^Q$: they can be non-trivial only in the two $(S_\parallel)^\perp$ directions.  Moreover, the $V^* = S_\parallel \oplus (S_\parallel)^\perp$ decomposition is dual to a decomposition of the charge lattice, $\L = \S \oplus \S^\perp$, into symplectic sublattices, where $\S$ is a rank-2 sublattice which includes all the BPS charges that become massless on $\cM_\text{metric}^Q$.  We see this as follows.\footnote{The following argument closely parallels one given in \cite{Argyres:2018zay} in the context of $\cN=2$ special Kahler geometry.}

At the fixed values of $\ba^\parallel$ corresponding to the coordinates of the point $P$, $\ba^\perp$ are a $\U(3)$ triplet of complex special coordinates on the transverse slice.  They need not be single valued: if $[\g] \in \pi_1(L^Q)$ is non-trivial, then there can be an associated non-trivial monodromy $M_{[\g]} \in \spdnz$. These are local monodromies of $P$ because the transverse slice is a cone, so $\g$ can be continuously ``shrunk'' to be arbitrarily close to $P$ by scaling towards the tip, $P$, of the cone. These monodromies generate a subgroup $\G(P) \subset \G(\cM)$ of the monodromy group which is a representation of $\pi_1(L^Q)$ in $\spdnz$. From the definition, $\bZ=Q^T\bA$, of the central charges, such a monodromy implies that if $Q$ is in the BPS spectrum, then so is $M^{-T}_{[\g]}Q$. Thus there will be a whole local EM duality orbit of $Q$, $\Phi := \G(P)^{-T} Q$, of charges of BPS states which have vanishing masses at $\cM_\text{metric}^Q$. Since the central charge is linear in the charges, if $Q^T\bA$ and $(Q')^T\bA$ both vanish on $\cM_\text{metric}^Q$, then $(\ell Q+m Q')^T\bA=0$ there as well for arbitrary integers $\ell, m$.  Thus algebraically $\cM_\text{metric}^Q$ is characterized by the integral span of $\Phi$, i.e., a fixed sublattice, $\S\subset\L$.  

Note that with respect to the real symplectic structure defined by the charge lattice and its Dirac pairing, complex conjugation maps $\S$ to itself.  Thus 
\begin{align}\label{ZonV}
Q^T\bA=Q^T\bar\bA=0
\quad\text{on $\cM_\text{metric}^\S$ for all $Q\in \S$}\, .
\end{align}
This means that at each point of $\cM_\text{metric}^\S$, $\bA$ takes values only  in the annihilator subspace of $\S$.  This is the subspace $\S^\text{ann} \subset V^*$ which is the kernel of the dual pairing with $\S\subset \L$.\footnote{In other words, $\S^\text{ann} := \{ v \in V^* \, |\, Q^T v=0\ \text{for all}\ Q\in \S\}$.  We do not use the usual notation, ``$\S^\perp$", for the annihilator of $\S$ since we are reserving $\S^\perp$ for the symplectic complement of $\S$ in $\L$.} Taking derivatives of \eqref{ZonV} in the $\ba^\parallel$ direction implies $Q^T\del_\parallel\bA = Q^T\delb_\parallel\bar\bA = 0$ on $\cM_\text{metric}^\S$ for all $Q\in \S$.  Thus the $2(n-1)$-dimensional symplectic subspace $S_\parallel\subset V^*$ spanned by $\del_\parallel \bA$ and $\delb_\parallel\bar\bA$ on $\cM_\text{metric}^\S$ is in the annihilator of $\S$: $S_\parallel \subset \S^\text{ann}$. This implies that $\S$ is at most rank 2. By assumption $0\neq Q\in \S$, so rank$(\S)\ge1$. But if $\S$ were rank 1 then all the massless states at $\cM_\text{metric}^\S$ would have commensurate charge vectors.  Thus there would be an EM duality frame in which they were all electrically charged with respect to a single low energy $\U(1)$ gauge factor.  This would give rise to an IR free $\cN=3$ $\U(1)$ gauge theory, which does not exist.

Thus $\S$ must have rank 2. It then follows\footnote{See section 4.2 of \cite{Argyres:2018zay} for the argument.} that
\begin{align}\label{SisWann}
S_\parallel = \S^\text{ann}\, ,
\end{align}
and $\S$ is a symplectic sublattice of $\L$.
%\footnote{The first two statements are straight forward, and an elementary proof of the last is as follows.  If $\S$ is rank 2, take $\be_j$, $j=1,2$, to be a basis of $\S$.  Let $\bs^j$, $j=1,\cdots,2n-2$, be a basis of $S_\parallel$.  Extend this to a basis of $V^*$, $\bs^a$, $a=1,\ldots,2n$, and let the dual basis of $V$ be $\bs_a$.  By definition of the induced Dirac pairing on $V^*$, if $\DD^{ab} := \vev{\bs^a, \bs^b}$, then $\vev{\bs_a, \bs_b} = (\DD^{-1})_{ab}$. Since $S_\parallel$ is symplectic $\DD^{ij} = \vev{\bs^i,\bs^j} \neq 0$  Since $\DD$ is antisymmetric and non-degenerate, $\vev{\bs_{2n-1}, \bs_{2n}} = (\DD^{-1})_{2n-1,2n} = \text{Pf}(\DD^{ij})/\text{Pf}(\DD) \neq0$.   Write $\be_i = e_i^a \bs_a$, so $0=\be_i^T \bs^j = e_i^a \bs_a^T \bs^j = e_i^j$ for $i=1,2$ and $j=1,\ldots,2n-2$, since $S_\parallel$ annihilates $\S$.  Then $\vev{\be_1,\be_2} = e_1^a e_2^b \vev{\bs_a, \bs_b} =  (e_1^{2n-1} e_2^{2n} - e_1^{2n} e_2^{2n-1}) \vev{\bs_{2n-1},\bs_{2n}}$.  But since $\vev{\bs_{2n-1}, \bs_{2n}}\neq0$, the  vanishing of right side would imply $\be_1 \parallel \be_2$, contradicting the assumption that $\S$ is rank 2.}
Thus the charge lattice splits into two symplectic sublattices, $\L = \S \oplus \S^\perp$, where $\S^\perp$ is the symplectic complement of $\S$. Recalling that $V = \L \otimes_\Z \C$ and $V^*$ is its linear dual, we have found that 
\begin{align}
    S_\parallel &= (\S^\perp \otimes \C)^* , &
    (S_\parallel)^\perp &= (\S \otimes \C)^* .
\end{align}
This decomposition of the charge lattice into symplectic sublattices which are dual to the decomposition of the special coordinates in directions parallel and transverse to $\cM_\text{metric}^\S$, together with the fact that $\cM_\text{metric}^\S$ is a complex linear subspace of $\cM$ in terms of the special coordinates, gives the required restriction of the rank-$n$ TSK structure of $\cM$ to a rank-$(n-1)$ TSK structure on $\cM_\text{metric}^\S$.  Namely, simply restrict $\bA$ to take values in the $\C^3 \otimes S_\parallel$ subspace, $\t$ to the $S_\parallel$ subspace, and the charge lattice and $\DD$ to the symplectic $\S^\perp$ sublattice. 

The physical interpretation of this splitting is not surprising: $\S$ is the space of electric and magnetic charges of a $\U(1)_\perp$ factor, for which some charged states become massless at $\cM_\text{metric}^\S$, while $\S^\perp$ is the space of electric and magnetic charges of $\U(1)^{n-1}_\parallel$, whose charged states are generically not massless at $\cM_\text{metric}^\S$.  This basis, $\U(1)^{n-1}_\parallel \times \U(1)_\perp$, of the $\U(1)^n$ vector multiplets reflects the splitting $V^* = S_\parallel \oplus (S_\parallel)^\perp$ of the dual charge space into symplectic subspaces.

Likewise, restricting to the complementary subspaces, $\S$ and $(S_\parallel)^\perp$, induces a TSK structure on the transverse slice through $P$. Indeed, the transverse slice through $P$ can be interpreted as the moduli space of a rank-1 $\cN=3$ SCFT describing the degrees of freedom which become massless at $\cM_\text{metric}^\S$, and the splitting derived above reflects the physically obvious fact that this rank-1 SCFT decouples in the IR from the other, heavy, charged states in the theory.

But the physical interpretation of the TSK structure inherited by $\cM_\text{metric}^\S$ is less clear. Nevertheless, this result imposes an interesting regularity on the geometry of TSK spaces, closely parallel to that seen in the structure of symplectic singularities. Just as we decomposed the rank-$n$ TSK space as $\cM = \cM_\text{smooth} \cup \cM_\text{metric}$ into a rank-$n$ TSK manifold and a $3(n-1)$-dimensional singular locus, since we now know that $\cM_\text{metric}$ is itself a rank-$(n-1)$ TSK space we can decompose it as $\cM_\text{metric} = \cM'_\text{smooth} \cup \cM'_\text{metric}$ and learn that $\cM'_\text{metric}$ is a rank-$(n-2)$ TSK space.

Iterating, we produce a decomposition of the original TSK space $\cM$ into the disjoint union of a series of TSK manifolds of decreasing rank, $\cM = \cM_\text{smooth} \cup \cM'_\text{smooth} \cup \cM''_\text{smooth} \cup \ldots$.  This necessarily terminates as the dimension decreases strictly at each step -- and it has to decrease by a number of complex dimensions a multiple of three, by the TSK constraint. This defines a \emph{stratification} and associated \emph{transverse slices}, of $\cM$ which will be formalized shortly in section \ref{sectionDefinitionTSKspace}. The strata are naturally partially ordered by inclusion under closure. This partial order can be graphically depicted in a \emph{Hasse diagram}, which is a tree graph in which the vertices represent the strata, and edges relate adjacent strata in the partial order. A transverse slice between two adjacent strata in the Hasse diagram is called an \emph{elementary slice}. Figures \ref{cones}b and \ref{cones}c give an example of a stratification and its Hasse diagram.

As described above, the transverse slices to this stratification also possess a TSK structure, interpreted physically as the moduli spaces of SCFTs at generic points on the strata. The elementary slices have complex dimension $3$, as this is the codimension of the solutions of $\bZ_Q=0$. 

This stratified structure, with transverse slices which are themselves characterized in a precise manner, is very reminiscent of the structure of nilpotent orbits of Lie algebras \cite{kraft1981minimal, kraft1982geometry, fu2017generic, Cabrera:2016vvv, Cabrera:2017njm} and more generally of symplectic singularities \cite{brieskorn1970singular, slodowy1980simple, beauville1999symplectic}, for which the existence of the stratification has been demonstrated in \cite{kaledin2006symplectic}. From the $\cN=2$ point of view, the hyperk\"ahler singular structure of the Higgs branch and its stratification has been studied recently in \cite{Bourget:2019aer}. 
%Since, the orbit of the Higgs branch under $\U(3)_R$ is the full moduli space $\mathcal{M}$ of an $\mathcal{N}=3$ theory, it is reasonable to expect that the stratification extends to the full TSK structure. 
A crucial difference with the geometries studied in \cite{Bourget:2019aer}, is that in the general singular hyperk\"ahler case considered there, the elementary slices can have arbitrary large (quaternionic) dimension, whereas in the TSK case, they always have the minimal dimension allowed for a TSK, i.e. three complex dimensions. In other words, the elementary slices --- or equivalently the edges of the Hasse diagram --- correspond to rank-1 $\cN=3$ geometries, which in turn are given by a subset of the Kodaira classification \cite{Aharony:2015oyb, Lemos:2016xke, Argyres:2019ngz}, see table \ref{tabCBrank1}. We will give examples after having given the formal definition of TSK spaces in section \ref{sectionDefinitionTSKspace}. 

\begin{table} 
    \centering
        \begin{tabular}{|c|c|c|}
    \hline 
        Kodaira & TSK & Corresponding $\cN\geq3$ theory\\
        \hline
        \hline
       $II^*$&$\C^3/\Z_6$& \begin{tabular}{c}
           $\cN=3$ S-fold\\
        $\Z_3$ gauging of $\SU(2)$ $\cN=4$\\
        $\Z_6$ gauging of $U(1)$ $\cN=4$ 
       \end{tabular} \\
        \hline
       $III^*$&$\C^3/\Z_4$&   \begin{tabular}{c}
            $\cN=3$ S-fold\\
            $\Z_2$ gauging of $\SU(2)$ $\cN=4$\\
      $\Z_4$ gauging of $U(1)$ $\cN=4$\\
       \end{tabular} \\
        \hline
      $IV^*$&$\C^3/\Z_3$&  \begin{tabular}{c}
      $\cN=3$ S-fold\\
      $\Z_3$ gauging of $U(1)$ $\cN=4$\\
       \end{tabular}   \\
        \hline
    \rcg  $I_0^*$&$\C^3/\Z_2$&  \begin{tabular}{c} $\SU(2)$ $\cN=4$ \\ $\Z_2$ gauging of $U(1)$ $\cN=4$ \end{tabular} \\ 
        \hline
    \rcg    $I_0$&$\C^3$&$U(1)$ $\cN=4$\\
        \hline
    \end{tabular}
    \caption{The list of allowed TSK geometries for rank-1 $\cN\geq3$ theories, with the associated Kodaira label. The green rows correspond to $\cN=4$ geometries. } 
    \label{tabCBrank1}
\end{table}

In summary, the singular geometry of the moduli space of $\cN=3$ SCFTs borrows aspects from both the hyperk\"ahler singular geometry of $\cN=2$ Higgs branches, namely the stratification, and from the special K\"ahler geometry of $\cN=2$ Coulomb branches, namely the fact that all the elementary slices have complex dimension 3.

\subsubsection*{Examples}

\paragraph{A. } Consider the rank one case, which is almost trivial. Here $\tau$ is a fixed number with positive imaginary part and the central charges are $Z^K = (q+p \tau)a^K$, with $(q,p) \in \mathbb{Z}^2$. As $(1,\tau)$ is an $\mathbb{R}$-basis of $\mathbb{C}$, $q+p \tau$ vanishes only for $p=q=0$. So the central charge for charged state vanishes when $a^1=a^2=a^3=0$. We recover that the origin is the only singular locus. 

\paragraph{B. } In the rank two case, which is less trivial, we illustrate the construction of this subsection. We have 
\begin{equation}
    Q = \left( \begin{array}{c}  p_1 \\ p_2  \\ q^1 \\ q^2 \end{array} \right) \, , \qquad \bA = \left( \begin{array}{ccc} a_D^{1,1} & a_D^{1,2} & a_D^{1,3} \\ a_D^{2,1} & a_D^{2,2} & a_D^{2,3} \\ a_1^1 & a_1^2 & a_1^3  \\ a_2^1 & a_2^2 & a_2^3  \end{array} \right) \, , 
\end{equation}
$\tau$ is a matrix given by equation (\ref{tauB}). The central charge reads
\begin{equation}
    Z^K_Q (\bA) = \left(q_1 + p_1 \varsigma + \frac{1}{2} p_2 \varsigma  \right) a_1^K + \left(q_2 + p_2 + p_2 \varsigma + \frac{1}{2} p_1 \varsigma  \right) a_2^K .
\end{equation}
Recall that the special coordinates $a^I_i$ range over $\C^6$ which covers six copies of a fundamental domain of the $S_3$ action \eqref{S3B} on $\C^6$.  The three pre-images of $\cM_\text{metric}$ \eqref{metsing} in the covering $\C^6$ are given by the three submanifolds (\ref{metsing1}), (\ref{metsing2}), and (\ref{metsing3}).  Since they are in different copies of an $S_3$ fundamental domain, only one of them need be taken to define the special coordinate chart in which we evaluate the central charge.  For concreteness, let us choose a chart so that \eqref{metsing1} describes $\cM_\text{metric}$.  Then the vanishing of $Z^K_Q(\bA)$ along this submanifold corresponds for charges of the form
\begin{equation}
 \S \, : \qquad    
 Q =  \bpm p \\ -p  \\ q \\ p-q \epm \, , \qquad 
 \text{for}\ (p,q) \in \Z^2 .
    %Q =  \left( \begin{array}{c}  p \\ 0  \\ 2q \\ q \end{array} \right) \, , \qquad 
    %Q =  \left( \begin{array}{c}  0 \\ p  \\ q \\ 2q-p \end{array} \right)
\end{equation}
This is the rank-2 symplectic sublattice $\Sigma$ introduced in the text. $\S^\perp$, the symplectic complement of $\S$, is then the span of
\begin{equation}
   \S^\perp \, : \qquad     
   \bpm p' \\ p' \\ q' \\ q'-p'  \epm \, , \qquad 
    \text{for}\ (p',q') \in \Z^2 .
      %\left( \begin{array}{c}  p' \\ -2p' \\ 0 \\ q' \end{array} \right) \, , %\qquad 
      %\left( \begin{array}{c}  2p' \\ p' \\ q' \\ -p' \end{array} \right)
\end{equation}
By (\ref{metsing1}), the special coordinates along $\cM_\text{metric}$ are given by $a_1^I=a_2^I \equiv a^I$. The $a^I$ are the components of the vector $\ba^\parallel$ introduced above. The metric (\ref{TSKmetric}) becomes $ \Im \, \varsigma \, (     \mathrm{d} a^I \mathrm{d}\ab_{I} + \mathrm{d}\ab_{I} \mathrm{d}a^I )$ up to an irrelevant numerical prefactor, and there is no further identification imposed on the coordinates. This means that the stratum has no metric singularity, and it is identified with $I_0$. 

\subsection{Conditions from SCFT locality and unitarity}

The $\cN=3$ superconformal multiplet structure for unitary theories (reviewed in section \ref{sec2.3.3}), together with the assumptions and conjectures relating 4d SCFT chiral rings to moduli space coordinate rings (discussed in section \ref{sec2.3}), provide a convincing set of relations between the SCFT operator algebra and the complex geometry of $\cM$.

What is the extent to which these relations can be used to constrain either the SCFT operator algebra or the TSK geometry of $\cM$? It turns out that the relations as yet provide only quite weak constraints in either direction. It is, however, unclear to us how much of this is due to the weakness of our current understanding of $\cN=3$ chiral rings and TSK spaces, and how much is due to these two structures capturing complementary information about the SCFT.

The chiral ring is the holomorphic coordinate ring of $\cM$, where holomorphic means with respect to a physical complex structure of $\cM$.  The gradings of the chiral ring by $\U(3)_R$ charges also encodes the $\U(3)_R$ action on $\cM$. Thus the chiral ring clearly captures information on the physical K\"ahler structures of $\cM$ and the action of its $\U(3)_R$ group of isometries. But the general structure of $\cN=3$ chiral rings described in section \ref{sec2.3.3} turns out to not give any additional constraints on TSK geometry beyond those developed in sections \ref{sec2.1} and \ref{sec2-TSK}.  For instance, the occurrence of dimension-2 chiral/anti-chiral operators from the stress-tensor multiplet follows from the existence of the invariant hermitian quadratic forms \eqref{TSKmetric4} on a TSK space, and the occurrence of the additional chiral and anti-chiral dimension-2 operators from the $\cN=4$ stress-tensor multiplet also follow from the existence of the invariant holomorphic quadratic form \eqref{N4quad1} on an $\cN=4$ TSK space.

On the other hand, some central aspects of TSK geometry, especially the charge lattice, its Dirac pairing, and the EM monodromy group, are invisible from the point of view of the chiral ring. Furthermore, the holomorphic coordinate ring of a TSK geometry with respect to its special complex structure does not seem to be captured by any $\cN=3$ chiral ring. But this ``extra'' structure on the TSK side does not seem to give rise to any general constraints on which combinations of $\cN=3$ chiral multiplets must or must not appear in SCFTs.

An argument of Aharony and Evtikhiev \cite{Aharony:2015oyb} does draw such conclusions, but only under the assumption of the existence of $\cN=2$ Coulomb branch subspaces of $\cM$. These subspaces are what we called the special slices of $\cM$ in section \ref{subspaces}.  A generalization of their argument goes as follows.

As discussed in section \ref{subspaces}, a special slice $\cC \subset \cM$ is a subspace fixed point-wise by a $\U(2)$ subgroup of $\U(3)_R$ and is holomorphic with respect to any of the physical complex structures of $\cM$. So if $\cC$ exists then there will be chiral ring operators corresponding to generators of the holomorphic coordinate ring of $\cC$.  With respect to a choice of $\cN=2$ subalgebra of the $\cN=3$ superconformal algebra, $\cC$ is a Coulomb branch, so the  subgroup of the $\cN=3$ isometry group which leaves it fixed is $\U(2) \simeq \SU(2)_R \times \U(1)_F$, where $\SU(2)_R$ is part of the $\cN=2$ $R$-symmetry group and $\U(1)_F$ is the $\cN=2$ flavor symmetry.

Decompose a general $\cN=3$ chiral multiplet, $\mathbf{X \Bb}_{(R_1,R_2),r}$, into $\cN=2$ multiplets. Using the general decomposition of $\suf(3)$ irreps in terms of $\suf(2) \oplus \uf(1)$ irreps given in, e.g., eqn.\ (6.16) of \cite{Cordova:2016emh}, along with the condition \eqref{N3crc} for a there to be a chiral ring operator in the multiplet, it is a straight forward computation to see that any $\SU(2)_R \times \U(1)_F$-neutral chiral ring operator must be in the $\U(3)_R$ irrep with weights $(R_1,R_2) = (0,R)$ and $r = r_*$ where $r_*$ is defined in \eqref{rstar}. From the shortening conditions summarized in table \ref{tabMultiplets}, the only $\cN=3$ chiral multiplets in this $\U(3)_R$ representation are the $\mathbf{B\Bb}_{(0,R)}$ multiplets.

We learn from this that if a special slice ($\cN=2$ Coulomb branch) exists, then $\mathbf{B\Bb}_{(0,R)}$ and $\mathbf{B\Bb}_{(R,0)}$ multiplets for some $R$ must appear in the spectrum of superconformal primaries in an $\cN=3$ SCFT, and furthermore, that these are the only primaries contributing chiral ring operators on the special slice. Since these operators have dimension $\D=R$ and $R\in \Z$, it follows that the special slice holomorphic coordinate ring generators necessarily have integer scaling dimensions.  

But, as discussed at length in section \ref{subspaces}, we do not have an argument (physical or mathematical) for the existence of special slices in TSK spaces of ranks greater than 1. It is possible that the stratified structure of TSK spaces and the fact that rank-1 TSK spaces do have special slices can be used to show the existence of special slices in general.  We will leave the exploration of this possibility to later work \cite{PAMN3II}.

\subsection{Proposal for a definition of a TSK space}
\label{sectionDefinitionTSKspace}

\begin{figure}[ht]
\centering
\begin{tikzpicture}
%% fig a
\begin{scope}[scale=.7, xshift=-5cm]
\draw[thick,draw=black!45,fill=black!10] (0,0) -- (1.5,-2.5) arc (375:165:1.5cm and .5cm) -- cycle;
\draw[thick,dashed,black!25] (1.5,-2.5) arc (15:165:1.5cm and .5cm);
\draw[rotate=120,thick,draw=black!45,fill=black!10] (0,0) -- (1.5,-2.5) arc (375:165:1.5cm and .5cm) -- cycle;
\draw[rotate=120,thick,dashed,black!25] (1.5,-2.5) arc (15:165:1.5cm and .5cm);
\draw[rotate=240,thick,draw=black!45,fill=black!10] (0,0) -- (1.5,-2.5) arc (375:165:1.5cm and .5cm) -- cycle;
\draw[rotate=240,thick,black!45] (1.5,-2.5) arc (15:165:1.5cm and .5cm);
\node[circle,fill=red,scale=.5] (br) at (0,0) {};
\draw[rotate=40,ultra thick,draw=red] (0,0) -- (3.15,0);
\draw[rotate=150,ultra thick,draw=red] (0,0) -- (2.15,0);
\draw[rotate=250,ultra thick,draw=red] (0,0) -- (3.15,0);
\draw[rotate=260,ultra thick,draw=red] (0,0) -- (3.15,0);
\node at (0,-4) {\bf (a)};
\end{scope}
%% fig b
\begin{scope}[scale=.7,xshift=+2cm]
\draw[thick,draw=black!45] (0,0) -- (0,-3.25);
%\draw[thick,dashed,black!25] (1.5,-2.5) arc (15:165:1.5cm and .5cm);
\draw[rotate=150,thick,draw=black!45,fill=black!10] (0,0) -- (1.5,-2.5) arc (375:165:1.5cm and .5cm) -- cycle;
\draw[rotate=150,thick,black!45] (1.5,-2.5) arc (15:165:1.5cm and .5cm);
\draw[rotate=210,thick,draw=black!45,fill=black!10] (0,0) -- (1.5,-2.5) arc (375:165:1.5cm and .5cm) -- cycle;
\draw[rotate=210,thick,black!45] (1.5,-2.5) arc (15:165:1.5cm and .5cm);
\node[circle,fill=red,scale=.5] (br) at (0,0) {};
\draw[ultra thick,draw=red] (0,0) -- (-.04,3.0);
\node[red] at (-.5,1.0) {$\cM_1$};
\node[red] at (+.43,-0.32) {$\cM_0$};
\node[black!75] at (+.43,-1.7) {$\cM_2$};
\node[black!75] at (1.3,2.3) {$\cM_4$};
\node[black!75] at (-1.3,2.3) {$\cM_3$};
\node at (0,-4) {\bf (b)};
\end{scope}
%% fig c
\begin{scope}[scale=.7,xshift=+9cm]
	\node [circle,draw,fill,inner sep=1pt,label=above:$\cM_4$] 
	at (0,2) [] (4) {};
	\node [circle,draw,fill,inner sep=1pt,label=above:$\cM_3$] 
	at (-2,2) [] (3) {};
	\node [circle,draw,fill,inner sep=1pt,label=right:$\cM_2$] 
	at (+1,0) [] (2) {};
	\node [circle,draw,fill,inner sep=1pt,label=left:$\cM_1$] 
	at (-1,0) [] (1) {};
	\node [circle,draw,fill,inner sep=1pt,label=below:$\cM_0$] 
	at (0,-2) [] (0) {};
	\draw (1) edge (4);
	\draw (1) edge (3);
	\draw (0) edge (1);
	\draw (0) edge (2);
\node at (0,-4) {\bf (c)};
\end{scope}
\end{tikzpicture}
\caption{Cartoons of scale-invariant moduli spaces, where each real dimension in the figure represents 3 complex dimensions.  The cones and lines are meant to extend to infinity; they are truncated here due to lack of space.  Figure (a) shows a bouquet of three cones: the conformal vacuum is the common tip of the cones, while at all other points conformality is spontaneously broken.  Each cone has sub-cones of metric singularities, shown in red. Figure (b) shows a space where $\cM_\text{smooth} = \cM_4 \cup \cM_3 \cup \cM_2$, the disjoint union of two dim${}_\C{=}6$ manifolds and one dim${}_\C{=}3$ manifold, and $\cM_\text{metric} = \cM_1 \cup \cM_0$ is the disjoint union of a dim${}_\C{=}3$ manifold $\cM_1$, and a 0-dimensional manifold $\cM_0$ (the conformal vacuum).  The strata of this space are the $\cM_j$ and their partial ordering by inclusion under closure is given in the Hasse diagram shown in figure (c).}
\label{cones}
\end{figure}
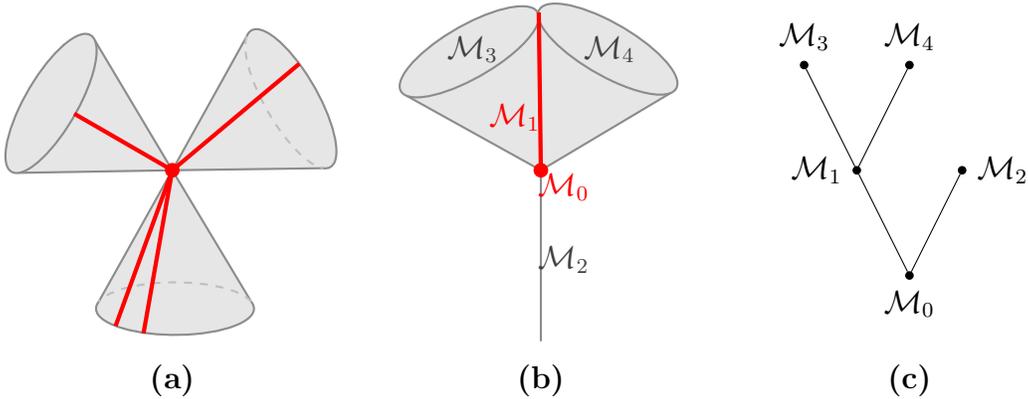

We end by suggesting here what we think are a reasonable minimum set of properties that a general TSK space should have, summarizing the basic ingredients above. Following \cite{goresky1988stratified}:

\begin{definition}\label{defn3}
A TSK space, $\cM$, is a complete metric space analytically stratified by TSK manifolds:
\begin{description}
    \item[(a)] $\cM = \bigcup_{j\in P} \cM_j$ for a finite partially ordered set $P$ of pairwise disjoint locally closed subsets $\cM_j$ called \emph{strata}.
    \item[(b)] $\cM_i \cap \bar{\cM_j} \neq \varnothing$ iff $\cM_i \subset \bar{\cM_j}$ iff $i\le j$.
    \item[(c)] $P$ is a graded poset\footnote{This means that $P$ has a rank function, ${\rm rank}:P\to\Z_{\ge0}$, compatible with the ordering such that if $j$ covers $i$ then ${\rm rank}(j)={\rm rank}(i)+1$. ``$j$ covers $i$'' (notated $j\gtrdot i$) means that $j>i$ and there is no $k$ such that $j>k>i$.} where dim$_\C(\cM_j)=3\,{\rm rank}(j)$. %Also $P$ has a minimum element of rank $0$.
    \item[(d)] Each stratum is a TSK manifold.  
    \item[(e)] Through any point of stratum $\cM_i$ and for all $j>i$ there exists a transverse slice in $\bar{\cM_j}$, $\cT_{i<j}$, which is a TSK space.
    \item[(f)] $\bigcup_{i<j} \cM_i = {\rm sing}(\bar\cM_j)$, the locus of metric or complex singularities of the metric completion of $\cM_j$.
    \item[(g)] ${\rm sing}(\bar\cM_j)$ is an analytic subspace of $\bar{\cM_j}$, and the special structure and metric on ${\rm sing}(\bar\cM_j)$ is the one induced by the metric completion of $\cM_j$. 
\end{description}
\end{definition}
See figure \ref{cones} for a visualization of some simple examples of such stratified spaces.

We recap the motivation for this definition:  The metric completeness of $\cM$ implies that its singular locus inherits a distance function.  Furthermore, together with a regularity assumption that the common zeros of the central charges for BPS states in the spectrum are nowhere dense in $\cM$ --- which seems reasonable since the central charges are linear in the special coordinates --- the condition that the singular locus occurs when the central charges \eqref{N3Z} all vanish for some charges in the BPS spectrum implies that the singular locus is an analytic subspace of $\cM$ of complex-codimension 3. We argued above that almost everywhere $\cM_\text{metric}$ inherits a TSK structure from that of $\cM$, basically by restriction. Induction on $P$ then implies the TSK stratification described above.

It might be useful to note that the condition that the singular locus is complex-analytic (with respect to any of the complex structures of $\cM$) endows $\cM$ with a Whitney stratification \cite{whitney1965tangents}.  This kind of stratification is more constrained and has nicer regularity properties than the looser topological stratification demanded above.

The condition that the stratification is graded ensures that there is a stratum of complex dimension $3m$ for every $0\le m\le n$ where $3n$ is the largest dimension of any stratum.
%, and the existence of a minimum element of $P$ ensures that the dimension-0 stratum (the superconformal vacuum) is unique. 
Transverse slices are the moduli spaces of the $\cN=3$ SCFT describing the massless degrees of freedom at each stratum, so are TSK spaces. Since the rank of $\cT_{i<j}$ as a TSK space is $\text{rank}(j)-\text{rank}(i)$, they can be defined using the above definition of a TSK by induction on $P$.

The elementary transverse slices, $\cT_{i\lessdot j}$, are the transverse slices where the index, $j$, of the enclosing stratum covers $i$ in $P$. We argued above that the possible geometries of the elementary slices is constrained to be one of the 5 possibilities listed in table \ref{tabCBrank1}.  This raises the possibility that all TSK geometries can be constructed in an algebraic way using the elementary transverse slices as ``building blocks''.  We will explore this idea in the second part of this study, \cite{PAMN3II}.  

We conclude by illustrating this proposed definition of TSK spaces using the simple examples developed above.

%%%%%%%%%
%%%%%%%%%
%%%%%%%%%

\subsubsection*{Examples}

\paragraph{A-} In the rank one examples, the Hasse diagrams are trivial and correspond directly to the geometries of table \ref{tabCBrank1}: 
\begin{equation}
    \begin{tikzpicture}
	\tikzstyle{hasse} = [circle, fill,inner sep=1pt];
		%I0
		\node [hasse] at (0,-1) [] (2) {};
		\node at (.5,-1) {$I_0$};
		\node at (0,-3) {$k=1$};
		%I0*
		\node [hasse] at (2,0) [] (11) {};
		\node [hasse] at (2,-2) [] (12) {};
		\node at (2.5,-1) {$I_0^\ast$};
		\draw (11) edge (12);
		\node at (2,-3) {$k=2$};
		%IV*
		\node [hasse] at (4,0) [] (21) {};
		\node [hasse] at (4,-2) [] (22) {};
		\node at (4.5,-1) {$IV^\ast$};
		\draw (21) edge (22);
		\node at (4,-3) {$k=3$};
		%III*
		\node [hasse] at (6,0) [] (31) {};
		\node [hasse] at (6,-2) [] (32) {};
		\node at (6.5,-1) {$III^\ast$};
		\draw (31) edge (32);
		\node at (6,-3) {$k=4$};
		%II*
		\node [hasse] at (8,0) [] (41) {};
		\node [hasse] at (8,-2) [] (42) {};
		\node at (8.5,-1) {$II^\ast$};
		\draw (41) edge (42);
		\node at (8,-3) {$k=6$};
		\end{tikzpicture}
\end{equation}
We label each edge of the Hasse diagram by the geometry of its associated elementary transverse slice. The case $k=1$ is not singular, so we don't have any transition. The unique stratum is a copy of $I_0$. In the other cases, the geometries are singular, the singular locus is the origin, and there is a single transverse slice. 

\paragraph{B-} For our rank two example, one can conjecture the following Hasse diagram: 
\begin{equation}\label{N4Su3}
    \begin{tikzpicture}
	\tikzstyle{hasse} = [circle, fill,inner sep=1pt];
		\node [circle, fill,inner sep=1pt,label=left:$\cM_2$] at (0,0) (1) {};
		\node [circle, fill,inner sep=1pt,label=left:$\cM_1$] at (0,-2) (2) {};
		\node [circle, fill,inner sep=1pt,label=left:$\cM_0$] at (0,-4) (3) {};
		\node at (.5,-1) {$I_0^\ast$};
		\node at (.5,-3) {$I_0$};
		\draw (1) edge (2);
		\draw (3) edge (2);
		\end{tikzpicture}
\end{equation}
Since the theory has rank two, we expect a diagram of height two. In addition, we computed the space of metric singularities in equation (\ref{metsing}). It is a complete intersection and an irreducible variety, so there is only one stratum, $\cM_1$, of complex dimension 3. As a consequence, the poset $P$ is totally ordered, and its Hasse diagram has the shape of a line. 

We still have to identify the geometry of the elementary slices. Here it is handy to use some physics intuition. The moduli space $\C^6/S_3$ corresponds to the moduli space of the $\SU(3)$ $\cN=4$ superYang-Mills theory. Then $\cM_1$ is where an $\SU(2)$ subgroup of the gauge group remains unbroken, and thus the transverse slice to the first stratum is the moduli space of an $\cN=4$ $\SU(2)$ sYM theory, which is the $I_0^* \equiv \C^3/\Z_2$ TSK space. 

The stratum $\cM_1$ itself can be instead identified by looking at the parametrization of $\cM_{metric}$ \eqref{metsing1}-\eqref{metsing2} in $\til\cM$. Recall that the orbifold group is $S_3=\Z_3\rtimes \Z_2$. The $\Z_3$ component interchanges \eqref{metsing1}-\eqref{metsing2} and thus to understand $\cM_1$ is sufficient to consider one of the subspaces, say \eqref{metsing1}. The $\Z_2$ instead fixes this subspace and thus $\cM_1\cong\C^3\equiv I_0$. A similar analysis can be carried directly in $\cM$ by looking at the $\cM_{\rm metric}$ parametrized by $u^I$ and $v^I$. The scaling dimension of the uniformizing parameter of the $(u^I)^3=(v^I)^2$ complete intersection singularity, is in fact 1.

It is interesting to notice that in this case $\cM_1$ has a (non-normal) complex singularity at the origin, but no intrinsic metric singularity. This singularity is instead present in the full space and it represents the locus where the full $\SU(3)$ gauge group is unbroken, that is why in our depiction in \eqref{N4Su3} a third stratum $\cM_0$ is present. The second part of this study, \cite{PAMN3II}, will contain a more detailed analysis of these structures beyond these simple examples. 

\acknowledgments

It is a pleasure to thank Amihay Hanany, Simeon Hellerman, Madalena Lemos, Matteo Lotito, Leonardo Rastelli and Travis Schedler for useful discussions.  PCA was supported in part by DOE grant DE-SC0011784 and by Simons Foundation Fellowship 506770. AB was supported by EPSRC grant EP/K034456/1 and STFC grant ST/P000762/1. MM was supported in part by NSF grants PHY-1151392 and  PHY-1620610. The completion of this work benefited from the 2019 Pollica summer workshop, which was supported in part by the Simons Foundation (Simons Collaboration on the Non-perturbative Bootstrap) and in part by the INF. This work also benefited from time spent at the Aspen Center for Physics, which is supported by National Science Foundation grant PHY-1607611.

%%%%%%%%%%
%%%%%%%%%%
%%%%%%%%%%

\appendix

\section{Notations and conventions}\label{appA}

We represent the weight space of $\mathfrak{u}(3)_R$. This space has natural coordinates $(r_1 , r_2 , r_3)$ which are related to $\mathbf{R} = (R_1,R_2)$ and $r$ by 
\begin{equation}
    \begin{cases}
    R_1 = r_1-r_2 \\ R_2 = r_2-r_3 \\ r = r_1+r_2+r_3 
    \end{cases}
    \Leftrightarrow 
    \begin{cases}
    r_1 = \frac{1}{3} (2R_1+R_2+r) \\ r_2 = \frac{1}{3} (-R_1+R_2+r) \\ r_3 = \frac{1}{3} (-R_1-2R_2+r)  
    \end{cases}
\end{equation}
According to (\ref{supercharges}), the supercharges $Q$ transform in a representation with weights 
\begin{equation}
  Q  \textrm{ : } (R_1,R_2,r) \in \{ (1,0,-1),(-1,1,-1),(0,-1,-1)\}  \, , 
\end{equation}
or equivalently 
\begin{equation}
  Q  \textrm{ : } (r_1,r_2,r_3) \in \left\{ \left(\frac{1}{3},-\frac{2}{3},-\frac{2}{3}\right),\left(-\frac{2}{3},\frac{1}{3},-\frac{2}{3}\right),\left(-\frac{2}{3},-\frac{2}{3},\frac{1}{3}\right)\right\}   \, . 
\end{equation}
We choose these weights as the basis of our root space. This makes it easy to see graphically how the supercharges act to generate the superconformal multiplets. This means that a weight will be represented by a point with coordinates $(q_1,q_2,q_3)$ given in terms of $(r_1,r_2,r_3)$ or $(R_1,R_2,r)$ by 
\begin{equation}
        \begin{cases}
    q_1 = \frac{1}{3} (r_1-2r_2-2r_3) = \frac{1}{3} (2R_1+R_2-r)  \\ q_2 = \frac{1}{3} (-2r_1+r_2-2r_3)  = \frac{1}{3} (-R_1+R_2-r)   \\ q_3 = \frac{1}{3} (-2r_1-2r_2+r_3)  = \frac{1}{3} (-R_1-2R_2-r)  .
    \end{cases}
\end{equation}

%%%%%%%%%%
%%%%%%%%%%
%%%%%%%%%%

\bibliographystyle{JHEP}

\begin{thebibliography}{10}

\bibitem{Seiberg:1997ax}
N.~Seiberg, {\it {Notes on theories with 16 supercharges}},  {\em Nucl. Phys.
  Proc. Suppl.} {\bf 67} (1998) 158--171,
  [\href{http://arxiv.org/abs/hep-th/9705117}{{\tt hep-th/9705117}}].
  [,158(1997)].

\bibitem{Cecotti:2015wqa}
S.~Cecotti, {\em {Supersymmetric Field Theories}}.
\newblock Cambridge University Press, 2015.

\bibitem{Argyres:2018zay}
P.~C. Argyres, C.~Long, and M.~Martone, {\it {The Singularity Structure of
  Scale-Invariant Rank-2 Coulomb Branches}},  {\em JHEP} {\bf 05} (2018) 086,
  [\href{http://arxiv.org/abs/1801.01122}{{\tt arXiv:1801.01122}}].

\bibitem{Caorsi:2018zsq}
M.~Caorsi and S.~Cecotti, {\it {Geometric classification of 4d $\mathcal{N}=2$
  SCFTs}},  {\em JHEP} {\bf 07} (2018) 138,
  [\href{http://arxiv.org/abs/1801.04542}{{\tt arXiv:1801.04542}}].

\bibitem{PAMN3II}
P.~Argyres, A.~Bourget, and M.~Martone, {\it {On the moduli spaces of 4d
  $\cN=3$ SCFTs II:Orbifold vs. non-Orbifold structure}},  {\em to appear.}

\bibitem{Seiberg:1994bz}
N.~Seiberg, {\it {Exact results on the space of vacua of four-dimensional SUSY
  gauge theories}},  {\em Phys. Rev.} {\bf D49} (1994) 6857--6863,
  [\href{http://arxiv.org/abs/hep-th/9402044}{{\tt hep-th/9402044}}].

\bibitem{Seiberg:1994rs}
N.~Seiberg and E.~Witten, {\it {Electric - magnetic duality, monopole
  condensation, and confinement in N=2 supersymmetric Yang-Mills theory}},
  {\em Nucl. Phys.} {\bf B426} (1994) 19--52,
  [\href{http://arxiv.org/abs/hep-th/9407087}{{\tt hep-th/9407087}}]. [Erratum:
  Nucl. Phys.B430,485(1994)].

\bibitem{Seiberg:1994aj}
N.~Seiberg and E.~Witten, {\it {Monopoles, duality and chiral symmetry breaking
  in N=2 supersymmetric QCD}},  {\em Nucl. Phys.} {\bf B431} (1994) 484--550,
  [\href{http://arxiv.org/abs/hep-th/9408099}{{\tt hep-th/9408099}}].

\bibitem{Xie:2015rpa}
D.~Xie and S.-T. Yau, {\it {4d N=2 SCFT and singularity theory Part I:
  Classification}},  \href{http://arxiv.org/abs/1510.01324}{{\tt
  arXiv:1510.01324}}.

\bibitem{Chen:2016bzh}
B.~Chen, D.~Xie, S.-T. Yau, S.~S.~T. Yau, and H.~Zuo, {\it {4D $\mathcal{N} =
  2$ SCFT and singularity theory. Part II: complete intersection}},  {\em Adv.
  Theor. Math. Phys.} {\bf 21} (2017) 121--145,
  [\href{http://arxiv.org/abs/1604.07843}{{\tt arXiv:1604.07843}}].

\bibitem{Wang:2016yha}
Y.~Wang, D.~Xie, S.~S.~T. Yau, and S.-T. Yau, {\it {$4d$ $\mathcal{N} = 2$ SCFT
  from complete intersection singularity}},  {\em Adv. Theor. Math. Phys.} {\bf
  21} (2017) 801--855, [\href{http://arxiv.org/abs/1606.06306}{{\tt
  arXiv:1606.06306}}].

\bibitem{Chen:2017wkw}
B.~Chen, D.~Xie, S.~S.~T. Yau, S.-T. Yau, and H.~Zuo, {\it {4d N=2 SCFT and
  singularity theory Part III: Rigid singularity}},
  \href{http://arxiv.org/abs/1712.00464}{{\tt arXiv:1712.00464}}.

\bibitem{Beem:2017ooy}
C.~Beem and L.~Rastelli, {\it {Vertex operator algebras, Higgs branches, and
  modular differential equations}},  {\em JHEP} {\bf 08} (2018) 114,
  [\href{http://arxiv.org/abs/1707.07679}{{\tt arXiv:1707.07679}}].

\bibitem{Argyres:2015ffa}
P.~Argyres, M.~Lotito, Y.~L{\"u}, and M.~Martone, {\it {Geometric constraints
  on the space of $ \mathcal{N} $ = 2 SCFTs. Part I: physical constraints on
  relevant deformations}},  {\em JHEP} {\bf 02} (2018) 001,
  [\href{http://arxiv.org/abs/1505.04814}{{\tt arXiv:1505.04814}}].

\bibitem{Argyres:2015gha}
P.~C. Argyres, M.~Lotito, Y.~L{\"u}, and M.~Martone, {\it {Geometric
  constraints on the space of $ \mathcal{N} $ = 2 SCFTs. Part II: construction
  of special Kahler geometries and RG flows}},  {\em JHEP} {\bf 02} (2018) 002,
  [\href{http://arxiv.org/abs/1601.00011}{{\tt arXiv:1601.00011}}].

\bibitem{Argyres:2016xua}
P.~C. Argyres, M.~Lotito, Y.~L{\"u}, and M.~Martone, {\it {Expanding the
  landscape of $ \mathcal{N} $ = 2 rank 1 SCFTs}},  {\em JHEP} {\bf 05} (2016)
  088, [\href{http://arxiv.org/abs/1602.02764}{{\tt arXiv:1602.02764}}].

\bibitem{Argyres:2016xmc}
P.~Argyres, M.~Lotito, Y.~L{\"u}, and M.~Martone, {\it {Geometric constraints
  on the space of $ \mathcal{N}$ = 2 SCFTs. Part III: enhanced Coulomb branches
  and central charges}},  {\em JHEP} {\bf 02} (2018) 003,
  [\href{http://arxiv.org/abs/1609.04404}{{\tt arXiv:1609.04404}}].

\bibitem{Caorsi:2018ahl}
M.~Caorsi and S.~Cecotti, {\it {Special Arithmetic of Flavor}},  {\em JHEP}
  {\bf 08} (2018) 057, [\href{http://arxiv.org/abs/1803.00531}{{\tt
  arXiv:1803.00531}}].

\bibitem{Argyres:2018urp}
P.~C. Argyres and M.~Martone, {\it {Scaling dimensions of Coulomb branch
  operators of 4d N=2 superconformal field theories}},
  \href{http://arxiv.org/abs/1801.06554}{{\tt arXiv:1801.06554}}.

\bibitem{Bourget:2018ond}
A.~Bourget, A.~Pini, and D.~Rodr\'iguez-G\'omez, {\it {The Importance of Being
  Disconnected, A Principal Extension for Serious Groups}},
  \href{http://arxiv.org/abs/1804.01108}{{\tt arXiv:1804.01108}}.

\bibitem{Argyres:2018wxu}
P.~C. Argyres and M.~Martone, {\it {Coulomb branches with complex
  singularities}},  {\em JHEP} {\bf 06} (2018) 045,
  [\href{http://arxiv.org/abs/1804.03152}{{\tt arXiv:1804.03152}}].

\bibitem{Chacaltana:2015bna}
O.~Chacaltana, J.~Distler, and A.~Trimm, {\it {Tinkertoys for the Twisted $E_6$
  Theory}},  \href{http://arxiv.org/abs/1501.00357}{{\tt arXiv:1501.00357}}.

\bibitem{Chacaltana:2017boe}
O.~Chacaltana, J.~Distler, A.~Trimm, and Y.~Zhu, {\it {Tinkertoys for the $E_7$
  Theory}},  {\em JHEP} {\bf 05} (2018) 031,
  [\href{http://arxiv.org/abs/1704.07890}{{\tt arXiv:1704.07890}}].

\bibitem{Chacaltana:2018zag}
O.~Chacaltana, J.~Distler, A.~Trimm, and Y.~Zhu, {\it {Tinkertoys for the $E_8$
  Theory}},  \href{http://arxiv.org/abs/1802.09626}{{\tt arXiv:1802.09626}}.

\bibitem{Aharony:2013hda}
O.~Aharony, N.~Seiberg, and Y.~Tachikawa, {\it {Reading between the lines of
  four-dimensional gauge theories}},  {\em JHEP} {\bf 08} (2013) 115,
  [\href{http://arxiv.org/abs/1305.0318}{{\tt arXiv:1305.0318}}].

\bibitem{Argyres:2016yzz}
P.~C. Argyres and M.~Martone, {\it {4d $ \mathcal{N} $ =2 theories with
  disconnected gauge groups}},  {\em JHEP} {\bf 03} (2017) 145,
  [\href{http://arxiv.org/abs/1611.08602}{{\tt arXiv:1611.08602}}].

\bibitem{Caorsi:2019vex}
M.~Caorsi and S.~Cecotti, {\it {Homological classification of 4d $ \mathcal{N}
  $ = 2 QFT. Rank-1 revisited}},  {\em JHEP} {\bf 10} (2019) 013,
  [\href{http://arxiv.org/abs/1906.03912}{{\tt arXiv:1906.03912}}].

\bibitem{Ferrara:1998zt}
S.~Ferrara, M.~Porrati, and A.~Zaffaroni, {\it {N=6 supergravity on AdS(5) and
  the SU(2,2/3) superconformal correspondence}},  {\em Lett. Math. Phys.} {\bf
  47} (1999) 255--263, [\href{http://arxiv.org/abs/hep-th/9810063}{{\tt
  hep-th/9810063}}].

\bibitem{Aharony:2015oyb}
O.~Aharony and M.~Evtikhiev, {\it {On four dimensional N = 3 superconformal
  theories}},  {\em JHEP} {\bf 04} (2016) 040,
  [\href{http://arxiv.org/abs/1512.03524}{{\tt arXiv:1512.03524}}].

\bibitem{Garcia-Etxebarria:2015wns}
I.~Garc{\'i}a-Etxebarria and D.~Regalado, {\it {$ \mathcal{N}=3 $ four
  dimensional field theories}},  {\em JHEP} {\bf 03} (2016) 083,
  [\href{http://arxiv.org/abs/1512.06434}{{\tt arXiv:1512.06434}}].

\bibitem{Aharony:2016kai}
O.~Aharony and Y.~Tachikawa, {\it {S-folds and 4d N=3 superconformal field
  theories}},  {\em JHEP} {\bf 06} (2016) 044,
  [\href{http://arxiv.org/abs/1602.08638}{{\tt arXiv:1602.08638}}].

\bibitem{Nishinaka:2016hbw}
T.~Nishinaka and Y.~Tachikawa, {\it {On 4d rank-one $ \mathcal{N}=3 $
  superconformal field theories}},  {\em JHEP} {\bf 09} (2016) 116,
  [\href{http://arxiv.org/abs/1602.01503}{{\tt arXiv:1602.01503}}].

\bibitem{Garcia-Etxebarria:2016erx}
I.~Garc{\'i}a-Etxebarria and D.~Regalado, {\it {Exceptional $ \mathcal{N}=3 $
  theories}},  {\em JHEP} {\bf 12} (2017) 042,
  [\href{http://arxiv.org/abs/1611.05769}{{\tt arXiv:1611.05769}}].

\bibitem{Lemos:2016xke}
M.~Lemos, P.~Liendo, C.~Meneghelli, and V.~Mitev, {\it {Bootstrapping
  $\mathcal{N}=3$ superconformal theories}},  {\em JHEP} {\bf 04} (2017) 032,
  [\href{http://arxiv.org/abs/1612.01536}{{\tt arXiv:1612.01536}}].

\bibitem{Bourton:2018jwb}
T.~Bourton, A.~Pini, and E.~Pomoni, {\it {4d $\mathcal{N}=3$ indices via
  discrete gauging}},  \href{http://arxiv.org/abs/1804.05396}{{\tt
  arXiv:1804.05396}}.

\bibitem{Bonetti:2018fqz}
F.~Bonetti, C.~Meneghelli, and L.~Rastelli, {\it {VOAs labelled by complex
  reflection groups and 4d SCFTs}},
  \href{http://arxiv.org/abs/1810.03612}{{\tt arXiv:1810.03612}}.

\bibitem{Cornagliotto:2017dup}
M.~Cornagliotto, M.~Lemos, and V.~Schomerus, {\it {Long Multiplet Bootstrap}},
  {\em JHEP} {\bf 10} (2017) 119, [\href{http://arxiv.org/abs/1702.05101}{{\tt
  arXiv:1702.05101}}].

\bibitem{Argyres:2019ngz}
P.~C. Argyres, A.~Bourget, and M.~Martone, {\it {Classification of all
  $\mathcal{N}\geq 3$ moduli space orbifold geometries at rank 2}},
  \href{http://arxiv.org/abs/1904.10969}{{\tt arXiv:1904.10969}}.

\bibitem{Seiberg:1996bd}
N.~Seiberg, {\it {Five-dimensional SUSY field theories, nontrivial fixed points
  and string dynamics}},  {\em Phys. Lett.} {\bf B388} (1996) 753--760,
  [\href{http://arxiv.org/abs/hep-th/9608111}{{\tt hep-th/9608111}}].

\bibitem{Seiberg:1996qx}
N.~Seiberg, {\it {Nontrivial fixed points of the renormalization group in
  six-dimensions}},  {\em Phys. Lett.} {\bf B390} (1997) 169--171,
  [\href{http://arxiv.org/abs/hep-th/9609161}{{\tt hep-th/9609161}}].

\bibitem{Intriligator:1997pq}
K.~A. Intriligator, D.~R. Morrison, and N.~Seiberg, {\it {Five-dimensional
  supersymmetric gauge theories and degenerations of Calabi-Yau spaces}},  {\em
  Nucl. Phys.} {\bf B497} (1997) 56--100,
  [\href{http://arxiv.org/abs/hep-th/9702198}{{\tt hep-th/9702198}}].

\bibitem{Hanany:1997gh}
A.~Hanany and A.~Zaffaroni, {\it {Branes and six-dimensional supersymmetric
  theories}},  {\em Nucl. Phys.} {\bf B529} (1998) 180--206,
  [\href{http://arxiv.org/abs/hep-th/9712145}{{\tt hep-th/9712145}}].

\bibitem{Gaiotto:2018yjh}
D.~Gaiotto, Z.~Komargodski, and J.~Wu, {\it {Curious Aspects of
  Three-Dimensional ${\cal N}=1$ SCFTs}},  {\em JHEP} {\bf 08} (2018) 004,
  [\href{http://arxiv.org/abs/1804.02018}{{\tt arXiv:1804.02018}}].

\bibitem{Baggio:2015vxa}
M.~Baggio, V.~Niarchos, and K.~Papadodimas, {\it {On exact correlation
  functions in SU(N) $ \mathcal{N}=2 $ superconformal QCD}},  {\em JHEP} {\bf
  11} (2015) 198, [\href{http://arxiv.org/abs/1508.03077}{{\tt
  arXiv:1508.03077}}].

\bibitem{Gerchkovitz:2016gxx}
E.~Gerchkovitz, J.~Gomis, N.~Ishtiaque, A.~Karasik, Z.~Komargodski, and S.~S.
  Pufu, {\it {Correlation Functions of Coulomb Branch Operators}},  {\em JHEP}
  {\bf 01} (2017) 103, [\href{http://arxiv.org/abs/1602.05971}{{\tt
  arXiv:1602.05971}}].

\bibitem{Karananas:2017zrg}
G.~K. Karananas and M.~Shaposhnikov, {\it {CFT data and spontaneously broken
  conformal invariance}},  {\em Phys. Rev.} {\bf D97} (2018), no.~4 045009,
  [\href{http://arxiv.org/abs/1708.02220}{{\tt arXiv:1708.02220}}].

\bibitem{Cordova:2016emh}
C.~Cordova, T.~T. Dumitrescu, and K.~Intriligator, {\it {Multiplets of
  Superconformal Symmetry in Diverse Dimensions}},
  \href{http://arxiv.org/abs/1612.00809}{{\tt arXiv:1612.00809}}.

\bibitem{Dolan:2002zh}
F.~A. Dolan and H.~Osborn, {\it {On short and semi-short representations for
  four-dimensional superconformal symmetry}},  {\em Annals Phys.} {\bf 307}
  (2003) 41--89, [\href{http://arxiv.org/abs/hep-th/0209056}{{\tt
  hep-th/0209056}}].

\bibitem{Seiberg:1994pq}
N.~Seiberg, {\it {Electric - magnetic duality in supersymmetric nonAbelian
  gauge theories}},  {\em Nucl. Phys.} {\bf B435} (1995) 129--146,
  [\href{http://arxiv.org/abs/hep-th/9411149}{{\tt hep-th/9411149}}].

\bibitem{Intriligator:1995id}
K.~A. Intriligator and N.~Seiberg, {\it {Duality, monopoles, dyons, confinement
  and oblique confinement in supersymmetric SO(N(c)) gauge theories}},  {\em
  Nucl. Phys.} {\bf B444} (1995) 125--160,
  [\href{http://arxiv.org/abs/hep-th/9503179}{{\tt hep-th/9503179}}].

\bibitem{Cachazo:2002ry}
F.~Cachazo, M.~R. Douglas, N.~Seiberg, and E.~Witten, {\it {Chiral rings and
  anomalies in supersymmetric gauge theory}},  {\em JHEP} {\bf 12} (2002) 071,
  [\href{http://arxiv.org/abs/hep-th/0211170}{{\tt hep-th/0211170}}].

\bibitem{Gadde:2011uv}
A.~Gadde, L.~Rastelli, S.~S. Razamat, and W.~Yan, {\it {Gauge Theories and
  Macdonald Polynomials}},  {\em Commun. Math. Phys.} {\bf 319} (2013)
  147--193, [\href{http://arxiv.org/abs/1110.3740}{{\tt arXiv:1110.3740}}].

\bibitem{Manenti:2019jds}
A.~Manenti, {\it {Differential operators for superconformal correlation
  functions}},  \href{http://arxiv.org/abs/1910.12869}{{\tt arXiv:1910.12869}}.

\bibitem{Wess:1992cp}
J.~Wess and J.~Bagger, {\em {Supersymmetry and supergravity}}.
\newblock Princeton University Press, Princeton, NJ, USA, 1992.

\bibitem{Weinberg:1980kq}
S.~Weinberg and E.~Witten, {\it {Limits on Massless Particles}},  {\em Phys.
  Lett.} {\bf 96B} (1980) 59--62.

\bibitem{Cordova:2016xhm}
C.~Cordova, T.~T. Dumitrescu, and K.~Intriligator, {\it {Deformations of
  Superconformal Theories}},  {\em JHEP} {\bf 11} (2016) 135,
  [\href{http://arxiv.org/abs/1602.01217}{{\tt arXiv:1602.01217}}].

\bibitem{Argyres:1996eh}
P.~C. Argyres, M.~R. Plesser, and N.~Seiberg, {\it {The Moduli space of vacua
  of N=2 SUSY QCD and duality in N=1 SUSY QCD}},  {\em Nucl. Phys.} {\bf B471}
  (1996) 159--194, [\href{http://arxiv.org/abs/hep-th/9603042}{{\tt
  hep-th/9603042}}].

\bibitem{hulek1998geometry}
K.~Hulek and G.~Sankaran, {\it The geometry of siegel modular varieties},  {\em
  arXiv preprint math/9810153} (1998).

\bibitem{Freed:1997dp}
D.~S. Freed, {\it {Special Kahler manifolds}},  {\em Commun. Math. Phys.} {\bf
  203} (1999) 31--52, [\href{http://arxiv.org/abs/hep-th/9712042}{{\tt
  hep-th/9712042}}].

\bibitem{LopesCardoso:2019mlj}
G.~Lopes~Cardoso and T.~Mohaupt, {\it {Special Geometry, Hessian Structures and
  Applications}},  \href{http://arxiv.org/abs/1909.06240}{{\tt
  arXiv:1909.06240}}.

\bibitem{Donagi:1995cf}
R.~Donagi and E.~Witten, {\it {Supersymmetric Yang-Mills theory and integrable
  systems}},  {\em Nucl. Phys.} {\bf B460} (1996) 299--334,
  [\href{http://arxiv.org/abs/hep-th/9510101}{{\tt hep-th/9510101}}].

\bibitem{joyce2000compact}
D.~D. Joyce, {\em Compact manifolds with special holonomy}.
\newblock Oxford University Press on Demand, 2000.

\bibitem{Shephard:1954}
G.~Shephard and J.~Todd, {\it
  {\href{https://cms.math.ca/10.4153/CJM-1954-028-3}{Finite unitary reflection
  groups}}},  {\em Canadian J. Math.} {\bf 6} (1954) 274.

\bibitem{Chevalley:1955}
C.~Chevalley, {\it
  {\href{http://www.jstor.org/stable/2372597?origin=crossref&seq=1#page_scan_tab_contents}{Invariants
  of finite groups generated by reflections}}},  {\em Amer. J. Math.} {\bf 77}
  (1955) 778--782.

\bibitem{kraft1981minimal}
H.~Kraft and C.~Procesi, {\it Minimal singularities in gln},  {\em Invent.
  math} {\bf 62} (1981), no.~3 503--515.

\bibitem{kraft1982geometry}
H.~Kraft and C.~Procesi, {\it On the geometry of conjugacy classes in classical
  groups},  {\em Commentarii Mathematici Helvetici} {\bf 57} (1982), no.~1
  539--602.

\bibitem{fu2017generic}
B.~Fu, D.~Juteau, P.~Levy, and E.~Sommers, {\it Generic singularities of
  nilpotent orbit closures},  {\em Advances in Mathematics} {\bf 305} (2017)
  1--77.

\bibitem{Cabrera:2016vvv}
S.~Cabrera and A.~Hanany, {\it {Branes and the Kraft-Procesi Transition}},
  {\em JHEP} {\bf 11} (2016) 175, [\href{http://arxiv.org/abs/1609.07798}{{\tt
  arXiv:1609.07798}}].

\bibitem{Cabrera:2017njm}
S.~Cabrera and A.~Hanany, {\it {Branes and the Kraft-Procesi transition:
  classical case}},  {\em JHEP} {\bf 04} (2018) 127,
  [\href{http://arxiv.org/abs/1711.02378}{{\tt arXiv:1711.02378}}].

\bibitem{brieskorn1970singular}
E.~Brieskorn, {\it Singular elements of semi-simple algebraic groups},  in {\em
  Actes du Congres International des Math{\'e}maticiens (Nice, 1970)}, vol.~2,
  pp.~279--284, 1970.

\bibitem{slodowy1980simple}
P.~Slodowy, {\it Simple singularities},  in {\em Simple Singularities and
  Simple Algebraic Groups}, pp.~70--102.
\newblock Springer, 1980.

\bibitem{beauville1999symplectic}
A.~Beauville, {\it Symplectic singularities},  {\em arXiv preprint
  math/9903070} (1999).

\bibitem{kaledin2006symplectic}
D.~Kaledin, {\it Symplectic singularities from the poisson point of view},
  {\em Journal f{\"u}r die reine und angewandte Mathematik (Crelles Journal)}
  {\bf 2006} (2006), no.~600 135--156.

\bibitem{Bourget:2019aer}
A.~Bourget, S.~Cabrera, J.~F. Grimminger, A.~Hanany, M.~Sperling, A.~Zajac, and
  Z.~Zhong, {\it {The Higgs Mechanism -- Hasse Diagrams for Symplectic
  Singularities}},  \href{http://arxiv.org/abs/1908.04245}{{\tt
  arXiv:1908.04245}}.

\bibitem{goresky1988stratified}
M.~Goresky and R.~MacPherson, {\it Stratified morse theory},  in {\em
  Stratified Morse Theory}, pp.~3--22.
\newblock Springer, 1988.

\bibitem{whitney1965tangents}
H.~Whitney, {\it Tangents to an analytic variety},  {\em Annals of mathematics}
  (1965) 496--549.

\end{thebibliography}
\providecommand{\href}[2]{#2}\begingroup\raggedright\endgroup

\end{document}